\documentclass[manuscript,authorversion]{acmart}
\usepackage{booktabs}
\usepackage{array}
\usepackage{graphicx}
\usepackage{subcaption}
\usepackage{acronym}
\usepackage[export]{adjustbox}
\usepackage{wrapfig}
\usepackage{multirow}
\usepackage{makecell}
\usepackage{amsmath}

\DeclareMathOperator{\rank}{RANK}
\DeclareMathOperator{\fom}{FoM}
\DeclareMathOperator{\cpb}{CpB}

\AtBeginDocument{
  \providecommand\BibTeX{{
    \normalfont B\kern-0.5em{\scshape i\kern-0.25em b}\kern-0.8em\TeX}}}

\setcopyright{acmcopyright}
\copyrightyear{2023}
\acmYear{2023}
\acmDOI{}

\begin{document}

\title{Systematic Review of Lightweight Cryptographic Algorithms}
\author{Mohsin Khan}
\email{mohsin.khan@uit.no}
\orcid{0000-0003-1815-8642}
\author{Elisavet Kozyri}
\email{elisavet.kozyri@uit.no}
\orcid{0000-0001-6890-4282}
\author{Håvard Dagenborg}
\email{havard.dagenborg@uit.no}
\orcid{0000-0002-1637-7262}
\affiliation{%
  \institution{\\UiT, The Arctic University of Norway}
  \streetaddress{Hansine Hansens veg 18, Tromsø}
  \city{Tromsø}
  \state{}
  \country{Norway}
  \postcode{9019}
}

\acrodef{IoT}{Internet of Thing}
\acrodef{WSN}{Wireless Sensor Network}
\acrodef{LWBC}{Lightweight Block Cipher}
\acrodef{LWSC}{Lightweight Stream Cipher}
\acrodef{LWECC}{Lightweight Elliptical Curve Cryptography}
\acrodef{LHC}{Lightweight Hybrid Cipher}
\acrodef{SPN}{Substitution Permutation Network}
\acrodef{FN}{Feistel Network}
\acrodef{GFN}{Generalized Feistel Network}
\acrodef{EGFN}{Extended GFN}
\acrodef{RFID}{Radio Frequency Identification}
\acrodef{NSA}{National Security Agency}
\acrodef{NIST}{National Institute of Standard and Technology}
\acrodef{MAC}{Message Authentication Code}
\acrodef{IACR}{International Association for Cryptologic Research}
\acrodef{ECRYPT}{European Network of Excellence for Cryptology}
\acrodef{LWAE}{Lightweight Authentic Encryption}
\acrodef{OLWSC}{Other Lightweight Stream Ciphers}
\acrodef{FoM}{Figure of Merit}
\acrodef{CpB}{Cycles per Byte}
\acrodef{GE}{Gate Equivalency}
\acrodef{FPGA}{Field Programmable Gate Array}
\acrodef{ASIC}{Application Specific Integrated Circuits}
\acrodef{LMD}{Lai-Massey Design}
\acrodef{ARX}{Addition Rotation XOR}
\acrodef{RISC}{Reduced Instruction Set Computer}
\acused{RICS}
\acrodef{OPF}{Optimal Prime Field}
\acrodef{LFSR}{Linear Feedback Shift Register}
\acrodef{NLFSR}{Non-linear Feedback Shift Register}
\acrodef{FCSR}{Feedback Carry Shift Register}

\acrodef{MITM}{Meet-in-the-middle}
\acrodef{SCA}{Side Channel Attack}
\acrodef{TMDTO}{Time-Memory-Data Tradeoff}

\begin{abstract}
The emergence of small computing devices and the integration of processing units into everyday objects has made lightweight cryptography an essential part of the security landscape. Conventional cryptographic algorithms such as AES, RSA, and DES are unsuitable for resource-constrained devices due to limited processing power, memory, and battery. This paper provides a systematic review of lightweight cryptographic algorithms and the appropriateness of different algorithms in different areas such as IoT, RFID, and wireless sensor networks. Using tabular analysis and graphical interpretation, we compare these algorithms in terms of performance, security, energy consumption, and implementation costs. An overview of the evolution of lightweight cryptography based on those design trade-offs is also provided. 
\end{abstract}



\maketitle

\section{Introduction}

Conventional cryptographic tools are often unsuitable for resource-limited devices, like \ac{IoT} sensors and RFID tags, due to their high CPU and memory requirements~\cite{THABIT2023100759}. With the release of the Simon and Speck lightweight block cipher families~\cite{beaulieu2015simon} in 2015, the \ac{NSA} acknowledged the need for more lightweight security tools to protect these devices from basic security concerns like data leakage, manipulation, and unauthorized access~\cite{papp2015embedded}. In 2017, the \ac{NIST} also recognized this necessity and started an initiative to seek lightweight cryptographic algorithms, assessing them by performance benchmarks~\cite{cryptographic_technology_group_2017}. 

This survey covers recent developments in lightweight cryptographic methods, which typically offer lower latency, execution time, memory footprints, and implementation complexity than conventional cryptographic methods, but at the cost of weaker security guarantees.
Though the systematization of lightweight cryptographic algorithms has been accomplished in the past, previous surveys and reviews have focused on only certain aspects or classes of ciphers. For instance,~\citet{hatzivasilis2018review},~\citet{sevin2021survey}, and~\citet{mohd2015survey} review \acp{LWBC} while~\citet{manifavas2016survey},~\citet{jassim2021survey}, and~\citet{noura2019lightweight} review \acp{LWSC}. \citet{dhanda2020lightweight} has reviewed 54 lightweight cryptographic algorithms keenly, but the elucidation of the security solution is limited to  \ac{IoT} devices. In addition to the security solution for IoT devices, we have included the lightweight algorithms that provide security solutions for \ac{WSN}, \ac{RFID}, embedded systems, and middlewares as well. A few other surveys and reviews evaluate a wide spectrum of ciphers in lightweight cryptography~\cite{bokhari2018comparative, sallam2018survey}. In addition to the wide spectrum, we include missed-out and recently published lightweight algorithms and fine-grained categorizations and explanations of lightweight cryptography.

This survey has performed a comparative study based on the summary, resilience against attacks, and proven vulnerabilities of lightweight algorithms falling in the same classification. The classification maintains a structural framework by considering the architecture of lightweight ciphers and chronological order based on their published year. Further evaluation and analysis are followed by comparing each algorithm's hardware and software implementation. The data has been collected in tabular forms for each classification and analyzed via graphs, clarifying the security analysis, performance, and implementation cost. Finally, the explanation of each algorithm is established on the application of \ac{IoT}, \ac{RFID}, \ac{WSN}, and middlewares that is distributed among the general categorizations of lightweight algorithms such as ultra-lightweight, low-cost lightweight, lightweight, and high-cost lightweight algorithms. We highlight the best-performing lightweight algorithms for both symmetric and asymmetric cipher classes based on hardware and software measurements.

\begin{figure} [t]
    \includegraphics[height=2.9cm, keepaspectratio]{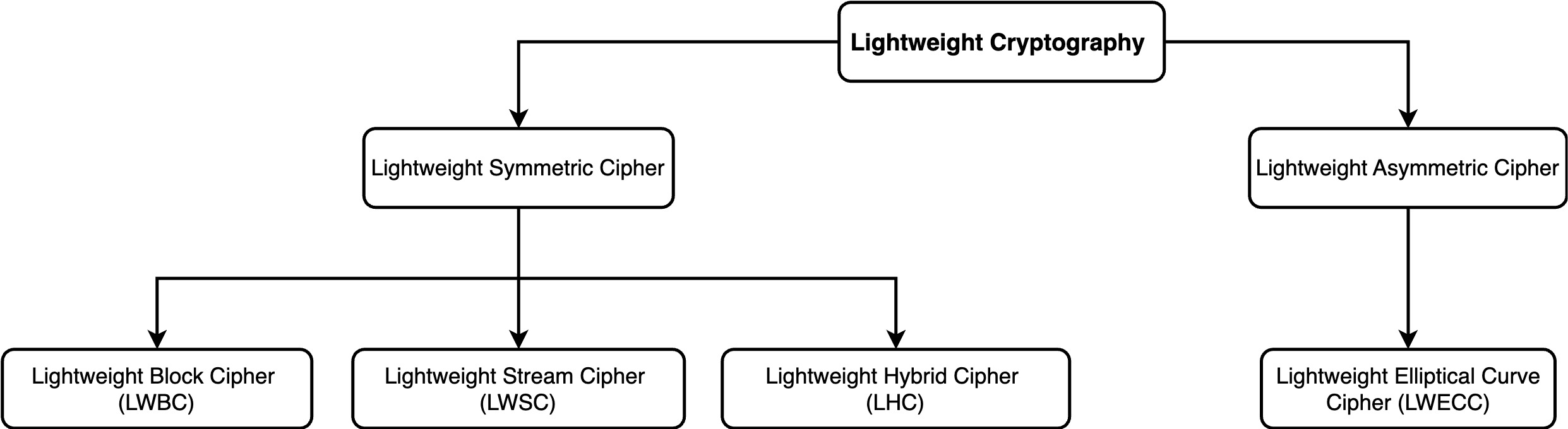}
   ~\caption{Classification of Lightweight Cryptography.}
    \Description {The branched classification of lightweight cryptography.}
    \label{fig:classification_LW_cryptography}
\end{figure}

\section{Classification of Lightweight Cryptography and Performance Metrics}

\begin{wrapfigure} {r} {0.37\textwidth}
    \includegraphics[width=0.93\linewidth, keepaspectratio]{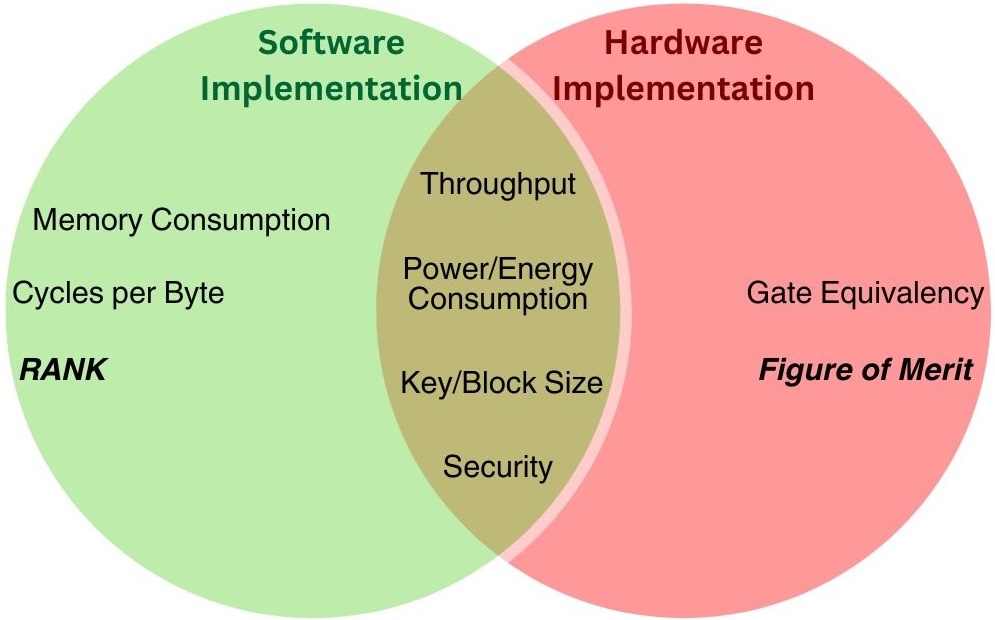}
   ~\caption{Venn Diagram Representing the Metrics Used in Software and Hardware Implementation.}
    \label{fig:venn_diagram}
    \Description{Venn Diagram representing the metrics used in software and hardware implementation.}
\end{wrapfigure}

The classification of lightweight cryptography is similar to that of traditional cryptography, as shown in Figure\ \ref{fig:classification_LW_cryptography}. The symmetric ciphers utilize a single key for both encryption and decryption processes, while the asymmetric ciphers employ two separate keys. In a hardware implementation, the cryptographic algorithm operates via specialized electronic circuitry or dedicated cryptographic hardware. For software implementations, the cryptographic algorithm is executed on a conventional general-purpose processor. The Venn diagram in Figure~\ref{fig:venn_diagram} illustrates qualitative and quantitative software and hardware implementation metrics. The following is a list of qualitative and quantitative metrics involved in hardware and software implementation:

\begin{itemize}

\item \emph{Throughput (Kbps)} is the measurement of the rate at which output is generated. So, it refers to the number of output bits over time. A higher throughput indicates faster cipher execution.

\item \emph{Power/Energy consumption ($\mu$J/$\mu$W)} is the amount of energy (joules) and power (watts) depleted for encryption and decryption. The lower power consumption and reduced energy usage lead to lower costs associated with executing the cipher.

\item \emph{Security} is a qualitative as well as quantitative metric. Qualitatively, it provides proven vulnerabilities and resistance against specific attacks for ciphers compiled in a comparative analysis table. Quantitatively, it is represented by block size and key size.

\begin{itemize}
    \item \emph{Block size (bits/bytes)} is the number of bits/bytes converted (encrypted/decrypted) by the cipher. Larger block sizes can improve security by increasing computational complexity and reducing patterns.

    \item \emph{Key size (bits/bytes)} is the length of the secret key in bits/bytes used to encrypt/decrypt the data. Larger key sizes can improve security by increasing the number of potential key combinations.
\end{itemize}

\end{itemize}

The following list includes metrics that are specifically related to the hardware implementation:

\begin{itemize}

\item \emph{\ac{GE}} is the unit of measurement for the complexity of digital circuits independent of the manufacturing method. GE is the physical area of a single NAND gate used in various CMOS technologies in lightweight cryptography~\cite{guria2021lightweight}. It is preferable to have a lower GE value as it is linked to reduced implementation costs. GE and area consumption are interchangeably used in this review.

\begin{table}[t]
\scriptsize
\caption{General Categorization of Lightweight Cryptographic Algorithms in Relation to the Lightweight Spectrum of \ac{GE} \cite{dhanda2020lightweight}.}
\label{tab:GE_table}
\begin{tabular}{l l c l}
\toprule
\textbf{Region} & \textbf{General Categorization} & \textbf{Gate Count}  & \textbf{Applications}    
\\ 
\midrule
Region-1 & Ultra-lightweight algorithms & $\leq1000$& Extremely Constraint Devices\\& \multicolumn{1}{l}{} & (RFID, Tiny IoT sensors and actuators, etc.) 
\\ 
\hline
Region-2 & Low-cost lightweight algorithms & 1000 - 2000 & Constraint Devices\\&  & (RFID, WSN, IoT Sensors, and actuators, etc.) 
\\ 
\hline
Region-3 & Lightweight algorithms & 2000 - 3000 & Limited Resource Devices\\&  & (Microcontrollers or embedded devices, etc.) 
\\ 
\hline
Region-4 & High-cost lightweight algorithms & $\geq3000$ & Moderate Resource Devices \\&  & (Embedded systems, middlewares, etc.) 
\\ 
\bottomrule
\end{tabular}
\end{table}

\item \emph{\ac{FoM}} is mathematically defined as nano bits per clock cycle per GE squared and represented in equation (\ref{eq:FoM}). The dynamic power in CMOS circuits is proportional to the area consumption. A higher \ac{FoM} value indicates increased effectiveness in terms of lightness, lower energy consumption, and diminished battery demand of the cipher~\cite{kim2020pipo, badel2010armadillo}.

\begin{equation} \label{eq:FoM} \fom =  \frac{ \text{throughput (nano bits)}}{ \text{clk} \times \text{GE}^2}  \end{equation}
\end{itemize}

The following list specifies the metrics that are exclusively associated with the software implementation:

\begin{itemize}

\item \emph{Memory consumption (bits)} is the amount of primary (RAM) and volatile memory (ROM) used by the cryptographic algorithm to encrypt or decrypt a specific amount of given data. Lower memory consumption is favored since it is associated with lower implementation costs. 

\item \emph{\ac{CpB}} is the number of clock cycles to process each byte of data used in a cryptographic algorithm. A higher \ac{CpB} implies that more clock cycles are required to process each byte, whereas a lower \ac{CpB} indicates that the system can quickly process data.

\item \emph{RANK} is similar to \ac{FoM} but uses \ac{CpB} in inverse instead of throughput~\cite{beaulieu2015simon}, as defined in equation (\ref{eq:RANK}). A high RANK indicates better overall software performance. 
\begin{equation} 
\label{eq:RANK} \rank =  \frac{10^6/ \cpb}{\text{ROM} + 2 \times \text{RAM}}  
\end{equation}
\end{itemize}

\section{Lightweight Symmetric Cipher}
The principle behind the lightweight symmetric cipher is identical to that of the regular symmetric cipher (the same key is used for both encryption and decryption), but the overhead is significantly reduced so that it works better on resource-constrained devices. There are several techniques that can be employed to create a lighter cryptographic algorithm. These include reducing the key size, block size, and number of rounds, implementing mathematical operations that require less computing resources, and minimizing the size of the internal state that holds intermediate values while the algorithm is being executed.

    \begin{figure}
        \centering
        \includegraphics[width=\linewidth, height=6cm, keepaspectratio]{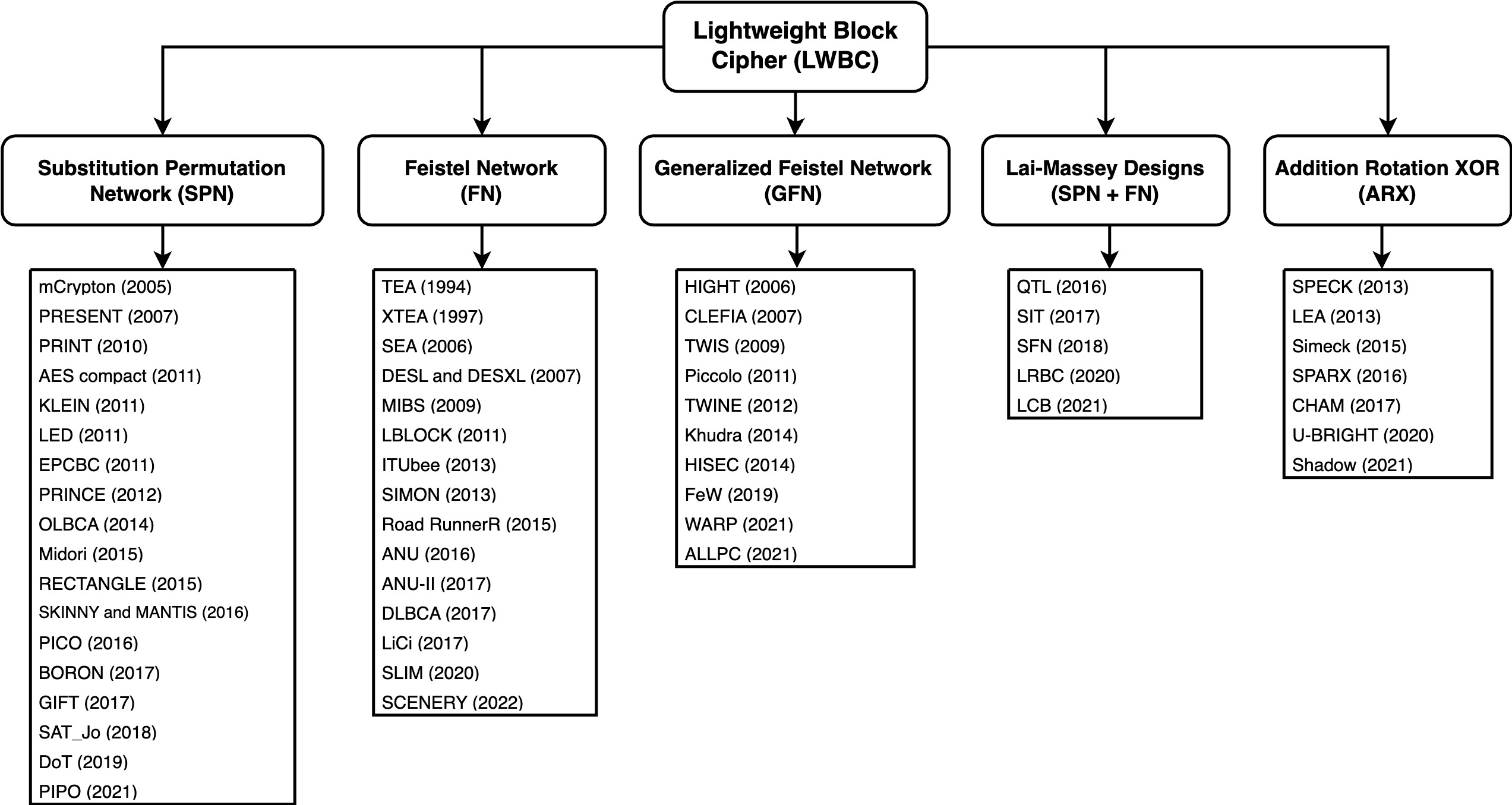}
       \caption{Classification of Lightweight Block Ciphers.}
        \Description{The Lightweight block ciphers distributed among different branches and associated ciphers}
        \label{fig:classification_LWBC_cryptography}
    \end{figure}
    
    \subsection{Lightweight Block Cipher (LWBC)}
    The block cipher splits the input into one or more blocks and encrypts each one using a symmetric key. Figure~\ref{fig:classification_LWBC_cryptography} shows the classification of \ac{LWBC} algorithms based on the architecture or core design techniques. Utilizing architecture-based classification allows for better veracity in the evaluation and examination of ciphers. By categorizing them based on their architectural design, it becomes easier to make more precise comparisons and conduct in-depth analyses.

        \subsubsection{\acf{SPN}}
        \acs{SPN} is a design framework used to create efficient cryptographic implementations of block ciphers that rely on multiple rounds of substitution (S-Boxes) and permutation (P-Boxes). Sufficient rounds of substitution and transposition are executed for enacting Claude Shannon's confusion and diffusion properties~\cite{shannon1949communication}. The confusion ensures that the key does not simply disclose the cipher text, and diffusion increases the repetition of the plaintext by disseminating it across rows and columns. Table~\ref{tab:CompAna_SPN} provides a comparison of lightweight \ac{SPN} ciphers based on the year of publication, security, and proven vulnerabilities.
        
        \begin{table}[t]
\renewcommand{\arraystretch}{1.4}
\scriptsize
\caption{Comparative Analysis of \ac{SPN} Block Ciphers.}
\label{tab:CompAna_SPN}
\begin{tabular}{l c l l}
\toprule
\textbf{Cipher} & \textbf{Year} & \textbf{Resistant Against Attacks}  & \textbf{Proven Vulnerabilities}
\\
\midrule
mCrypton~\cite{lim2005mcrypton}& 2005 & Linear; Differential & Related-key~\cite{park2009security}; \ac{MITM}~\cite{hao2015meet}
\\
\hline
PRESENT~\cite{bogdanov2007present} (ISO-29192) & 2007 & Differential; Algebraic; Integral~\cite{knudsen2002integral}; Related-key~\cite{biham1994new}; Slide~\cite{biryukov2000advanced} & Linear~\cite{cho2010linear}; Weak key~\cite{ohkuma2009weak}; Biclique~\cite{jeong2012biclique}; Saturation~\cite{collard2009statistical}
\\
\hline
PRINT~\cite{knudsen2010printcipher} & 2010 & Differential; Linear; Algebraic; Related-key; Statistical saturation & Invariant subspace~\cite{leander2011cryptanalysis}
\\
\hline
AES compact~\cite{moradi2011pushing} & 2011 & Differential Power Analysis (DPA)~\cite{kocher1999differential} & 
\\
\hline
KLEIN~\cite{gong2011klein} & 2011 & Linear; Differential; Algebraic; Integral; Side channel & Biclique~\cite{ahmadian2015biclique}; Truncated differential~\cite{lallemand2015cryptanalysis}
\\
\hline
LED~\cite{guo2011led} & 2011 & Differential; Linear; Algebraic; Cube; Slide; Integral; Rotational & \ac{MITM}~\cite{isobe2012security}; 
Biclique~\cite{jeong2012biclique}
\\
\hline
EPCBC~\cite{guo2011led}  & 2011 & Differential; Linear; Related-key; Integral; Slide; Algebraic & Weak keys~\cite{walter2013optimizing}
\\
\hline
PRINCE~\cite{borghoff2012prince} & 2012 & Differential; Linear; Slide; Biclique & Zero-correlation~\cite{cheng2017multidimensional}; Truncated differential~\cite{zhao2015truncated}
\\
\hline
OLBCA~\cite{aldabbagh2014olbca} & 2014 & Differential; Integral; Boomerang &
\\
\hline
Midori~\cite{banik2015midori}l & 2015 & Differential; Linear; Boomerang; \ac{MITM}; Algebraic; Slide & Related-key~\cite{gerault2016related}; Impossible differential~\cite{chen2017impossible}
\\
\hline
RECTANGLE~\cite{zhang2015rectangle} & 2015 & Differential; Linear; Statistical saturation; Integral; Slide; Related-key & Related-key
differential (19-r)~\cite{shan2014related}
\\
\hline
SKINNY~\cite{beierle2016skinny} & 2016 & Differential; Linear; \ac{MITM}; Integral; Slide; Algebraic & Zero-correlation; Related-tweakey differential~\cite{sadeghi2018cryptanalysis}
\\
\hline
MANTIS~\cite{beierle2016skinny} & 2016 & Related-tweak linear and differential; Invariant subspace &
\\
\hline
PICO~\cite{bansod2016pico} & 2016 & Differential; Biclique; Algebraic & Differential trails (21-r)~\cite{kumar2020optimal}; Key-recovery (11-r)~\cite{liu2020integral}
\\
\hline 
BORON~\cite{bansod2017boron} & 2017 & Differential; Zero-correlation; Biclique; Algebraic; Related-key; Slide &
\\
\hline
GIFT~\cite{banik2017gift} & 2017 & Differential; Linear; Integral; \ac{MITM}; Algebraic & Biclique~\cite{han2019unbalanced}; Related-key~\cite{cao2019related}
\\
\hline
SAT\_Jo~\cite{shantha2018sat_jo} & 2018 &  Differential; Linear; Algebraic~\cite{joshitta2019security} & Integral~\cite{qiu2021integral}
\\
\hline
DoT~\cite{patil2019dot} & 2019 & Linear; Biclique & Differential~\cite{kumar2022full}
\\
\hline
PIPO~\cite{kim2020pipo} & 2021 & Differential; Linear; Algebraic; Statistical; Invariant; Slide & Differential fault~\cite{lim2022differential}; Key recovery (8/10-r)~\cite{kim2022integral}
\\
\bottomrule
\multicolumn{4} {r@{}}{\textbf{r} represents rounds}
\end{tabular} 
\end{table}

        
          We aim to comprehensively explain the development of the lightweight \ac{SPN} ciphers based on the design modifications related to substitution, permutation, and key scheduling methods. \ac{NIST} declared the AES standard competition in 1997~\cite{nechvatal2001report}, and Crypton~\cite{lim1998crypton} was submitted but could not make it to the finalists. Later, mCrypton~\cite{lim2005mcrypton} was created by building the basic block of non-linear substitution, comprised of four 4-bit S-Boxes that reduce the block size. 
         
          The PRESENT cipher acquired recognition due to its compact hardware implementation. As a result, the cipher was ISO 29192 standardized~\cite{ISO29192}. The PRESENT's design achieved compactness using scan flip-flops and a single $4\times4$ bit S-Box, with a linear mixing layer for permutation. The scan flip-flops provide an efficient solution by combining a D flip-flop with a 2 to 1 multiplexer. The KLEIN~\cite{gong2011klein} cipher was developed to address the security weaknesses in PRESENT and the inefficiency of the software implementation. The cipher is designed with an involutive S-Box and non-linear permutation to save implementation costs for its inverse. GIFT~\cite{banik2017gift} is an improved version of PRESENT that addresses its weaknesses, such as linear hulls while reducing costs. The cost is reduced by using a simplified round function and a key schedule that involves bit permutation, designed in conjunction with the S-Box.
         
         AES serves as the key component for certain lightweight \ac{SPN} ciphers due to its mathematical efficiency and the broad spectrum of security features. The AES compact algorithm~\cite{moradi2011pushing} was developed utilizing scan flip-flops and the compact S-Box representation proposed by Canright~\cite{canright2005very}. The LED~\cite{guo2011led} cipher was designed based on AES, focusing on software implementation. The structure of the S-Boxes in LED is inherited from PRESENT, and the simplicity of the key schedule allows for a lower bound on the minimum number of S-Boxes. MANTIS~\cite{beierle2016skinny} is a cipher based on AES with a low latency and tweakable structure. It has a round function and tweak-scheduler designed for minimal rounds without compromising security.
         
         Some lightweight cryptographic algorithms were developed with specific devices or hardware in mind. For instance, the EPCBC cipher~\cite{guo2011led} was developed for Electronic Product Code scanning, using a $4\times4$ bit S-Box and an extension of PRESENT's P-Box for encryption. The PRINT cipher~\cite{knudsen2010printcipher} was developed for IC printing using an optimal 3-bit S-Box, where the output bits are permuted. 
         
        Specific \ac{SPN} lightweight ciphers have been developed with regard to certain design constraints, such as low latency and reduced energy consumption. The PRINCE cipher~\cite{borghoff2012prince} achieves low-latency cryptography by reusing the same S-Box 16 times instead of using different ones each time. The Midori cipher~\cite{banik2015midori} optimizes the cell permutation layers to enhance the diffusion speed and maximize the number of active S-Boxes per round, reducing energy consumption. The BORON cipher~\cite{bansod2017boron} applies a key schedule modeled after PRESENT alongside a non-linear S-Box layer that leads to an increased number of active S-Boxes and reduced energy consumption. The OLBCA cipher~\cite{aldabbagh2014olbca} improves overall performance through the incorporation of twelve 4-bit S-Boxes and optimized usage of active S-Boxes.
          
        Recent work on lightweight \ac{SPN} ciphers has focused on efficient implementation in both software and hardware. For instance, RECTANGLE~\cite{zhang2015rectangle} uses the bit-slice technique to implement S-Box with 12 basic logical instructions and a P-layer with three rotations. The bit-slice technique expresses a mathematical function using basic logical operations applied individually and in parallel to each bit of the input data. The PICO cipher~\cite{bansod2016pico} uses a compact key scheduling method inspired by SPECK and does not include any non-linear layers in its design, reducing the complexity of the cipher. SAT\_Jo~\cite{shantha2018sat_jo} was analyzed for security in Healthcare \ac{IoT}, but the cipher was designed for a wide range of applicability. The cipher implements a $4\times4$ bit S-Box over a Galois field of order $2^4$ and is designed to reduce the area through low overhead for multiple functions. The DoT cipher~\cite{patil2019dot} applies bit permutation, block shuffling, and circular shifting. The 8-bit permutation layer with a high diffusion mechanism results in a minimum flash memory requirement. 
         
        In addition to software and hardware implementation, the new generation of \ac{SPN} ciphers aims to provide resistance against side-channel attacks. SKINNY~\cite{beierle2016skinny} was designed using a compact S-Box, a new sparse diffusion layer, and a lightweight key schedule. The systematic organization of states by SKINNY facilitates the efficient computation of bounds on active S-Boxes and resistance against side-channel attacks. The PIPO cipher~\cite{kim2020pipo} was developed to provide an optimum lightweight cipher with higher-order masking implementations. The cipher employs a small number of non-linear operations to create an 8-bit S-Box for achieving effective higher-order masking. 


        \begin{wrapfigure} {r} {0.6\textwidth}
            \includegraphics[width=1\linewidth, height=5cm, keepaspectratio]{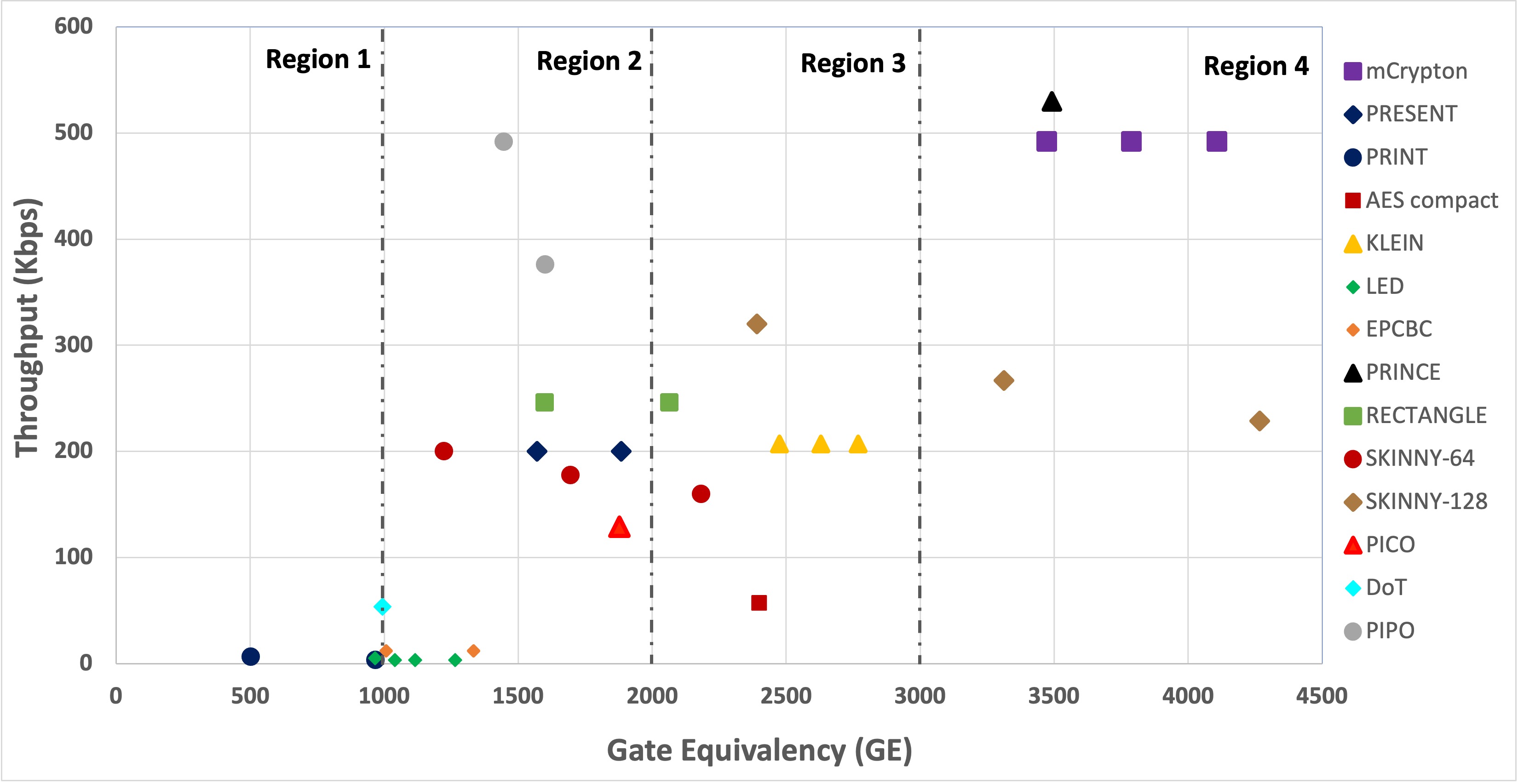}
           ~\caption{Throughput vs Gate Equivalency (GE) of SPN Block Ciphers.}
            \label{fig:Throughput_GE_SPN}
            \Description{The throughput and GE of SPN block ciphers.}
        \end{wrapfigure}  

        The hardware implementation represented in Figure~\ref{fig:Throughput_GE_SPN} offers the comparison and general performance evaluation of the \ac{SPN} block ciphers regarding the area consumption and throughput. The pertinent hardware implementation data manifests that mCrypton \cite{lim2005mcrypton} and PRINCE \cite{borghoff2012prince} provide the highest throughput of 492.3, and 529.9 Kbps, respectively (see Table~\ref{tab:HW_imp_SPN} in appendix for detailed measurements). Meanwhile, their comparable area consumption is higher. The PRINT cipher is among the most hardware and energy-efficient. Our analysis revealed that PRINT cipher and EPCBC have lower throughput than other ciphers, but this lower throughput is appropriate for the hardware or applications they were designed for. Among hardware and software-designed ciphers, RECTANGLE, SKINNY, PICO, DoT, and PIPO have a better balance of area consumption and throughput than KLEIN and LED, which have large area consumption and low throughput, respectively. PIPO achieves an optimum balance between area consumption and throughput.
        
        \begin{figure} [t]
            \includegraphics[width=\textwidth, height=5.4cm, keepaspectratio]{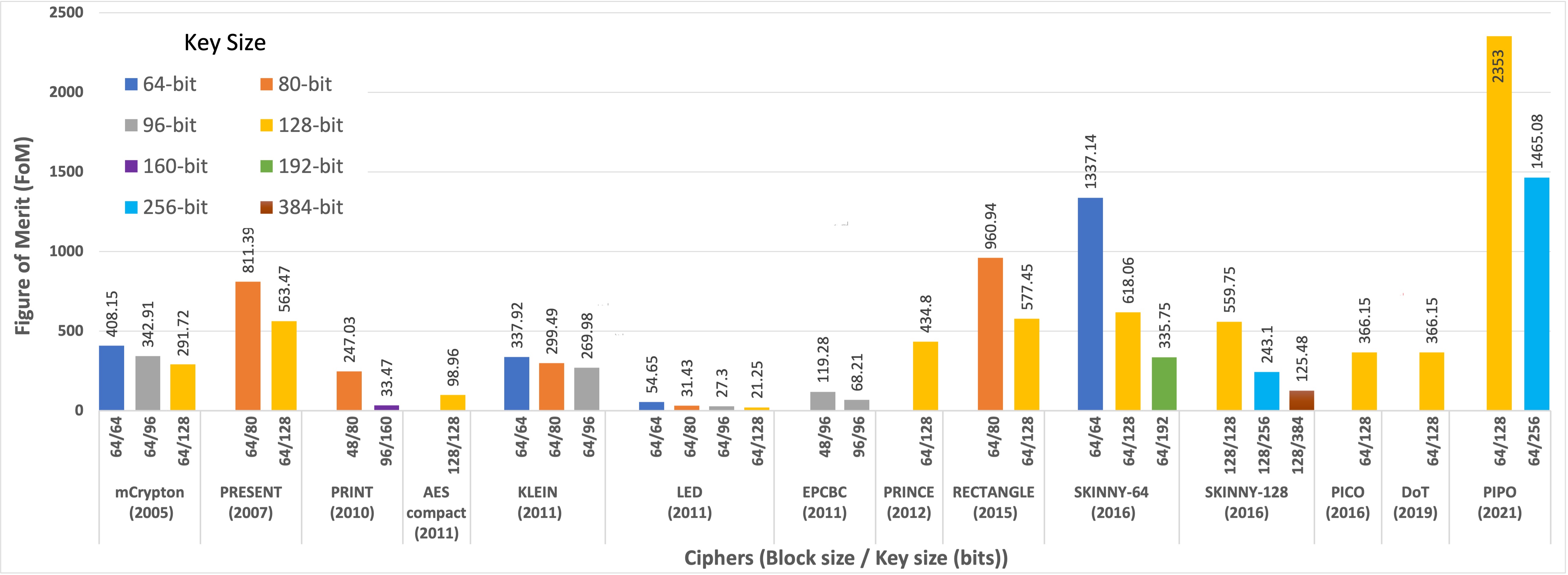}
           ~\caption{Figure of Merit (FoM) of SPN Block Ciphers.}
            \label{fig:FoM_SPN}
            \Description{Figure of Merit (FoM) of SPN block ciphers}
        \end{figure}
        
        Figure \ref{fig:Throughput_GE_SPN} also compares the overall performance of the \ac{SPN} ciphers relative to the suitability of these ciphers to their applications. The applications' suitability across different GE spectrums is outlined in Table \ref{tab:GE_table}. The ciphers such as PRINT, DoT, and the 64-bit variant of LED in region~1 are ultra-lightweight ciphers. Among those ciphers, DoT is better suited for \ac{IoT}, while the rest are better suited for RFID or WSN. In region~2 of the graph, most ciphers are suitable for \ac{IoT} and RFID, such as PRESENT, RECTANGLE, PICO, and PIPO. Thus, those ciphers lie in the general categorization of low-cost lightweight ciphers. The third region includes ciphers that use over 2000 GE, such as KLEIN, AES compact, SKINNY-128, RECTANGLE, and the 192-bit variant of SKINNY-64. The ciphers in region~4, such as mCrypton, PRINCE, SKINNY-128 (256/384-bit), would have been suitable for the extreme resource constraint devices while they were published, but nowadays, the size of the embedded devices and sensors are very small. The ciphers are appropriate for devices with moderate resources, like embedded systems and middleware.
        
        Figure~\ref{fig:FoM_SPN} illustrates the overall hardware performance of \ac{SPN} ciphers through \ac{FoM}. The \ac{FoM} of PRESENT is the highest among the \ac{SPN} ciphers developed before PRINCE, with LED having the lowest FoM. The ciphers developed after PRINCE, such as RECTANGLE and SKINNY, show better overall hardware performance than PRESENT. PIPO shows the best hardware performance by a significant margin.
        
        \begin{wrapfigure} {r} {0.45\textwidth}
            \includegraphics[width=1\linewidth, height=5.4cm]{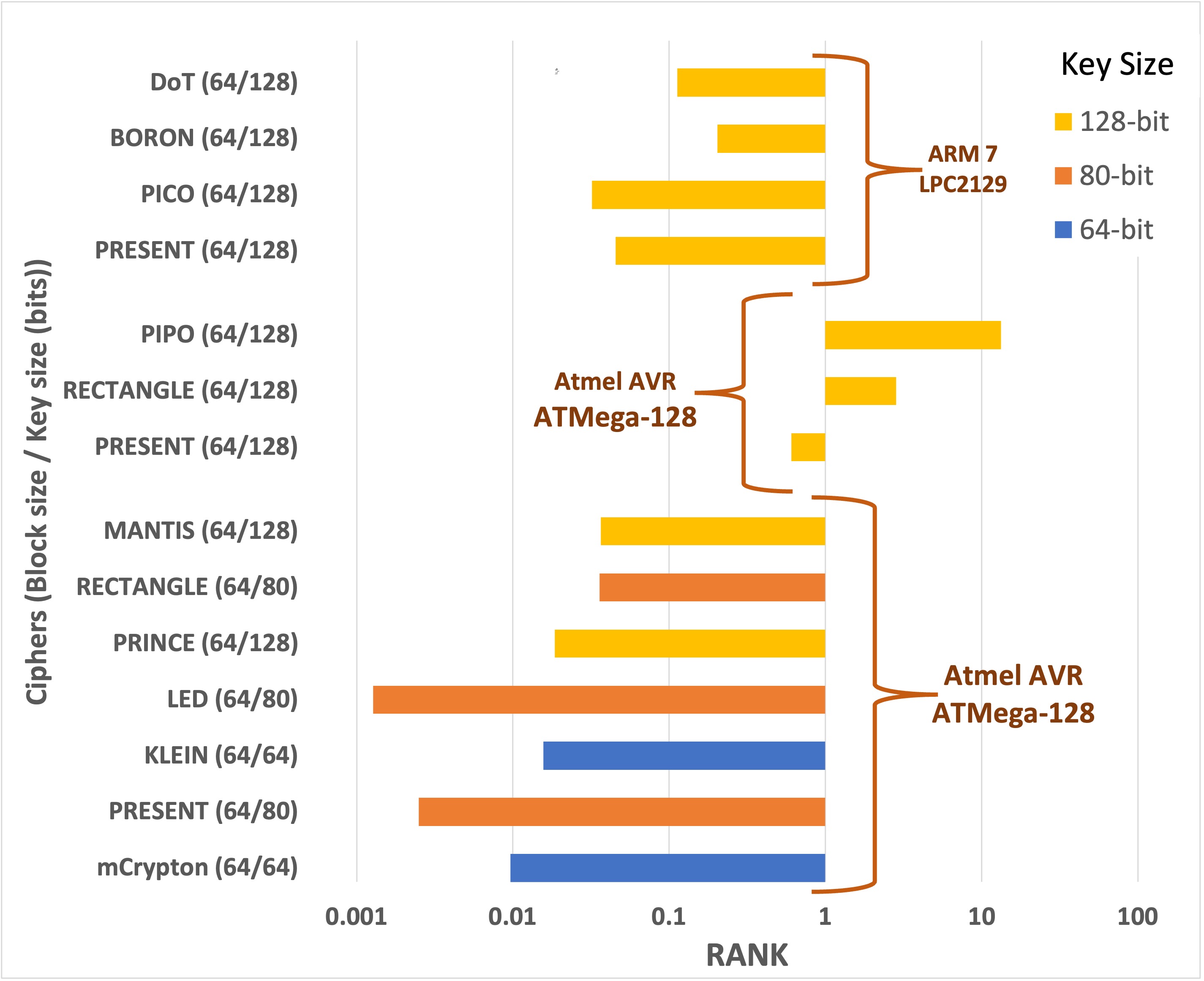}
           ~\caption{RANK of SPN Block Ciphers.}
            \label{fig:RANK_SPN}
            \Description{Calculated RANK of SPN block ciphers.}
        \end{wrapfigure}        
        The software performance of \ac{SPN} block ciphers is examined in this study through three distinct implementations, with each implementation executing the PRESENT cipher to enable a fair and equitable comparison. The RANK of \ac{SPN} ciphers is shown in a logarithmic bar graph in Figure~\ref{fig:RANK_SPN}. The first two implementations have been executed on AVR ATMega-128~\cite{sevin2021survey, kim2020pipo} running at 8 MHz with a difference in RAM and code size of the algorithms, and the third implementation has been performed on ARM7-LPC2129~\cite{bansod2016pico, bansod2017boron, patil2019dot}. In the first implementation, MANTIS has the best RANK due to the balance in code size, RAM, and \ac{CpB} (see Appendix Table~\ref{tab:SW_imp_SPN}). RECTANGLE has less \ac{CpB} than MANTIS, but the code size and RAM are approximately double. Also, the throughput and energy consumption are best provided by RECTANGLE. In the second implementation, PIPO has the highest RANK. The \ac{CpB}, code size, and RAM of PIPO is considerably less than PRESENT and RECTANGLE. In the third implementation, PICO has the lowest RANK, even less than the PRESENT, while BORON provides the best RANK. After evaluating all three software implementations, it is evident that PIPO has superior software performance. This is demonstrated by comparing the RANK margin between PIPO and PRESENT, as well as the PRESENT and other considered \ac{SPN} ciphers.

        \begin{table}[t]
\renewcommand{\arraystretch}{1.5}
\scriptsize
\caption{Comparative Analysis of \ac{FN} Block Ciphers.}
\label{tab:CompAna_FN}
\begin{tabular}{l c l l}
\toprule
\textbf{Cipher} & \textbf{Year} & \textbf{Resistant Against Attacks}  & \textbf{Proven Vulnerabilities}
\\
\midrule
TEA \cite{wheeler1994tea} & 1994 & Differential &  Related-key; Key-recovery~\cite{bogdanov2012zero}; Truncated differential~\cite{hong2004differential}
\\ 
\hline
XTEA \cite{needham1997tea} & 1997 &  Related-key & Key-recovery~\cite{bogdanov2012zero}; Truncated differential~\cite{hong2004differential}
\\ 
\hline
SEA \cite{standaert2006sea} & 2006 & Linear; Differential; Related-key; Algebraic &
\\ 
\hline
DESL/DESXL \cite{leander2007new} & 2007 & Differential; Algebraic; Davis Murphy \cite{davies1995pairs} &
\\
\hline
MIBS \cite{izadi2009mibs} & 2009 & Differential; Linear; Algebraic; Related-key & Biclique~\cite{faghihi2016biclique}
\\ 
\hline
LBLOCK \cite{wu2011lblock} & 2011 & Differential; Linear; Related-key &  Biclique~\cite{wang2012security}; Cube~\cite{li2013cube}
\\ 
\hline
ITUbee \cite{karakocc2013itubee} & 2013 & Differential; Linear; \ac{MITM}; Related-key; Impossible differential & Differential fault~\cite{fu2016differential}
\\ 
\hline
SIMON \cite{cryptoeprint:2013:404} & 2013 & Algebraic~\cite{raddum2015algebraic} & Power analysis~\cite{yoshikawa2016power}; Differential~\cite{biryukov2015differential}
\\ 
\hline
RoadRunneR \cite{baysal2015roadrunner} & 2015 & Differential; Linear; \ac{MITM}; Side channel & Truncated differentials (5/7-r)~\cite{yang2016truncated}
\\ 
\hline
ANU-I \cite{bansod2016anu} & 2016 & Linear; Differential; \ac{MITM}; Algebraic; Key collision; Biclique &  Related-key~\cite{sasaki2018related}, Optimal differential trials (7-r)~\cite{kumar2020optimal}
\\ 
\hline
ANU-II \cite{dahiphale2017anu} & 2017 & Linear; \ac{MITM}; Algebraic; Key collision, Biclique & Differential~\cite{fan2022differential}
\\ 
\hline
DLBCA \cite{aldabbagh2017design} & 2017 & Differential; Boomerang &
\\ 
\hline
LiCi \cite{patil2017lici} & 2017 & Linear; Differential; Zero correlation; Biclique &
\\ 
\hline
SLIM \cite{aboushosha2020slim} & 2020 & Linear; Differential & Optimal
differential trails~\cite{chan2023differential} 
\\ 
\hline
SCENERY \cite{feng2022scenery} & 2022 & Linear; Differential; Sliding; Related-key \\
\bottomrule
\multicolumn{4} {r@{}}{\textbf{r} represents rounds}
\end{tabular} 
\end{table}
        \subsubsection{Feistel Network (FN)}
        
        In 1973, Horst Feistel and Don Coppersmith proposed the \ac{FN} structure, where each input block is divided into two parts. In each iteration, one part of the block applies the same functions or complex mathematical operations (substitution and permutation), and the resultant output is combined with the other part of the block. The process is repeated for a certain number of iterations. The decryption process utilizes the same circuitry as encryption with a reverse of the subkeys. Table~\ref{tab:CompAna_FN} outlines the strengths and vulnerabilities of lightweight \ac{FN} ciphers based on published cryptanalysis results.


        We discuss the design techniques used in developing \ac{FN} ciphers, including substitution, permutation, key schedule, and round functions. In 1994, the TEA cipher~\cite{wheeler1994tea} was developed in Cambridge computer laboratories using a simple key schedule, minimal cryptographic operations, and compact code representation. TEA was, however, found vulnerable to key-related attacks~\cite{ kelsey1997related}. The XTEA cipher~\cite{needham1997tea} was developed to address these vulnerabilities by increasing the computation complexity and revising the round function.
        
        Several ciphers were developed with an emphasis on facilitating software implementation. For instance, the SEA cipher~\cite{standaert2006sea} provides parameterized block and key sizes, which enable adaptation of the cipher to specific hardware and security requirements. SEA focuses on low-cost encryption for processors with limited instructions by utilizing minimal elementary operations. The ITUbee cipher~\cite{karakocc2013itubee} offers energy-efficient implementation by eliminating the need for key scheduling. The cipher utilizes 8-bit S-Boxes and performs simple table lookup operations.
        
        Certain cryptographic algorithms were designed for hardware implementation. The SIMON cipher~\cite{cryptoeprint:2013:404} employs feistel permutations in its round functions and a nonlinear function that requires minimal hardware. New lightweight cipher variants of DES were developed, known as DESL and DESXL~\cite{leander2007new}, which use a serialized efficient hardware architecture and only one S-Box. The MIBS cipher uses an \ac{SPN} round function and a $4\times4$-bit S-Box to enhance hardware efficiency. The DLBCA cipher was focused on providing better resistance against differential and boomerang attacks in constrained environments. The cipher uses eight 4-bit S-Boxes to enable lightweight hardware implementation.

        Advancements in cryptography have led to the creation of lightweight \ac{FN} ciphers optimized for both software and hardware implementation, such as the LBLOCK~\cite{wu2011lblock} and the SLIM~\cite{aboushosha2020slim} ciphers. The implementation cost of the LBLOCK cipher is reduced by minimizing S-Box size and the number of S-Boxes per round. The SLIM cipher achieves lightweightness through elementary internal operations and the use of a compact 4-bit S-Box four times in parallel.

        The bit-slice technique can be used to achieve both hardware and software implementations. The RoadRunneR cipher~\cite{baysal2015roadrunner} can be implemented with a small number of bitwise operations due to its efficient linear function and bit-slice S-Box. Similarly, the SCENERY cipher~\cite{feng2022scenery} also utilizes the bit-slice technique and minimizes implementation costs by using eight $4\times4$ bit S-Boxes in parallel and a single component linear layer.
        
       In addition to being compact, some ciphers prioritize power efficiency in both their hardware and software implementations such as ANU-I~\cite{bansod2016anu}, ANU-II~\cite{dahiphale2017anu}, and LiCi~\cite{patil2017lici}. ANU uses a $4\times4$ bit S-Box and a strong permutation layer to minimize the area consumption. ANU-II was created to improve performance and resistance to the weaknesses found in ANU-I~\cite{ sasaki2018related}. ANU-II improves efficiency by avoiding multiple S-Boxes and by introducing shift operators to activate a maximum number of S-Boxes in minimal rounds. The LiCi cipher uses a compact and energy-efficient S-Box design, along with a strategic utilization of shift operators.

         \begin{wrapfigure} {r} {0.5\textwidth}
            \includegraphics[width=1\linewidth, height=5cm, keepaspectratio]{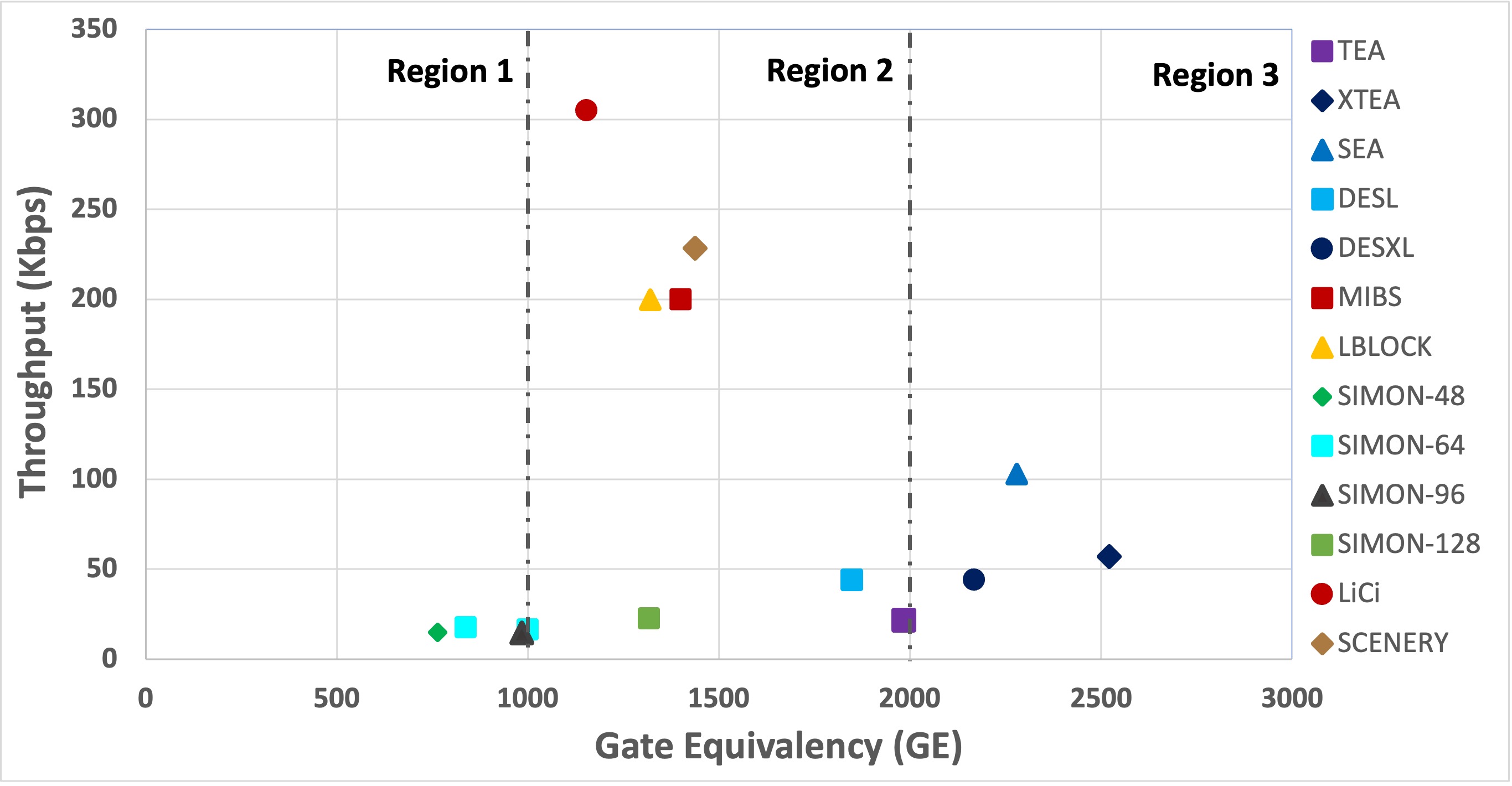}
           ~\caption{Throughput vs Gate Equivalency (GE) of \ac{FN} Block Ciphers.}
            \label{fig:Throughput_GE_FN}
            \Description{The throughput and Gate Equivalency of \ac{FN} block ciphers.}
        \end{wrapfigure}
        
        The hardware implementation of the earlier generation of \ac{FN} ciphers, such as TEA, XTEA, SEA, DESL, and DESXL, as seen in Figure~\ref{fig:Throughput_GE_FN}, has an area consumption in the range of 1800 to 2500 GE. Among these, SEA provides the highest throughput of 103 Kbps (see Appendix Table~\ref{tab:HW_imp_FN} for detailed measurements). The ciphers from the last few years, such as LiCi and SCENERY, have not only tried to minimize the area consumption but also to maximize the throughput in comparison to the other \ac{FN} ciphers. When comparing \ac{FN} ciphers with \ac{SPN} ciphers, we find that \ac{FN} ciphers provide an overall lower area consumption, while \ac{SPN} ciphers provide better throughput.


        Figure~\ref{fig:Throughput_GE_FN} also indicates application areas of the different ciphers depending on their throughput and GE. Region~1 of the figure includes the ultra-lightweight SIMON family of ciphers, which are suitable only for \ac{RFID} and \ac{WSN}-like applications due to their low throughput. Region~2 includes LiCi and SCENERY, which are more suitable for \ac{IoT}, while MIBS and LBLOCK are better suited for \ac{RFID}. TEA and DESL are outdated for current resource constraint devices, but they serve as the foundation for current research on lightweight \ac{FN} ciphers. Region~3 includes XTEA, SEA, and DESXL. These are older ciphers developed for applications such as passive \ac{RFID} and \ac{WSN}.
        
        \begin{wrapfigure} {r} {0.6\textwidth}
            \includegraphics[width=1\linewidth, height=7cm, keepaspectratio]{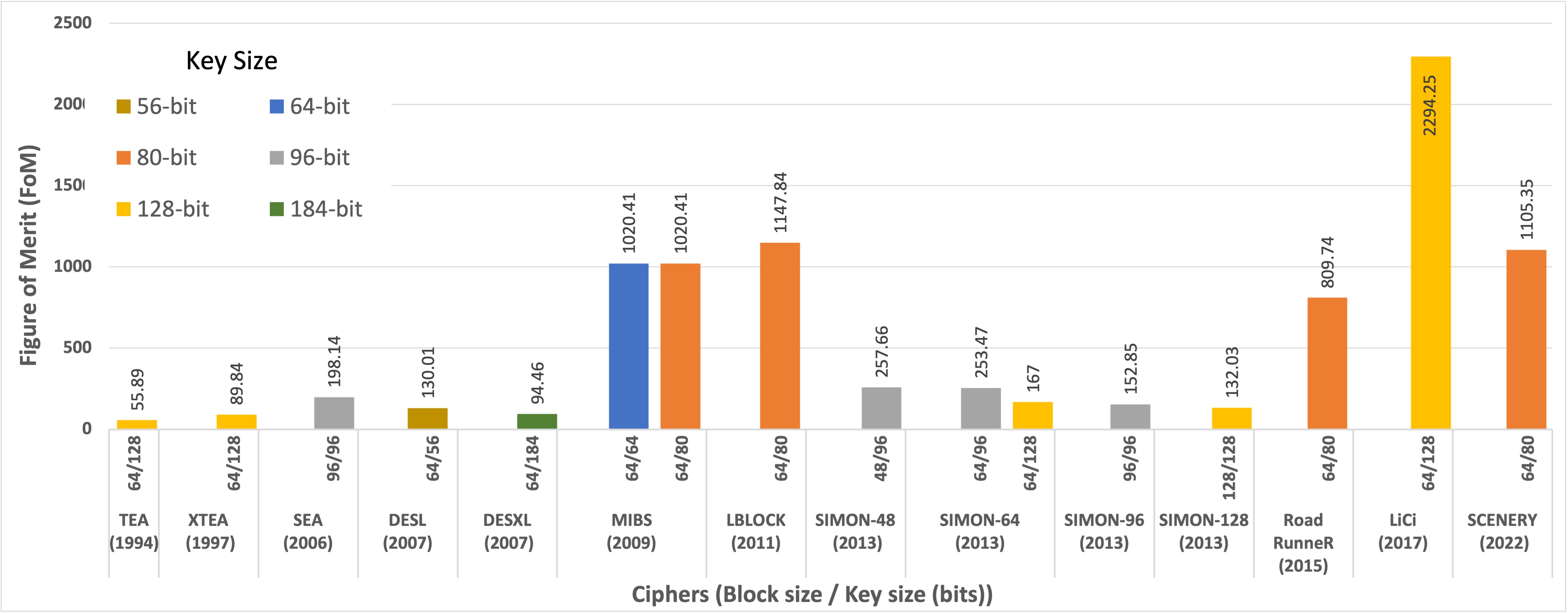}
           ~\caption{Figure of Merit (FoM) of \ac{FN} Block Ciphers.}
            \label{fig:FoM_FN}
            \Description{Calculated Figure of Merit (FoM) of FN block ciphers}
        \end{wrapfigure}
         Analysis of Figure~\ref{fig:FoM_FN} illustrates that the TEA, XTEA, SEA, DESL, and DESXL ciphers have the lowest \ac{FoM} among the \ac{FN} ciphers. The \ac{FN} ciphers developed after 2007 such as LBLOCK and MIBS, provide better \ac{FoM} scores. The Simon family of ciphers has a subpar overall hardware performance. The best \ac{FoM} is provided by LiCi, LBlock, and SCENERY in descending order.\\
         
\begin{wrapfigure} {r} {0.55\textwidth}
            \includegraphics[width=1\linewidth, height=7cm, keepaspectratio]{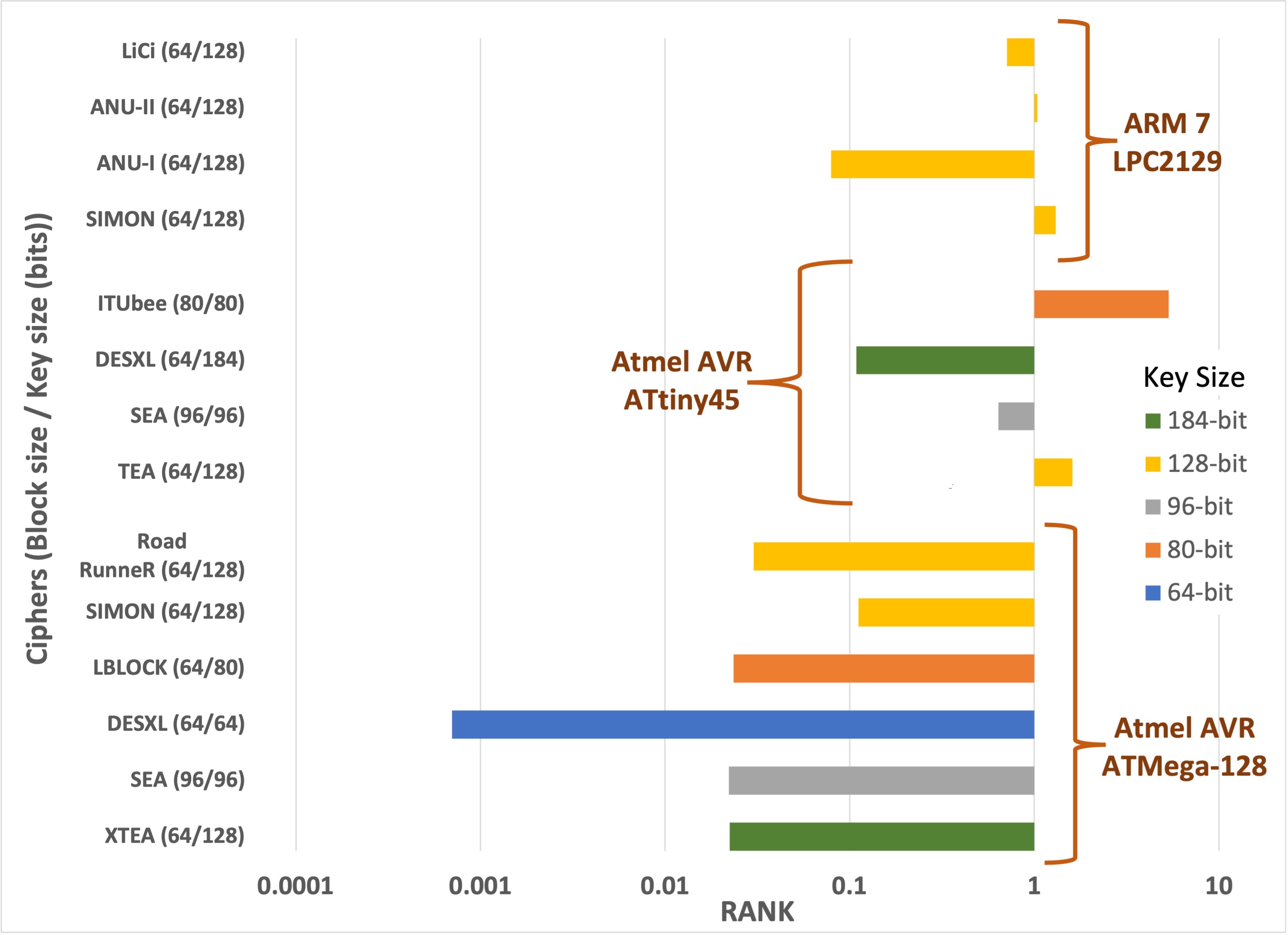}
           ~\caption{RANK of \ac{FN} Block Ciphers.}
            \label{fig:RANK_FN}
            \Description{Calculated RANK of FN block ciphers.}
        \end{wrapfigure}

        
        Figure~\ref{fig:RANK_FN} compares the RANK of the various \ac{FN}-based block ciphers grouped by the three distinct microcontroller platforms. They are implemented and evaluated on: ARM~7-LPC2119~\cite{patil2017lici, dahiphale2017anu}, Atmel AVR ATtiny45~\cite{karakocc2013itubee}, and Atmel AVR ATMega128~\cite{sevin2021survey}. The SIMON and DESXL ciphers are implemented on multiple platforms and included multiple times in the figure for comparison across three distinct microcontroller platforms.  
        
        For the ARM~7-LPC2119 microcontroller, SIMON has the best overall software performance due to its balance between the lowest number of cycles, optimum code size, and RAM requirement (see Appendix Table~\ref{tab:SW_imp_FN} for detailed measurements). DESXL has the highest number of \ac{CpB} as well as code size and thus the lowest RANK. For the Atmel ATtiny45 microcontroller, ITUbee has the highest RANK, while DESXL has the lowest. This shows that ITUbee performs better in software than DESXL, SEA, and TEA. On the Atmel ATMega-128 microcontroller, SIMON has the best RANK and best throughput. We observe a significant difference between ANU-I and ANU-II in RANK and throughput. ANU-II demonstrates better RANK than LiCi as well. After evaluating all three implementations, it can be inferred that DESXL has the least optimal software performance, while SIMON and ITUbee exhibit the most efficient software performance among \ac{FN} block ciphers.

        \subsubsection {\acf{GFN}}
        The traditional \ac{FN} splits the plaintext into 2$\mathit{n}$-bit parts applying a function to half of the split part before adding it to the other half, whereas the \ac{GFN} splits the plaintext into greater than or equal to 2$\mathit{n}$-bit sub-blocks and potentially applies different functions to each sub-block. Table~\ref{tab:CompAna_GFN} enumerates the security capabilities and limitations of \ac{GFN} ciphers. 

        \begin{table}[t]
\renewcommand{\arraystretch}{1.5}
\scriptsize
\caption{Comparative Analysis of \ac{GFN} Block Ciphers.}
\label{tab:CompAna_GFN}
\begin{tabular}{l c l l}
\toprule
\textbf{Cipher} & \textbf{Year} & \textbf{Resistant Against Attacks}  & \textbf{Proven Vulnerabilities}
\\
\midrule
HIGHT \cite{hong2006hight}(ISO-1803303) & 2006 & Differential; Linear; Algebraic, Slide, Saturation, Boomerang & Impossible differential \cite{lu2007cryptanalysis}; Related key \cite{ozen2009lightweight}; Biclique\cite{hong2011biclique}
\\ 
\hline
CLEFIA \cite{shirai2007128}(ISO-29192) & 2007 & Differential; Linear; Algebraic; Saturation Related-key & Impossible differential~\cite{tsunoo2008impossible}; Zero-correlation \cite{bogdanov2014zero}
\\ 
\hline
TWIS \cite{ojha2009twis} & 2009 & &  Differential~\cite{su2011full}; Key recovery~\cite{koccak2012cryptanalysis}
\\ 
\hline
Piccolo \cite{shibutani2011piccolo} & 2011 & Differential; Linear; Algebraic; Boomerang; Slide; \ac{MITM} & Biclique \cite{wang2012biclique}
\\ 
\hline
TWINE \cite{suzaki2012textnormal} & 2012 & Differential; Linear; Related-key; Slide; Saturation & Impossible differential~\cite{wei2019related}; Biclique~\cite{ccoban2012biclique} 
\\ 
\hline
Khudra \cite{kolay2014khudra} & 2014 & Linear; Differential; Boomerang; Side; Related-key & \ac{MITM}; Biclique~\cite{zheng2019security}
\\ 
\hline
HISEC \cite{aldabbagh2014hisec} & 2014 & Differential; Boomerang; Integral
\\ 
\hline
FeW \cite{kumar2014few} & 2019 & Differential; Linear; Impossible differential; Related-key &
\\ 
\hline
WARP \cite{banik2020warp} & 2021 & Differential; Linear; \ac{MITM}; Invariant subspace & Boomerang; Related-key~\cite{teh2022differential}
\\ 
\hline
ALLPC \cite{cheng2021allpc} & 2021 & Linear; Differential; Zero-correlation & Distinguishing~\cite{cheng2021allpc}\\ 
\bottomrule
\end{tabular} 
\end{table}
        
        We will describe the progression of \ac{GFN} ciphers based on their design modifications for proving compact implementation. Early work on \ac{GFN} ciphers include the HIGHT~\cite{hong2006hight} cipher, which was presented at the 2006 Workshop on Cryptographic Hardware and Embedded Systems (CHES 2006). The cipher achieves low-cost hardware implementation by using a lightweight round function and generating subkeys on the fly. The TWINE cipher~\cite{suzaki2012textnormal} aims to attain hardware efficiency with minimal hardware-oriented design choices that help in efficient software implementation. The implementation is simplified by using a single 4-bit S-Box and a key schedule without bit permutation.
                   
        In the race for lightweight cryptography, Sony developed their own lightweight cryptographic algorithm, CLEFIA~\cite{shirai2007128}, offering versatility by providing efficient use in hardware and software implementations. The F-functions are smaller and can be processed concurrently without requiring an inverse. Influenced by the design of CLEFIA, the TWIS cipher~\cite{ojha2009twis} uses a diffusion matrix to generate lighter round keys and key whitening steps that maintain security.

        In addition to cost reduction, certain ciphers prioritize security. For instance, the Piccolo cipher~\cite{shibutani2011piccolo} can achieve a high level of security while maintaining a notably compact hardware implementation. The area is reduced by implementing a permutation-based key schedule, while the diffusion property between rounds is improved using 8-bit permutation. The implementation of HISEC cipher~\cite{aldabbagh2014hisec} shows better resistance to differential, integral, and boomerang attacks. The FeW cipher~\cite{kumar2014few} uses a novel technique of Feistel-M structure, enhancing the security margins and efficient software implementation. The Feistel-M structure involves mixing in the round function between two Feistel branches of a 4-branch generalized Feistel structure.
        
        The aforementioned ciphers, such as HIGHT, CLEFIA, and Piccolo, utilize type-2 \ac{GFN}. However, they have slow diffusion, leading to the design of WARP~\cite{banik2020warp} and ALLPC~\cite{cheng2021allpc} focusing on faster diffusion. The diffusion enhancement for the WARP cipher is achieved by incorporating an improved shuffle over 32 nibbles to the type-2 \ac{GFN}. The same enhancement for ALLPC cipher is achieved by adding some linear operations in the structure of type-1 \ac{GFN}. 
        
        A lightweight design that optimizes area consumption on \ac{ASIC} implementations does not necessarily reduce area footprints on \acp{FPGA} chips: \acp{ASIC} should have fewer gates, while \acp{FPGA} should have fewer lookup tables. The Khudra~\cite{kolay2014khudra} cipher reduces the number of lookup tables by using a shift register-based key schedule and balancing lookup tables and flip-flops. The recursive architecture of the cipher reduces the number of required S-Boxes and resources for permutation.

        \begin{wrapfigure} {r} {0.6\textwidth}
            \includegraphics[width=1\linewidth, height=5cm, keepaspectratio]{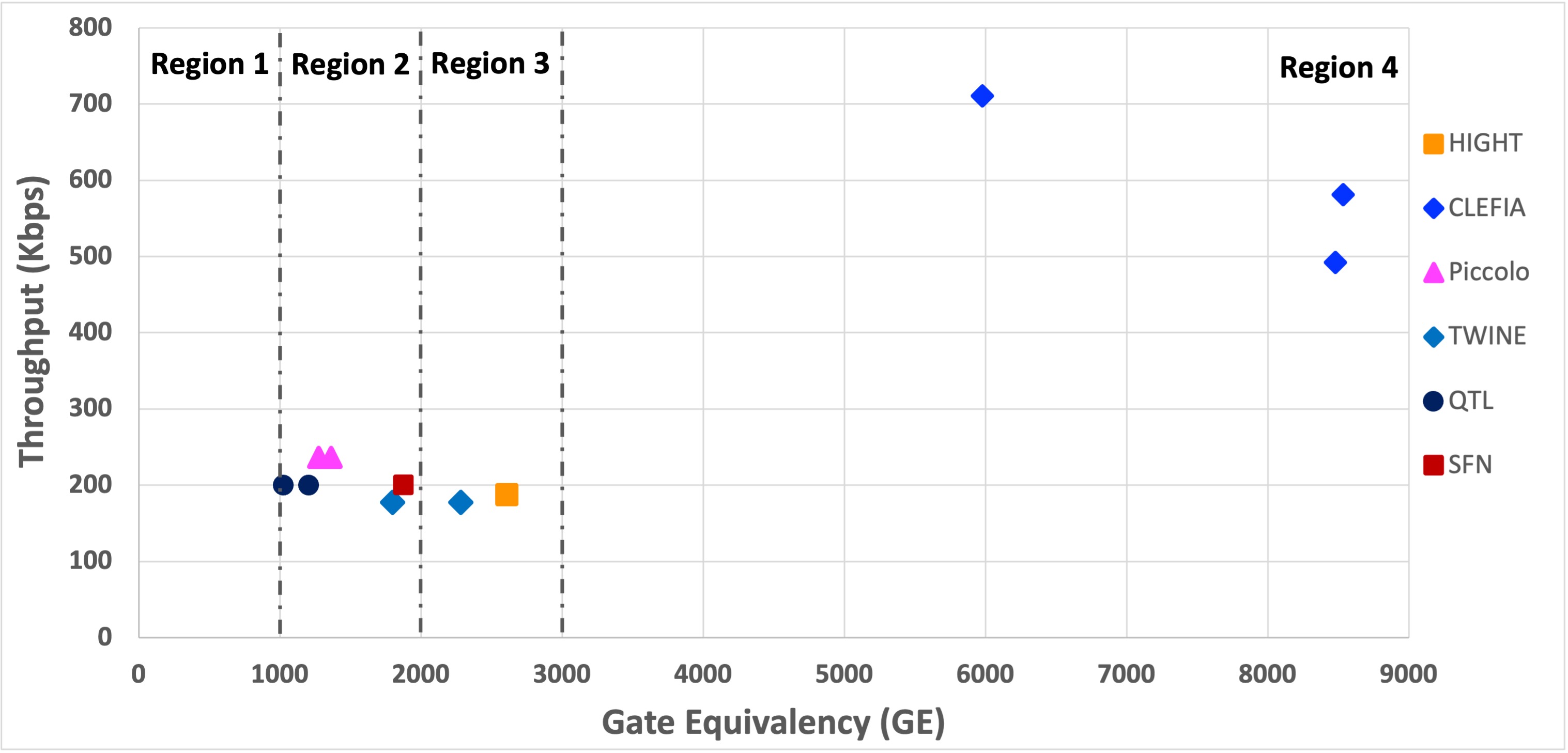}
           ~\caption{Throughput vs Gate Equivalency (GE) of GFN and LMD Block Ciphers.}
            \label{fig:Throughput_GE_GFN_LMD}
            \Description{The throughput and Gate Equivalency of GFN and LMD block ciphers}
        \end{wrapfigure}
         The hardware implementation values represented in Figure~\ref{fig:Throughput_GE_GFN_LMD} illustrate that CLEFIA has the largest area consumption. However, it is worth noting that this cipher has the highest throughput compared to the other ciphers (see Appendix Table~\ref{tab:HW_imp_GFN_LMD}). TWINE and Khudra have lower area consumption and slightly lower throughput than HIGHT. Piccolo offers the best combination of throughput and \ac{GE}. More recent ciphers, such as WARP and ALLPC, have incredibly low GE among \ac{GFN} ciphers, but available documentation does not provide throughput measurement. 

        
        Figure \ref{fig:Throughput_GE_GFN_LMD} also characterizes that most of the ciphers in \ac{GFN} lie in region~2 and region~3, except CLEFIA. CLEFIA is delineated in region~4 and seems to be far from classifying as a lightweight cryptographic algorithm. HIGHT, Piccolo, and TWINE are low-cost lightweight algorithms, suitable for resource-constrained devices, such as \ac{IoT} senors and \ac{RFID} devices. 
        \begin{table}[t]
\renewcommand{\arraystretch}{1.5}
\scriptsize
\caption{Comparative Analysis of LMD Block Ciphers.}
\label{tab:CompAna_LMD}
\begin{tabular}{l c l l}
\toprule
\textbf{Cipher} & \textbf{Year} & \textbf{Resistant Against Attacks}  & \textbf{Proven Vulnerabilities}
\\
\midrule
\centering{QTL \cite{li2016qtl}} & 2016 &  Differential; Linear; Algebraic & Related-key; Key recovery~\cite{ccoban2017cryptanalysis}
\\ 
\hline
\centering{SIT \cite{usman2017sit}} & 2017 & Differential; Linear; Square;  Related-key; Interpolation
\\ 
\hline
\centering{SFN \cite{li2018sfn}} & 2018  & Differential; Linear; Related-key; Slide; Impossible differential; Algebraic; Integral & Related key distinguisher; \ac{MITM}~\cite{sadeghi2018cryptanalysis}
\\ 
\hline
\centering{LRBC \cite{biswas2020lrbc}} & 2020 & Linear; Differential; Side channel 
\\ 
\hline
\centering{LCB \cite{roy2021lcb}} & 2021 &  Linear; Differential; Key related; Slide; Side channel; Structural & Distinguishing~\cite{chan2023differential}\\
\bottomrule
\end{tabular} 
\end{table}
        \begin{figure} [t]
            \includegraphics[width=\textwidth, height=4cm, keepaspectratio]{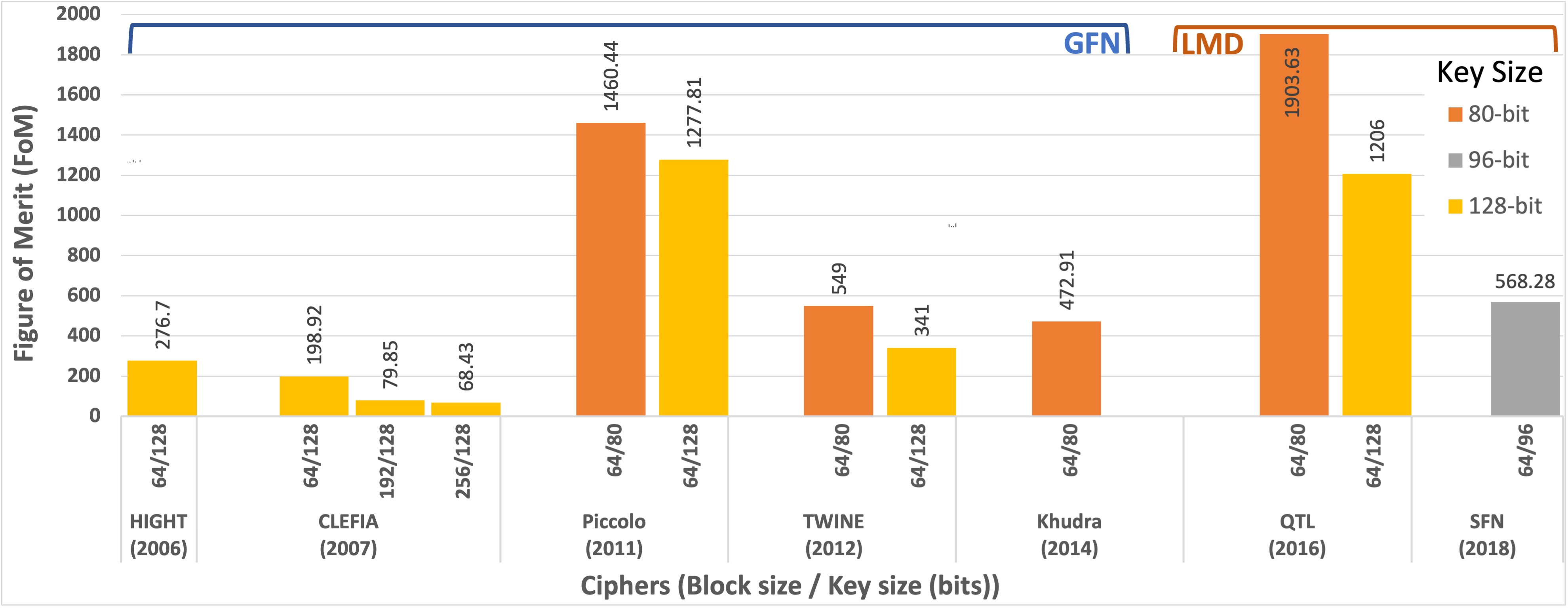}
           ~\caption{Figure of Merit (FoM) of GFN and LMD Block Ciphers.}
            \label{fig:FoM_GFN}
            \Description{FoM of GFN block ciphers}
        \end{figure}
        
        The bar graph in Figure~\ref{fig:FoM_GFN} shows that Piccolo has the highest overall hardware performance among \ac{GFN} ciphers. TWINE has slightly better overall hardware performance than Khudra, while CLEFIA is the least-performing cipher.
        

        \begin{wrapfigure} {r} {0.55\textwidth}
            \includegraphics[width=1\linewidth, height=4cm, keepaspectratio]{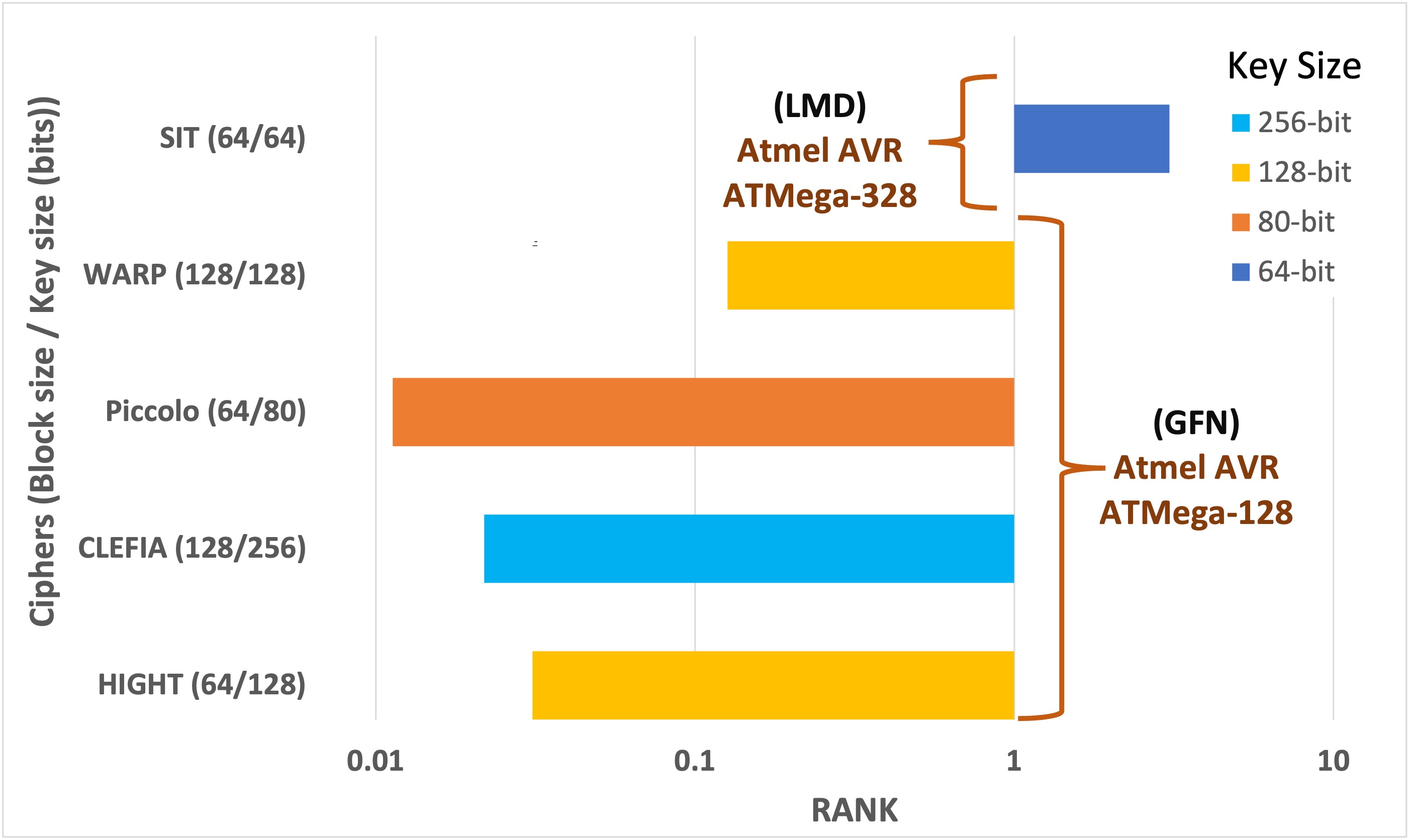}
           ~\caption{RANK of GFN and LMD Block Ciphers}
            \label{fig:RANK_GFN_LMD}
            \Description{RANK of GFN and LMD block ciphers}
        \end{wrapfigure}
        The software implementation of \ac{GFN} ciphers has been carried out on ATMEGA-128~\cite{sevin2021survey}. The graphical representation provided in Figure~\ref{fig:RANK_GFN_LMD} is plotted on RANK. The code size, RAM, and \ac{CpB} of WARP are the lowest (see Appendix Table~\ref{tab:SW_imp_GFN_LMD} for detailed measurements). Thus, the overall software performance is highest among \ac{GFN} ciphers. Piccolo has a smaller code size than CLEFIA and a lower RAM consumption than both CLEFIA and HIGHT. However, Piccolo's \ac{CpB} among \ac{GFN} ciphers is greater, resulting in lower overall software performance. 
        
        \subsubsection{\acf{LMD}}     
        \acs{LMD} combines the beneficial properties of \ac{FN} and \ac{SPN} to increase the level of confusion and diffusion in a cipher. This design was initially employed by the IDEA cipher~\cite{8711697}. Since then, only a few other lightweight ciphers have been developed with this design. A comparative analysis of \ac{LMD} ciphers based on their security strengths and proven vulnerabilities, can be found in Table~\ref{tab:CompAna_LMD}.

        The LMD design reduces the area utilization and improves diffusion. For instance, the QTL cipher~\cite{li2016qtl} addressed the slow diffusion problem by integrating \ac{GFN} structure with fast diffusion benefits of \ac{SPN} structure. The exclusion of the key schedule in QTL reduces area and energy consumption in hardware implementation. 
        
        The SIT cipher~\cite{usman2017sit} was developed specifically for \ac{IoT} networks, emphasizing resistance to \ac{MITM} attacks. SIT uses \ac{FN} structure, incorporating substitution-diffusion functions within its design. The SFN cipher~\cite{usman2017sit} combines \ac{SPN} and \ac{FN} structures for efficient encryption and decryption in a single program. The LRBC cipher~\cite{biswas2020lrbc} uses the same concept as that of SFN, with the difference that the algorithm is particularly designed through the integration of \ac{FN} and linear-SPN with a simple key combination. A year after the release of LRBC, a new version named LCB~\cite{roy2021lcb} was released, which requires less chip area and provides enhanced security with minimal computational delay. The reduction in the chip area was achieved by minimizing the number of logical operations. 
        

        Though only a handful of ciphers adapt the \ac{LMD} design, the ones we found show remarkable performance in hardware-related metrics such as GE, throughput, and \ac{FoM}, as can be found in Figure~\ref{fig:FoM_GFN} and appendix Table~\ref{tab:HW_imp_GFN_LMD}. Particularly, LRBC and LCB have an astonishingly low area consumption of under 300 GE, which was never achieved before. Unfortunately, throughput metrics are missing for these ciphers. SFN utilizes the highest area consumption of 1896 GE among all \ac{LMD} ciphers that lie within the low-cost lightweight categorization. QTL consumes less area than SFN cipher, resulting in a better \ac{FoM}.
    

         QTL and SFN are LMD ciphers with published GE and throughput values. Figure~\ref{fig:Throughput_GE_GFN_LMD} illustrates that QTL and SFN are located in region~2 and explains the appropriateness of the ciphers for \ac{IoT} sensors, RFID, and WSN.

        Software implementations of LMD ciphers have been carried out on the ATMEGA-328 microcontroller~\cite{usman2017sit}. Since there is missing data for these implementations, we were only able to draw conclusions based on \ac{CpB} (see Appendix Table~\ref{tab:SW_imp_GFN_LMD}), which shows that SIT has lower \ac{CpB} and thus faster execution than \ac{SPN} and \ac{GFN} ciphers. Given that \ac{GFN} and \ac{LMD} implementations were executed on 8-bit microcontrollers, a comprehensive comparison indicates that SIT performs better than most of the \ac{GFN} ciphers, with a significant difference in overall software performance. 
        
        \subsubsection{\acf{ARX}}
        
        The hardware and especially software implementation of \acs{ARX} ciphers are comparatively fast and cheap as these ciphers comprise only three elementary operations. Table~\ref{tab:CompAna_ARX} provides a comparative analysis of the security vulnerabilities and strengths of the \ac{ARX} ciphers.

        \begin{table}[t]
\renewcommand{\arraystretch}{1.5}
\scriptsize
\caption{Comparative Analysis of ARX Block Ciphers.}
\label{tab:CompAna_ARX}
\begin{tabular}{l c l l}
\toprule
\textbf{Cipher} & \textbf{Year} & \textbf{Resistant Against Attacks}  & \textbf{Proven Vulnerabilities}
\\
\midrule
SPECK \cite{cryptoeprint:2013:404} & 2013 & &  Differential; Rectangle~\cite{abed2013cryptanalysis}; Differential fault~\cite{tupsamudre2014differential}
\\ 
\hline
LEA \cite{hong2013lea} & 2013 & Differential; Linear; Slide; Boomerang; Integral & Zero correlation (13/14-r)~\cite{zhang2016zero}; Differential (9-r)~\cite{dwivedi2018differential}
\\
\hline
Simeck \cite{yang2015simeck} & 2015 & Differential; Linear; Algebraic; \ac{MITM}; Slide; Rotational; Related-key & Zero correlation (15/17-r); Impossible differential (12/13-r)~\cite{sadeghi2018improved}
\\ 
\hline
SPARX \cite{dinu2016design} & 2016 & Differential; Linear; Single trail; Integral & Power \ac{SCA}~\cite{ramesh2019side}
\\ 
\hline
CHAM \cite{koo2017cham} & 2017 & Differential; Linear; Integral; Boomerang;  Biclique; Slide; Rotational & Gradient statistical~\cite{ryabko2019gradient}
\\ 
\hline
Shadow \cite{guo2021shadow} & 2021 & Impossible differential; Biclique\\
\bottomrule
\multicolumn{4} {r@{}}{\textbf{r} represents rounds}
\end{tabular} 
\end{table}
        \begin{figure} [t]
            \includegraphics[width=\textwidth, height=3.5cm, keepaspectratio]{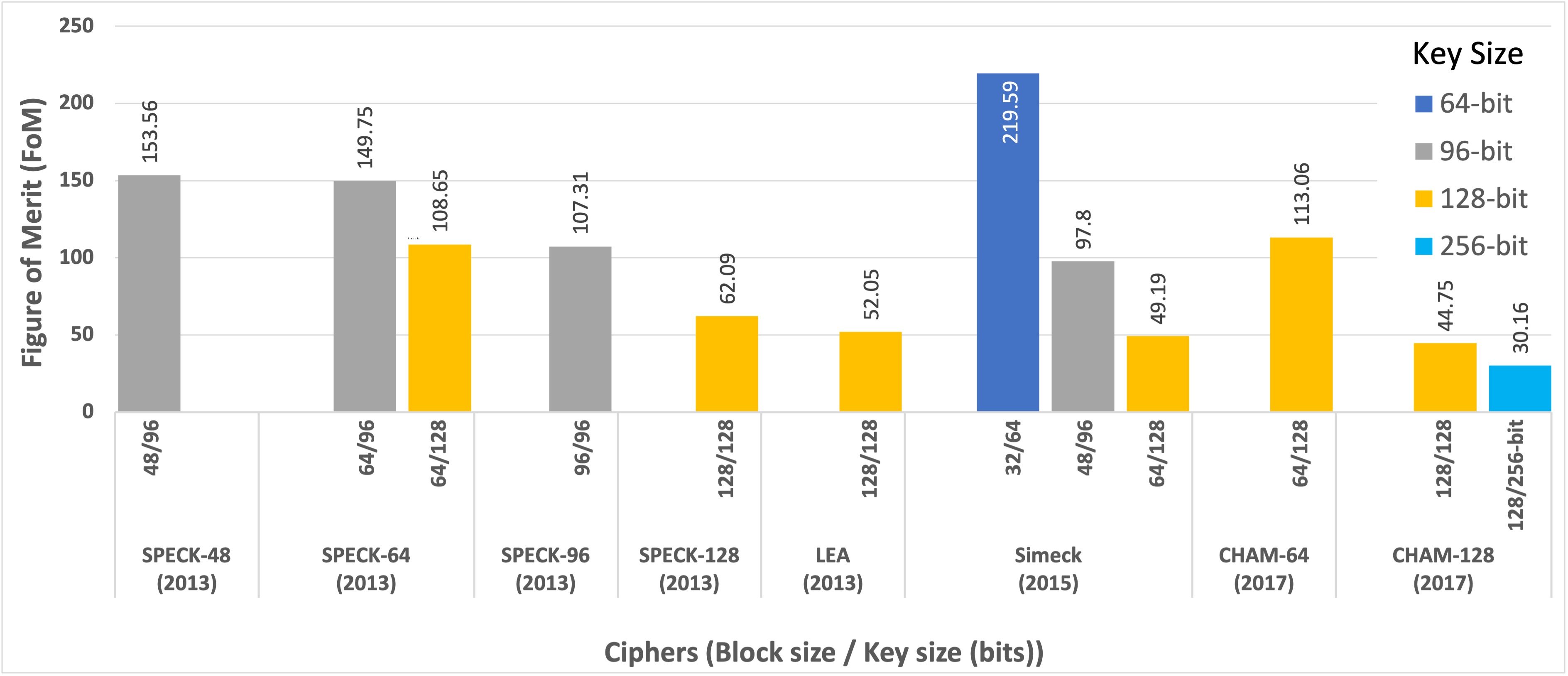}
           ~\caption{Figure of Merit (FoM) of ARX Block Ciphers.}
            \label{fig:FoM_ARX}
            \Description{FoM of ARX block ciphers}
        \end{figure}
        %
        Designing an \ac{ARX} cipher that is secure against single-trail differential and linear cryptanalysis is difficult. SPARX cipher~\cite{dinu2016design} solves the issue by using an extensive linear permutation instead of bit rotations based on the Long Trail Strategy (LTS)~\cite{dinu2016design}. LTS uses light linear layers in combination with computationally intensive S-Boxes.         
        
        \ac{ARX} designs are typically optimized for software implementation, such as SPECK cipher~\cite{cryptoeprint:2013:404} developed by \ac{NSA}. The efficient software implementation for SPECK was achieved by incorporating modular addition for nonlinearity. The LEA cipher~\cite{hong2013lea} was introduced as an improvement to SPECK, having slightly better performance in software and verified security, except in 64-bit processors~\cite{hong2013lea}. The efficiency was improved by parallel execution of the simple key-schedule procedures using the same round functions in the last round. The CHAM cipher~\cite{koo2017cham} was introduced as an improvement over LEA. The optimization was accomplished by employing stateless on-the-fly key scheduling and fewer round keys. The Simeck cipher~\cite{yang2015simeck} was created by amalgamating the foremost features of SIMON and SPECK. The cipher achieves an efficient hardware implementation by utilizing SIMON's updated round function and \ac{LFSR}-based constants in the key schedule.
                    
         The UBRIGHT~\cite{sehrawat2020ultra} and Shadow~\cite{guo2021shadow} ciphers were designed to address the diffusion problem. UBRIGHT enhances diffusion by using modulo operation for non-linearity and applying round permutation at the end of each round. Shadow improved diffusion by introducing a new logical combination method of \ac{GFN} and \ac{ARX} operations. 
        
        
        \begin{wrapfigure} {r} {0.48\textwidth}
            \includegraphics[width=1\linewidth, keepaspectratio]{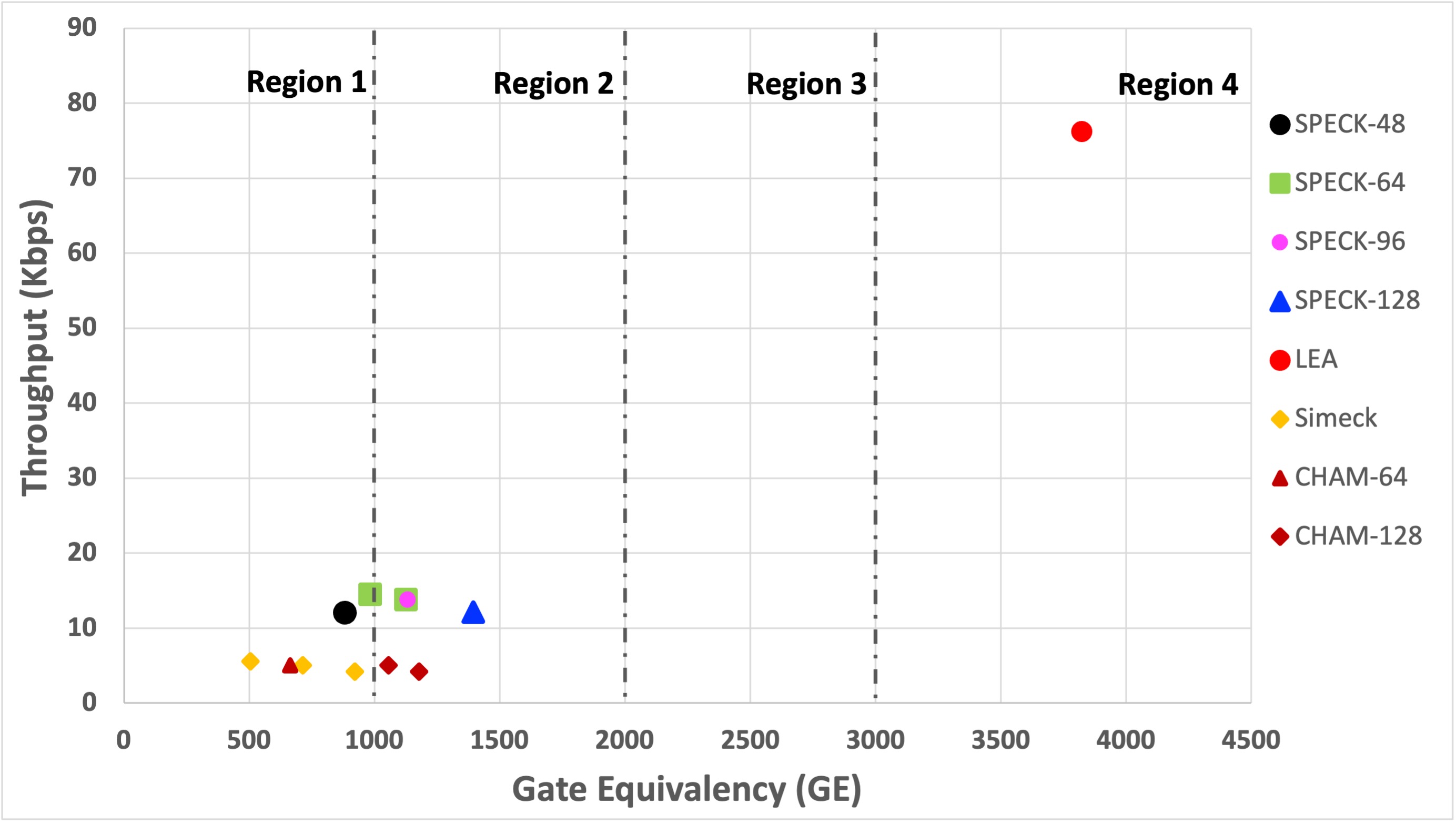}
           ~\caption{Throughput vs Gate Equivalency (GE) of ARX Block Ciphers.}
            \label{fig:Throughput_GE_ARX}
            \Description{The throughput vs. GE of ARX block ciphers}
        \end{wrapfigure}        
        The hardware implementation analysis shown in Figure~\ref{fig:Throughput_GE_ARX} elucidates that LEA has the highest area consumption but offers much higher throughput than other \ac{ARX} ciphers. Although SPECK is renowned for having low area consumption, ciphers created after 2013, such as Simeck and CHAM, have further reduced area consumption (see Appendix Table~\ref{tab:HW_imp_ARX}). However, throughput has suffered as a consequence. While the CHAM-128 and Simeck-64 ciphers offer lower GE, it is important to note that they also provide the lowest throughput. In terms of throughput, the SPECK family performs better than Simeck.

    
        The applicability of these ciphers within the distinct spectrum of GE, as shown in Figure~\ref{fig:Throughput_GE_ARX}, demonstrates that ARX-based lightweight ciphers appear to have less than 1500 GE area consumption, except for Shadow-64 and LEA. Region~1 contains ciphers like SPECK-48, SPECK-64, Simeck, and CHAM-64, which make them ultra-lightweight ciphers. The ciphers are suitable for use in extremely resource-constrained devices, such as 8-bit microcontrollers and RFIDs. Despite possessing larger key sizes indicating better security, SPECK-96, SPECK-128, and CHAM-128 are situated within the range of 1000 to 1500 GE and lie within the category of low-cost lightweight ciphers. The ciphers are better suited for RFID, WSN, and \ac{IoT}. SPARX and Shadow are not included in Figure \ref{fig:Throughput_GE_ARX} due to missing data. SPARX is developed for microcontrollers widely used in IoT~\cite{dinu2016design}, while Shadow is developed for low-power, versatile IoT sensors.
        
        The graphical analysis provided by Figure~\ref{fig:FoM_ARX} illuminates that the overall hardware performance (\ac{FoM}) of LEA, Simeck-64, and CHAM-128 is the least. Simeck-32 provides the best \ac{FoM} because of its smaller block size. While Simeck was developed as an improvement to SPECK, the SPECK family provides better \ac{FoM} and is ranked second.

        
        \begin{wrapfigure} {r} {0.4\textwidth}
            \includegraphics[width=1\linewidth, height=3cm, keepaspectratio]{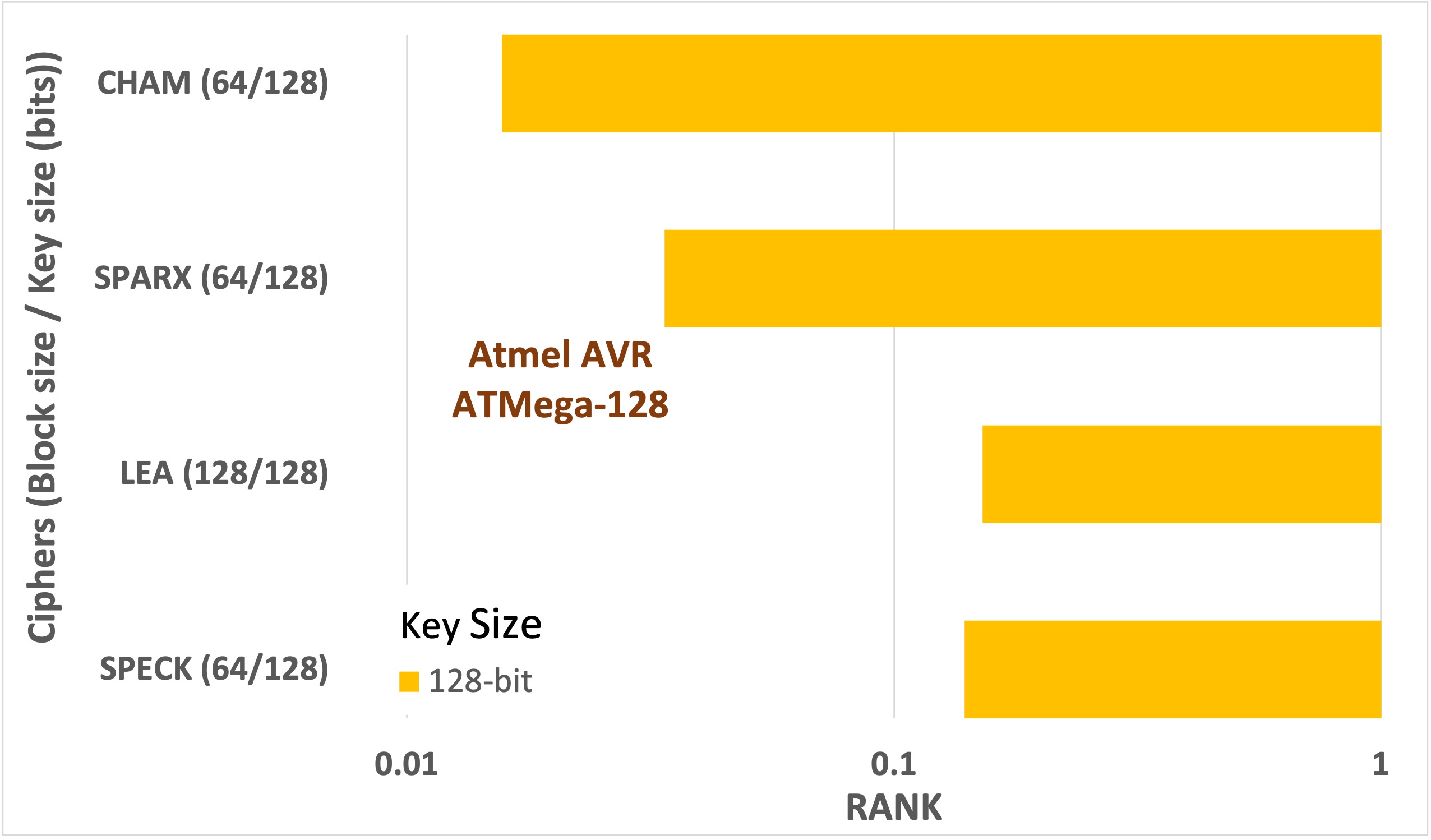}
           ~\caption{RANK of ARX Block Ciphers.}
            \label{fig:RANK_ARX}
            \Description{RANK of ARX block ciphers}
        \end{wrapfigure}
        The software implementation of ARX ciphers has been performed on an Atmel AVR ATMega-128~\cite{sevin2021survey}. The RANK analysis from Figure~\ref{fig:RANK_ARX} confirms that LEA has the highest overall software performance and SPECK is ranked second by a slight margin. Both the SPECK and LEA ciphers offer an improved balance of \ac{CpB}, code size, and RAM usage. Furthermore, both have significantly lower energy consumption and higher throughput (see Appendix Table~\ref{tab:SW_imp_ARX} for detailed measurements). CHAM has the least RANK since the \ac{CpB} is the highest.

    \begin{figure} [t]
        \includegraphics[width=\textwidth, height=5.2cm, keepaspectratio]{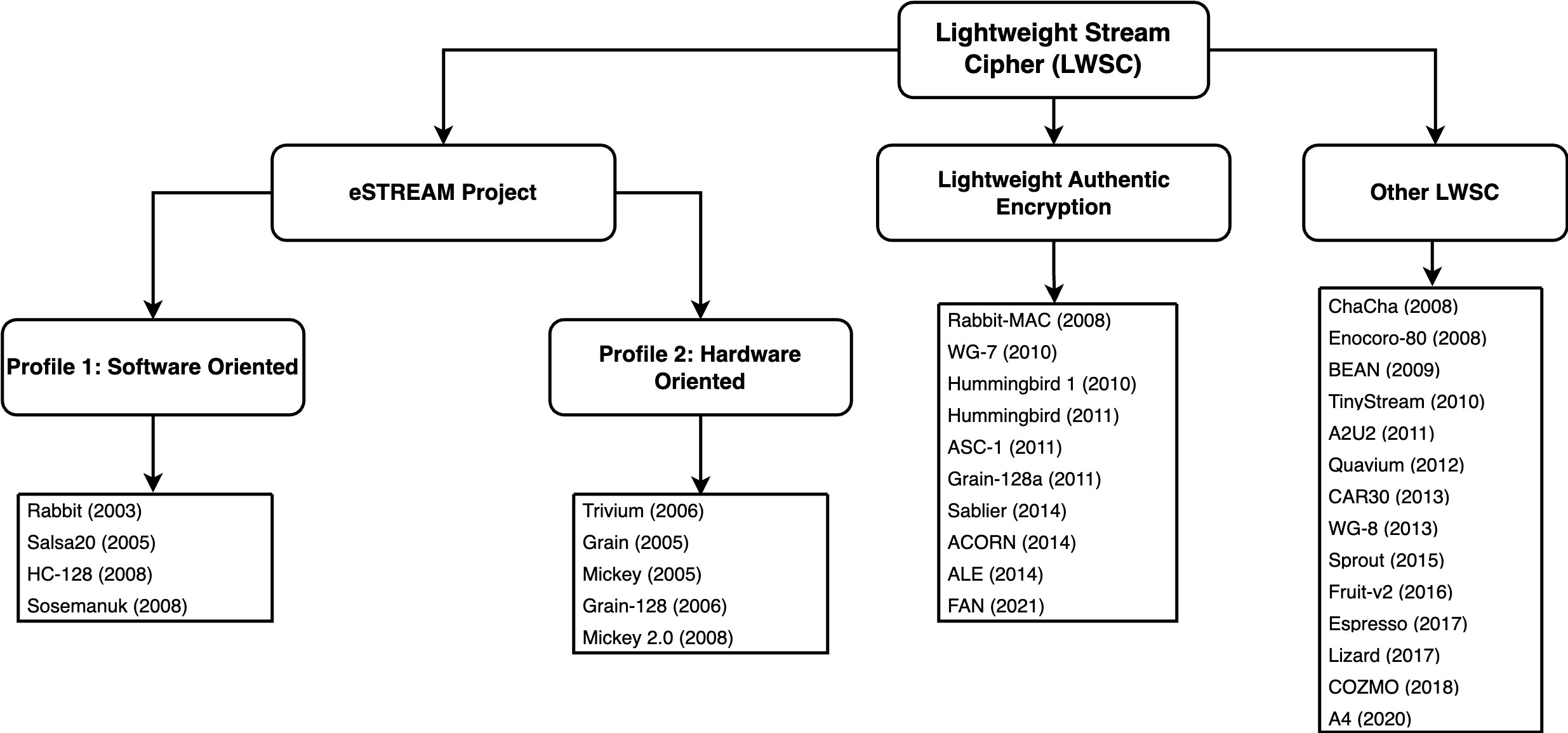}
       ~\caption{Classification of Lightweight Stream Ciphers.}
        \label{fig:classification_LWSC_cryptography}
        \Description{The Lightweight stream ciphers distributed among different branches.}
    \end{figure}
    
    \subsection{Lightweight Stream Cipher (LWSC)}
    
        Stream ciphers combine a pseudorandom keystream with individual bits or bytes in a data stream to produce the ciphertext. Since most stream ciphers combine multiple designs and structures, the design-oriented classification approach used for \acp{LWBC} is insufficient. For instance, the Rabbit, CAR30, and Tinystream ciphers use different methods for generating keystreams. Rabbit uses a chaotic map function, CAR30 uses Cellular Automata, and Tinystream uses the Tree Parity Machine (TPM) approach. For \ac{LWSC}, we therefore, adopt the classification proposed by \citet{manifavas2016survey}, as shown in Figure~\ref{fig:classification_LWSC_cryptography}.

        \subsubsection{eSTREAM: The ECRYPT Stream Cipher Project}
                    
            The \ac{ECRYPT} Stream Cipher Project (eSTREAM) was an effort initialized in 2004 by the Information Societies Technology Programme of the European Commission to find new and more efficient stream ciphers. The submitted ciphers were evaluated through three phases and distributed among Profile~1 (software-oriented) and Profile~2 (hardware-oriented)~\cite{robshaw2008estream}. Profile~1 includes ciphers for software applications with high throughput, whereas Profile~2 includes ciphers for hardware applications with limited resources.

        \begin{table}[t]
\renewcommand{\arraystretch}{1.5}
\scriptsize
\caption{Comparative Analysis of eSTREAM (software-oriented) Ciphers.}
\label{tab:CompAna_eSTREAM_Profile1}
\begin{tabular}{l c l l }
\toprule
\textbf{Cipher} & \textbf{Year} & \textbf{Resistant Against Attacks}  & \textbf{Proven Vulnerabilities}
\\
\midrule
Rabbit \cite{boesgaard2003rabbit} & 2003 & Divide and conquer; Guess and determine; Correlation & Distinguising~\cite{lu2008cryptanalysis}
\\ 
\hline
Salsa20 \cite{bernstein2005salsa20} & 2005 & Differential; Algebraic; Weak-key; Related-key & Related-key~\cite{shao2012related}; Truncated Differential (5-r)~\cite{crowley2005truncated}
\\ 
\hline
HC-128 \cite{wu2008stream} & 2008 & Correlation; Distinguishing; Algebraic & Differential fault; Side-channel~\cite{raizada2015some}
\\ 
\hline
Sosemanuk \cite{berbain2008sosemanuk} & 2008 & Distinguishing; Algebraic; \ac{TMDTO} &  Correlation~\cite{lee2008cryptanalysis}; Guess and determine~\cite{tsunoo2006evaluation}\\
\bottomrule
\multicolumn{4} {r@{}}{\textbf{r} represents rounds}
\end{tabular} 
\end{table}
        
        \begin{table}[t]
\renewcommand{\arraystretch}{1.5}
\scriptsize
\caption{Comparative Analysis of eSTREAM (hardware-oriented) Ciphers.}
\label{tab:CompAna_eSTREAM_Profile2}
\begin{tabular}{l c l l}
\toprule
\textbf{Cipher} & \textbf{Year} & \textbf{Resistant Against Attacks}  & \textbf{Proven Vulnerabilities}
\\
\midrule
Trivium \cite{canniere2008trivium} (ISO-29192-3) & 2006 & Guess and determine; Algebraic; Re-synchronization & Differential fault~\cite{hojsik2008differential}; Algebraic side channel~\cite{kazmi2017algebraic}
\\ 
\hline
Grain v1 \cite{hell2007grain} & 2005 & & Differential fault~\cite{siddhanti2017differential}; Related-key~\cite{canniere2008analysis}; Distinguishing \cite{khazaei2005distinguishing}
\\ 
\hline
Grain-128 \cite{hell2006stream} & 2006 & Linear approximations; Algebraic; \ac{TMDTO} & Related-key \cite{canniere2008analysis}; Dynamic cube
\\ 
\hline
MICKEY 1.0 \cite{babbage2005stream} & 2005 & Algebraic  & \ac{TMDTO}; Weak keys
\\ 
\hline
MICKEY 2.0 \cite{babbage2008mickey} & 2008 & Resynchronization; Algebraic; \ac{TMDTO}; Guess and determine & Slide synchronization; Related-key~\cite{ding2013cryptanalysis}\\
\bottomrule
\end{tabular} 
\end{table}       
        
        The finalists of Profile~1 (software-oriented) ciphers were Rabbit~\cite{boesgaard2003rabbit}, Salsa20~\cite{bernstein2005salsa20}, HC-128~\cite{wu2008stream}, and Sosemanuk~\cite{berbain2008sosemanuk}.
        The comparative analysis associated with their security strengths and vulnerabilities is provided in Table~\ref{tab:CompAna_eSTREAM_Profile1}. The design of the Rabbit cipher was inspired by the randomness-like properties of real-valued chaotic maps, which are exponentially sensitive to small perturbations but restricted to fixed-point values. The chaotic maps are structured on coupled non-linear maps. The Salsa20 cipher has a core of a hash function used in a counter mode as a stream cipher~\cite{bernstein2005salsa20}. Developing a fast and lightweight primitive of Salsa20 is made simpler by using basic operations and avoiding the use of S-Boxes. The HC-128 cipher was designed for super-scalar microprocessors by providing a higher degree of parallelism in certain operations. The efficiency of the cipher was improved by reducing internal state and static data.      
        
        The hardware-oriented ciphers in Profile 2 comprise Trivium~\cite{canniere2008trivium}, Grain~\cite{hell2007grain}, and Mickey~\cite{babbage2005stream, babbage2008mickey}. A comparison of the ciphers, including their security strengths and weaknesses, can be found in Table~\ref{tab:CompAna_eSTREAM_Profile2}. The Trivium cipher was designed for cost-efficient hardware implementation and was selected as an international standard under ISO/IEC 29192-3~\cite{ISO29192}. The cipher achieves compactness by reducing word size, replacing S-Boxes, and employing a bit-oriented scheme. The Grain-v1 cipher is structured upon dual shift registers and a nonlinear output function. The speed of the cipher can be optimized through the provision of extra hardware resources. Grain-v1 was found to be susceptible to certain attacks~\cite{canniere2008analysis}, which led to the development of Grain-128. To prevent attacks, the size of the internal state was increased by enlarging the linear and non-linear shift registers to 128 bits. The Mickey cipher uses irregular clocking of shift registers to provide randomness. Mickey~2.0 was introduced as a sequel to Mickey~1.0 with improved resistance over \ac{TMDTO} and weak key attacks~\cite{babbage2008mickey} by increasing the stages of the linear and non-linear shift register. 
        
        
        \begin{wrapfigure} {r} {0.45\textwidth}
            \includegraphics[width=1\linewidth, height=5cm, keepaspectratio]{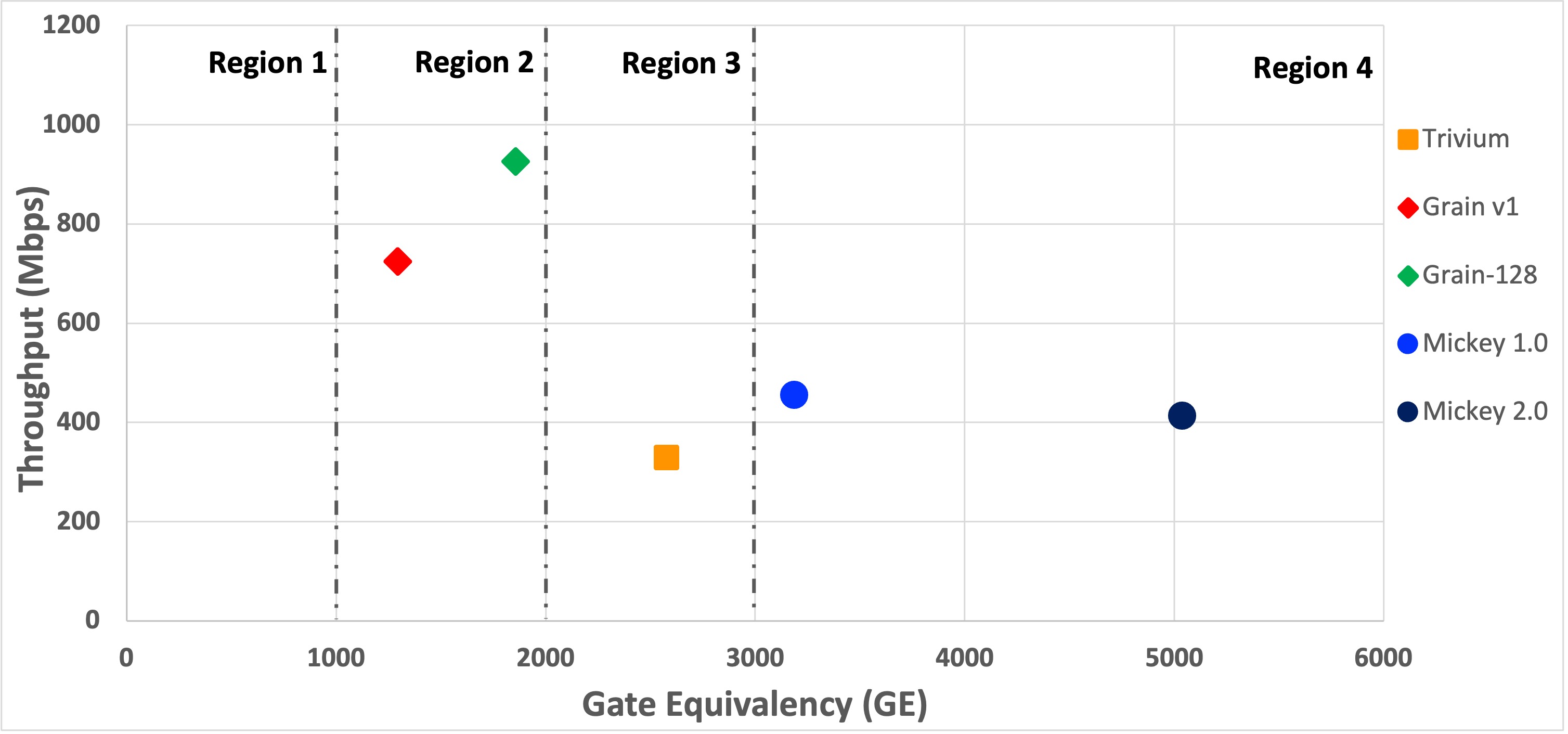}
           ~\caption{Throughput vs. Gate Equivalency (GE) of eSTREAM (hardware-oriented) Ciphers.}
            \label{fig:Throughput_GE_eSTREAM}
            \Description{Throughput vs Gate Equivalency of eSTREAM hardware-oriented ciphers.}
        \end{wrapfigure}
         The comparison of hardware-oriented (Profile 2) eSTREAM cipher implementations in Figure~\ref{fig:Throughput_GE_eSTREAM} shows that Grain v1 has the lowest GE while Mickey 2.0 has the highest. Grain-128 cipher provides an impressive throughput of 925.9 Mbps at a clock rate of 10 MHz (see Appendix Table~\ref{tab:HW_imp_eSTREAM} for detailed measurements). The area consumption of Trivium falls in between Grain and Mickey families. However, both provide a higher throughput than Trivium.



        Figure~\ref{fig:Throughput_GE_eSTREAM} also compares the throughput and GE of eSTREAM (Profile 2) ciphers with respect to their applicability. The Grain ciphers lie in region~2 and are thereby low-cost lightweight algorithms. The Grain family is suitable for RFID, WSN, and IoT devices. Trivium lies in region~3 and is suitable for limited resource devices such as microcontrollers or embedded systems. The Mickey family is located in region~4 and consists of high-cost lightweight algorithms suitable for embedded systems or middleware.
        
        \begin{wrapfigure} {r} {0.55\textwidth}
            \includegraphics[width=1\linewidth, height=5cm, keepaspectratio]{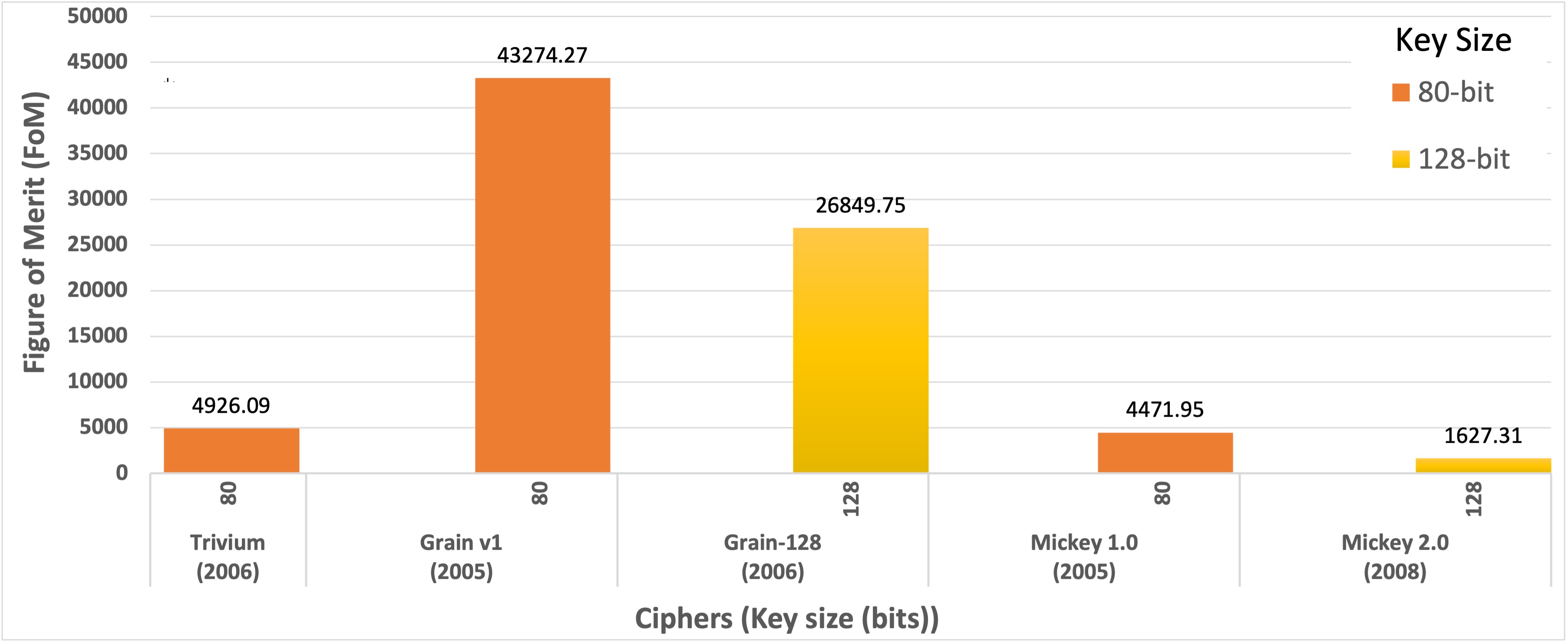}
           ~\caption{Figure of Merit (FoM) of eSTREAM (hardware-oriented) Ciphers.}
            \label{fig:FoM_eSTREAM}
            \Description{FoM of eSTREAM hardware-oriented ciphers}
        \end{wrapfigure}
        According to the bar graph of \ac{FoM} in Figure~\ref{fig:FoM_eSTREAM}, it becomes apparent that the Mickey family has the lowest overall hardware performance. The Grain family has the best overall hardware performance. Grain-v1 outperforms Grain-128 significantly within the Grain family, and Trivium performs slightly better than Mickey 1.0.

        \begin{figure} [t]
            \includegraphics[width=\textwidth, height=3.7cm, keepaspectratio]{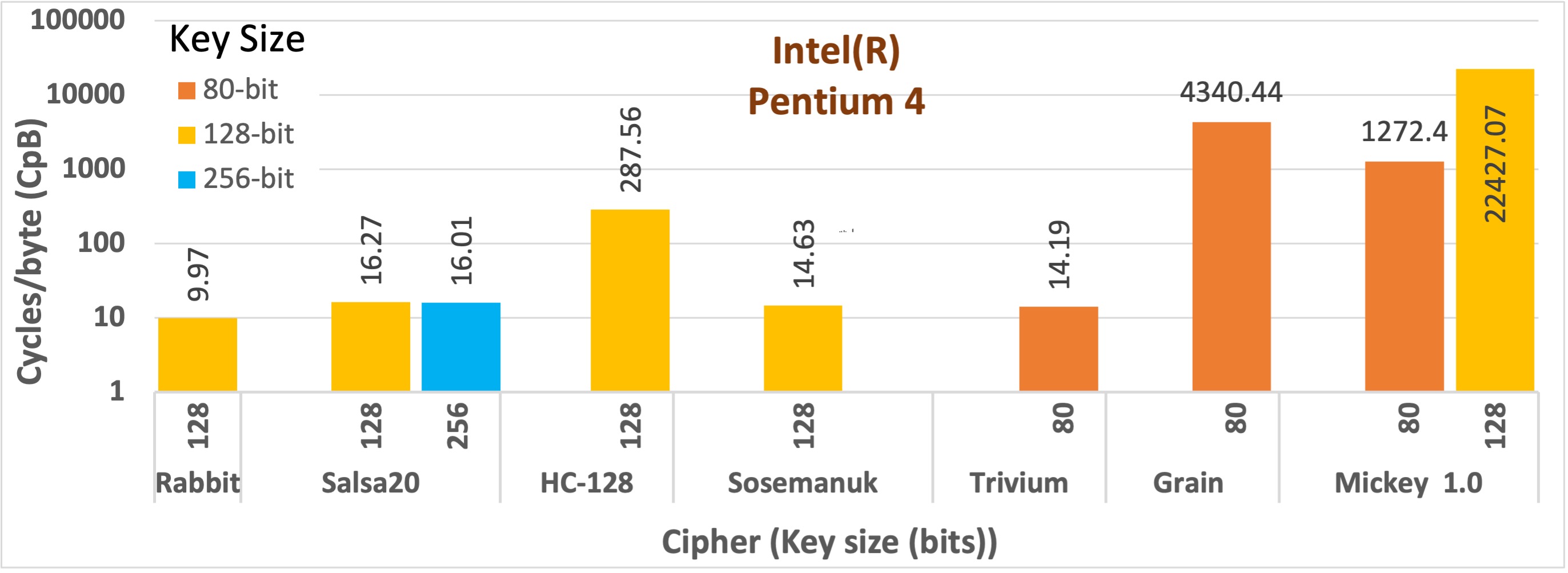}
           ~\caption{Cycles per byte (CpB) of eSTREAM Ciphers.}
            \label{fig:CpB_eSTREAM}
            \Description{Cycles per byte (CpB) of eSTREAM ciphers}
        \end{figure}      
        The software implementation of eSTREAM ciphers was conducted on an Intel Pentium 4 system~\cite{jassim2021survey}. The graphical interpretation of RANK has been provided in Figure~\ref{fig:CpB_eSTREAM}. We are evaluating performance based on \ac{CpB} because of the absence of RAM and code size values. The graphical interpretation confirms that the eSTREAM (Profile 1) ciphers have weak software implementation except for Trivium. In eSTREAM (Profile 2) ciphers, HC-128 has the highest \ac{CpB} but is still better than Grain and Mickey from eSTREAM (Profile 1) ciphers. The Rabbit cipher has the lowest \ac{CpB}, which means faster execution speed (see Appendix Table~\ref{tab:SW_imp_eSTREAM}). 

        
        \subsubsection{\acf{LWAE}}
        Authenticated encryption is a cryptographic procedure that ensures confidentiality, integrity, and authenticity at the same time. The authentication is provided by generating authentication tags or \acp{MAC}. The encryption processes the plaintext and generates the ciphertext with an authentication tag, whereas the process of decryption is integrated with integrity validation, i.e., the plaintext is retained, and the authentication tag is matched with the ciphertext, producing an error in an unmatched situation. Table~\ref{tab:CompAna_LWAE} enumerates the related strengths and weaknesses of \ac{LWAE} ciphers.

            
         The \ac{LWAE} ciphers have improved over time by utilizing various design techniques, effectively mitigating several vulnerabilities by ensuring confidentiality, integrity, and authenticity. The Grain-128a~\cite{gren2011grain} cipher was an improvement to the Grain-128 cipher, extended to support optional authentication. The utilization of regularly clocked shift registers within the cipher provides an advantage in performance enhancement and resistance to \ac{SCA}. The ASC-1~\cite{jakimoski2011asc} encryption algorithm was designed with 4 rounds of AES-128, and the leak extraction from each round is used as a key to encrypt the data. The cipher performs encryption and authentication in a single pass, resulting in a faster execution.  

        The Sablier~\cite{zhang_shi_xu_yao_li_2014} and ACORN~\cite{wu_2014} ciphers were submitted to the CAESAR competition for authenticated ciphers, announced at 
        the Early Symmetric Crypto workshop in 2013~\cite{zhang2018survey}. The Sablier cipher is an efficient hardware-oriented cipher with a simple non-linear function and linear transformation. Its efficiency is enhanced by a new bitwise internal structure for keystream generation and parallel computation.  ACORN was selected as one of the winners in the CAESAR competition due to its lightweight features and resistance against \ac{SCA}. 
        
        The ALE cipher~\cite{bogdanov2013ale} is efficient in hardware as well as software due to its development from the advantageous components of ASC-1~\cite{jakimoski2011asc}. The use of parallel computation and single-pass authentication allows for a quick and efficient implementation in a compact manner. The FAN cipher~\cite{jiao2021fan} uses a novel approach of Continuous-Key-Use (CKU), supporting key storage in non-volatile memory for initialization and encryption. The bit-slice technique and byte-wise operations facilitate efficient implementation in hardware and software.

        Specific lightweight authentic stream ciphers were meticulously designed for distinct applications and operational requirements. For instance, the Rabbit-MAC~\cite{tahir2008rabbit} cipher was specifically developed to enhance link-layer security in \acp{WSN}. The implementation cost of Rabbit-MAC is reduced by using basic arithmetic operations to achieve strong non-linear mixing of the internal state. The WG-7 cipher~\cite{luo2010lightweight} was designed for RFID tags that offer security features like untraceability and protection against tag and reader impersonation. It utilizes basic logic operations and lookup tables instead of S-Boxes, resulting in lower resource consumption.

        \begin{table}[t]
\renewcommand{\arraystretch}{1.3}
\scriptsize
\caption{Comparative Analysis of LWAE Stream Ciphers.}
\label{tab:CompAna_LWAE}
\begin{tabular}{l c l l }
\toprule
\textbf{Cipher} & \textbf{Year} & \textbf{Resistant Against Attacks}  & \textbf{Proven Vulnerabilities}
\\
\midrule
Rabbit-MAC~\cite{tahir2008rabbit} & 2008 & Divide and conquer; Guess and determine; Correlation &
\\ 
\hline
WG-7~\cite{luo2010lightweight} & 2008 & \ac{TMDTO}; Differential; Cube; Correlation & Distinguishing; Algebraic; Key-recovery~\cite{orumiehchiha2012cryptanalysis}
\\ 
\hline
Grain-128a~\cite{gren2011grain} & 2011 & Linear approximations; Algebraic; \ac{TMDTO}~\cite{hell2006stream} & Differential fault~\cite{banik2012differential}; Related key chosen IV~\cite{ding2013related}
\\ 
\hline
Sablier~\cite{zhang_shi_xu_yao_li_2014} & 2014 & Differential; \ac{TMDTO}; Cube; Guess and determine; Rotational & State recovery; Key recovery~\cite{feng2014cryptanalysis}
\\ 
\hline
ACORN \cite{wu_2014} & 2014 & Statistical & State recovery; Key recovery; State collision; Cube~\cite{yang2019cube}\\ 
\hline
FAN \cite{jiao2021fan} & 2021 & Correlation; Algebraic; Guess-and-determine; \ac{TMDTO}; Related-key; Cube &\\
\bottomrule
\end{tabular} 
\end{table}
        
        The juxtaposition of hardware implementation data presented in Table~\ref{tab:HW_imp_eSTREAM} and Table~\ref{tab:HW_imp_LWAE} in the appendix signifies that most of the \ac{LWAE} ciphers utilize less area than eSTREAM ciphers. Grain~128a has the lowest area consumption, while ASC-1 has the highest among \ac{LWAE} ciphers. Although we have included Hummingbird-2 in the \ac{LHC} section, we also place it here due to its support for authentication. Hummingbird-2 cipher requires less area than Grain-128, ASC-1, and ALE.

        Grain-128a and Sablier lie in the low-cost lightweight cipher categorization shown in Table~\ref{tab:GE_table} and are thereby useful in environments like RFID, \ac{WSN}, and \ac{IoT}. Hummingbird-2, ALE, and FAN lie in lightweight cipher categorization and can be used in limited-resource devices such as embedded systems and microcontrollers. ASC-1 is a high-cost lightweight algorithm, more suitable for middleware and embedded systems. Note that because throughput values for most of the \ac{LWAE} ciphers are missing from the literature, performance graphs, similar to those we have provided for the other cipher classes, are not included for \ac{LWAE} ciphers. Instead, our analysis only relies on the GE measurements.
        
        \subsubsection{Other Lightweight Stream Ciphers}
        
        \begin{table}[t]
\renewcommand{\arraystretch}{1.2}
\scriptsize
\caption{Comparative Analysis of Other Lightweight Stream Ciphers.}
\label{tab:CompAna_OLWSC}
\begin{tabular}{l c l l }
\toprule
\textbf{Cipher} & \textbf{Year} & \textbf{Resistant Against Attacks}  & \textbf{Proven Vulnerabilities}
\\
\midrule
ChaCha~\cite{bernstein2008chacha} & 2008 & & Rotational~\cite{barbero2022rotational}; Chosen IV (7-r)~\cite{maitra2016chosen}
\\ 
\hline
Enocoro-80~\cite{watanabe2008enocoro} (ISO/IEC29192) & 2008 & Linear; Distinguishing; Guess-and-determine; \ac{TMDTO};  & Related-key~\cite{watanabe2010update}; Slide~\cite{ding2015slide}
\\ 
\hline
BEAN \cite{kumar2009bean} & 2009 & & Distinguisher; Key recovery~\cite{aagren2011cryptanalysis}
\\ 
\hline
A2U2 \cite{david2011a2u2} & 2011 & & Chosen-IV; Guess and determine~\cite{abdelraheem2011cryptanalysis}
\\ 
\hline
QUAVIUM \cite{tian2012quavium} & 2012 & Correlation; Algebraic; Differential &
\\ 
\hline
CAR30 \cite{das2013car30} & 2013 & Algebraic; Cube; Inversion; \ac{TMDTO}; Chosen IV; Guess and determine &
\\ 
\hline
WG-8 \cite{fan2013wg} & 2013 & Algebraic; Correlation; Differential; Distinguishing; \ac{TMDTO} & Related key~\cite{ding2014cryptanalysis}; Distinguishing~\cite{rostami2019cryptanalysis}
\\ 
\hline
Pandaka \cite{chen2014pandaka} & 2014 & Ciphertext only; Known plaintext; \ac{TMDTO} & Desynchronization; Guess and determine~\cite{yarom2015evaluation}
\\ 
\hline
Sprout \cite{armknecht2015lightweight} & 2015 & Algebraic; Chosen IV; Linear; Dynamic cube; Weak key-IV & Key-recovery~\cite{lallemand2015cryptanalysis}; Guess-and-determine~\cite{esgin2016practical}
\\ 
\hline
Fruit-v2 \cite{ghafari2016fruit} & 2016 & \ac{TMDTO}; Related-key; Linear; Cube; Algebraic; Weak key-IV & Fast correlation~\cite{wang2019fast}
\\ 
\hline
Espresso \cite{dubrova2017espresso} & 2017 & Linear; \ac{TMDTO}; Chosen IV; Differential; Weak keys &  Algebraic; Rønjom-Helleseth~\cite{yao2021cryptanalysis}
\\ 
\hline
Lizard \cite{hamann2017lizard} & 2017 & Algebraic; Linear; Guess and determine; Cube; IV collisions. & Key recovery~\cite{banik2017some}, Differential~\cite{siddhanti2017differential}, \ac{TMDTO}~\cite{maitra2017tmdto}
\\ 
\hline
A4 \cite{mohandas2020a4} & 2020 & \ac{MITM}; Algebraic; Differential; Correlation; Exhaustive
\\ 
\bottomrule
\multicolumn{4} {r@{}}{\textbf{r} represents rounds}
\end{tabular} 
\end{table} 
        
        This section introduces lightweight stream ciphers inspired by the aforementioned ciphers, such as Salsa20, Grain, and Trivium, with improvements in performance, security, and cost-effectiveness. In addition, it also includes other lightweight stream ciphers that utilize distinct architectural and design approaches such as Welsh-Gong transformation, dynamic key dependence approach, Cellular Automata, etc. The ciphers encompassed in this classification are comparatively analyzed based on security strengths and weaknesses in Table~\ref{tab:CompAna_OLWSC}.

        The ciphers represented in the following section provide an eminent improvement over their predecessors in terms of either security, performance, or both. The ChaCha cipher~\cite{bernstein2008chacha} adapts the same basic design principles as the Salsa20 eSTREAM cipher but with increased diffusion in each round for faster execution. The quarter-round of ChaCha and its rotation distances were designed to facilitate rapid diffusion through bits. The Enocoro cipher~\cite{watanabe2008enocoro} was developed in response to the hardware-oriented eSTREAM ciphers, but its PRNG is inspired by PANAMA~\cite{daemen1998fast}, making it suitable for software and hardware implementation. The hardware implementation of Enocoro is further improved by including an extra substitution alongside its \ac{SPN}.
        
        The Grain cipher is regarded as the most hardware-efficient cipher among eSTREAM Profile~2 ciphers, and its design has served as an inspiration for certain other ciphers. The BEAN cipher~\cite{kumar2009bean} is a variation of the Grain stream cipher with a difference of both shift registers being replaced by \acp{FCSR} with the addition of S-Box for better diffusion. The Sprout cipher~\cite{armknecht2015lightweight} was submitted at Fast Software Encryption (FSE, 2015), organized by \ac{IACR}. The cipher is based on the Grain structure but with a shorter internal state and fixed values instead of registers to reduce area consumption. The Grain family has weaknesses in initialization procedures, and Sprout's round key function gives rise to certain vulnerabilities~\cite{lallemand2015cryptanalysis}. As a result, the Fruit-v2 cipher~\cite{ghafari2016fruit} was developed as a successor to Sprout and Grain to minimize the weaknesses. The cipher improves performance with a new round key design, longer keystream, and lighter feedback function.   

        The Quavium~\cite{tian2012quavium} and COZMO~\cite{bonnerji2018cozmo} ciphers were developed based on the structure of the Trivium. The Quavium cipher was designed on the basic structure of Trivium and extended into a scalable form by using the Trivium shift registers in a coupling connection. The COZMO encryption method was developed by combining Trivium and A5/1 algorithms with an emphasis on improving key randomness. The modified A5/1 cipher utilizes the Trivium algorithm-generated key stream to leverage the strengths of both algorithms for increased speed and efficiency. 


        Certain ciphers were developed using distinct architectures, such as Cellular Automata and Welsh-Gong transformations. The CAR30 cipher~\cite{das2013car30}~\cite{fan2013wg} is based on Cellular Automata (CA) which provides efficient performance in hardware and software. The speed and throughput are improved by generating an entire keystream block at each iteration and parallel transformation of state bits in each cycle. The WG-8 cipher~\cite{fan2013wg} was introduced to address the weaknesses discovered in WG-7~\cite{orumiehchiha2012cryptanalysis}. This was achieved by increasing the tap positions in \ac{LFSR} and inheriting the randomness properties provided by Welsh-Gong transformations. As a result, the new WG structure provides improved randomness of keys and stronger security.


        Certain ciphers were designed to fulfill the security requirements of particular hardware or specific applications. The TinyStream cipher~\cite{chen2010tinystream} uses the Trusted Platform Module (TPM) specifically developed for \ac{WSN}. In the process of generating a secure keystream, 128 random bits can be produced for each rotation of the state. The A2U2 cipher is a hardware-oriented cipher that was specifically developed to provide security in printed ink RFID tags. The compact design is achieved through a short-length register, \ac{LFSR}-based counter, and non-linear boolean functions. The Pandaka cipher~\cite{chen2014pandaka} was specifically designed for RFID tags with a novel concept of creating a keystream, passing it to a reader, and allowing the reader to immediately and covertly inform the tags of the keystream. The Espresso encryption algorithm was presented as a 5G wireless communication solution to balance reduced hardware costs with fast execution. This was achieved by reducing the size of feedback functions and incorporating maximum parallelizability in the Galois configuration. Lizard~\cite{hamann2017lizard} was submitted in FSE and developed for passive RFID tags with low power consumption. The hardware efficiency was improved using Grain's structure with changes such as smaller state, larger key, and double loading of the secret key during initialization. The A4 encryption algorithm uses a single \ac{LFSR} and \ac{FCSR}, resulting in faster and less complex processing. The cipher has two separate algorithms, one for the sender and one for the receiver.

        The LSC~\cite{noura2019lightweight} and LESCA~\cite{noura2023lesca} ciphers utilized the dynamic key dependence approach for secure and efficient implementation. The security of LSC is improved by generating a unique dynamic key for each input message using a secret key and a nonce. The performance is enhanced by employing simple operations such as Xorshift PRNG while omitting block permutation and diffusion operations. The LESCA cipher utilizes a unique method that adjusts keys dynamically to balance efficient performance and security. The cipher's compactness is achieved through simple logical operations, a lightweight PRNG, and a permutation function.

        \begin{figure}
            \includegraphics[width=\textwidth, height=3cm, keepaspectratio]{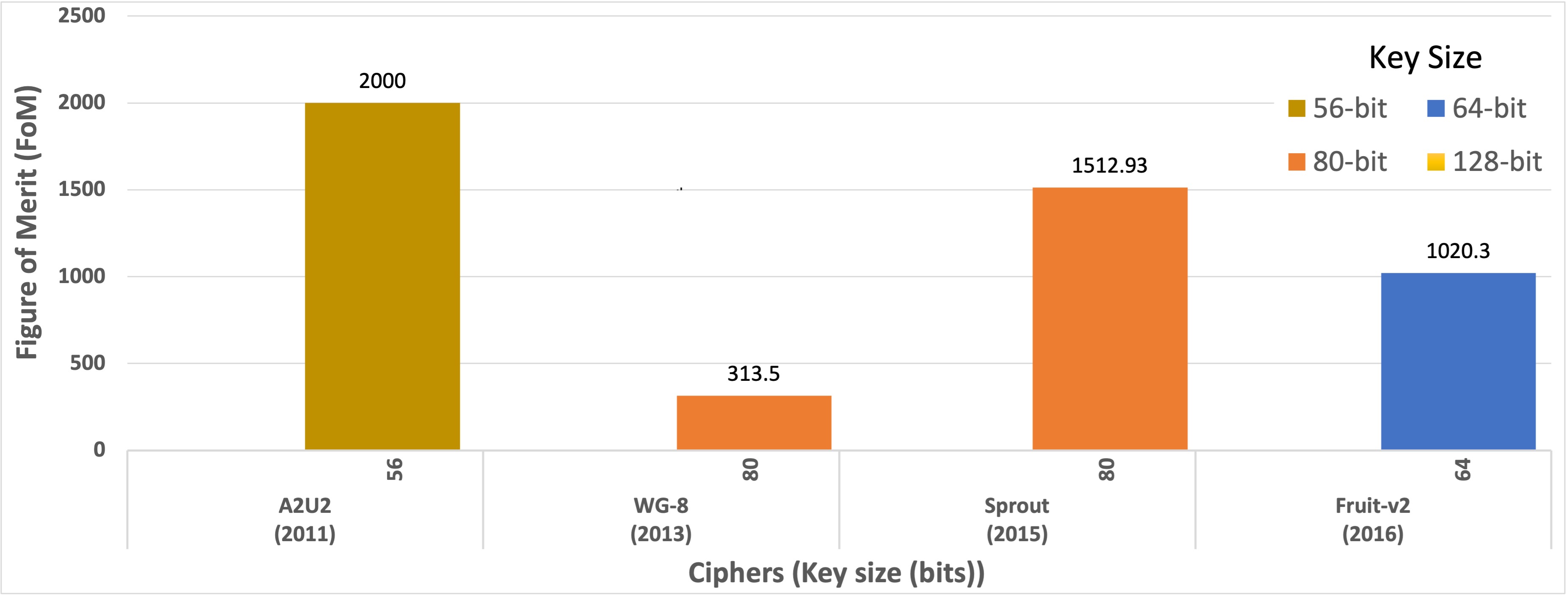}
           ~\caption{Figure of Merit (FoM) of Other Lightweight Stream Ciphers.}
            \label{fig:FOM_OLWSC}
            \Description{Figure of Merit (FoM) of other lightweight stream ciphers}
        \end{figure}
        
        \begin{wrapfigure} {r} {0.5\textwidth}
            \includegraphics[width=1\linewidth, height=4cm, keepaspectratio]{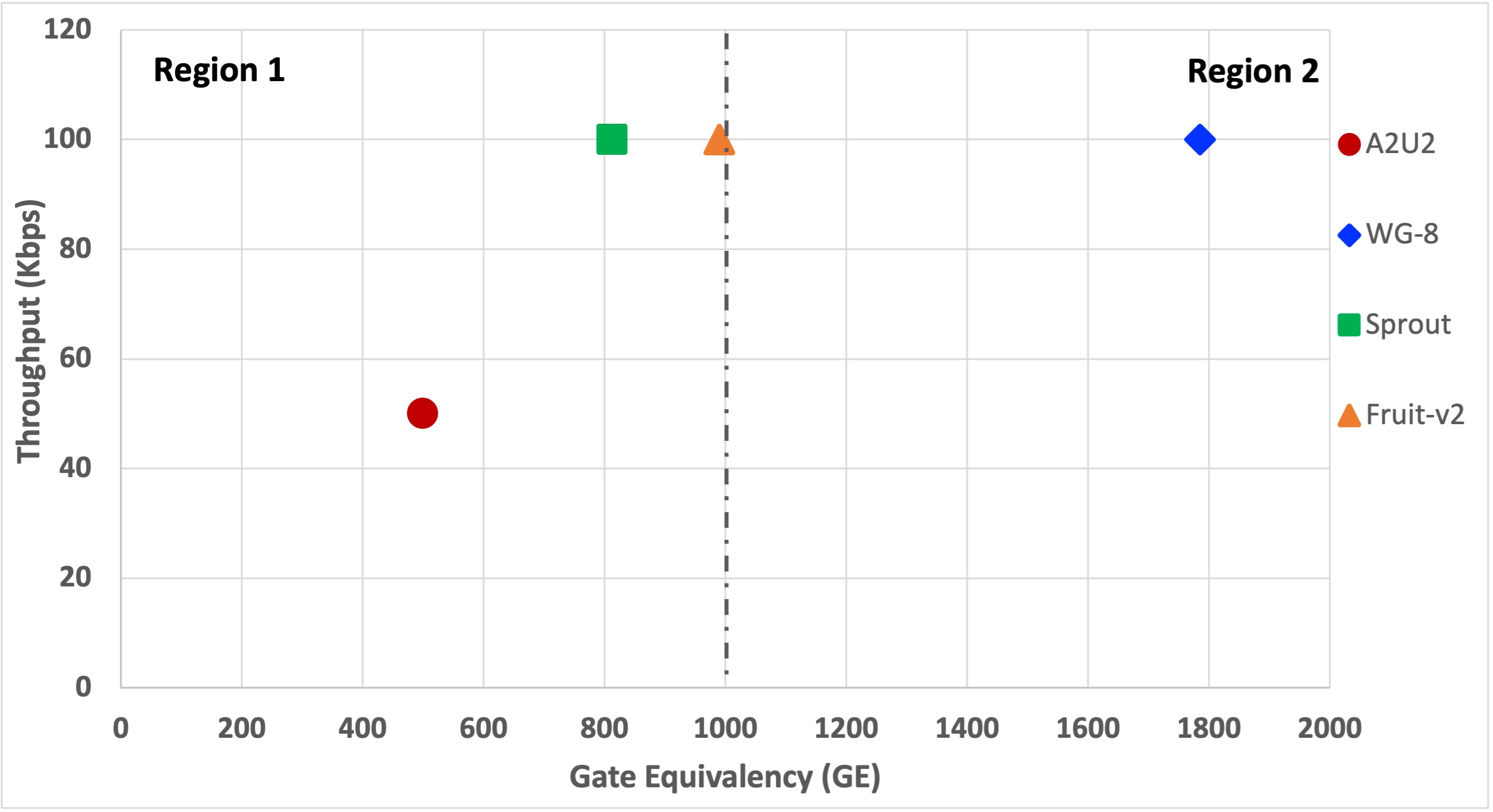}
           ~\caption{Throughput vs. Gate Equivalency (GE) of OLWSC.}
            \label{fig:Throughput_GE_OLWSC}
            \Description{Throughput vs Gate Equivalency of OLWSC}
        \end{wrapfigure}
        
        As per the hardware implementation data (see Appendix Table~\ref{tab:HW_imp_OLWSC}), it can be observed that there has been a general reduction in the area consumption of the ciphers. A2U2 has the lowest area consumption with a throughput of 50 Kbps. Quavium has substantial area consumption but still performs better than Trivium. The average area consumption of the ciphers in this region lies within 1500 GE. The eSTREAM (Profile 2) ciphers have significantly greater throughput, but the GE is also considerably greater. Figure~\ref{fig:FOM_OLWSC} and~\ref{fig:Throughput_GE_OLWSC} only show A2U2, WG-8, Sprout, and Fruit-v2 ciphers as they have both GE and throughput values. The dot plot representation illustrates that the mentioned ciphers have remarkably low area consumption and optimal throughput. Also, Figure~\ref{fig:FOM_OLWSC} illustrates that A2U2 provides the best overall hardware performance, and the least is provided by WG-8. The Sprout and Fruit v2 ciphers exhibit a much better \ac{FoM} as compared to the WG-8 cipher.

        Figure~\ref{fig:Throughput_GE_OLWSC} also provides applicability categorization of A2U2, WG-8, and Sprout. A2U2 and Sprout lie in region~1, while WG-8 lies in region~2. Thus, WG-8 is appropriate for WSN, RFID, and IoT applications, whilst Sprout is better suited for RFID or small IoT sensors.  Based only on the factor of area consumption, ChaCha, A2U2, Pandaka-16, Sprout, and Fruit-v2 are ultra-lightweight ciphers and designed for extremely constrained devices such as RFID and small \ac{IoT} sensors. LESCA is specifically designed for real-time \ac{IoT}. WG-8, Pandaka-32, and Expresso are low-cost lightweight ciphers that are developed for devices such as RFID, WSN, and \ac{IoT} sensors.

         
        \begin{table}[t]
\renewcommand{\arraystretch}{1.5}
\scriptsize
\caption{Comparative Analysis of LHC Ciphers.}
\label{tab:CompAna_LHC}
\begin{tabular}{l c l l }
\toprule
\textbf{Cipher} & \textbf{Year} & \textbf{Resistant Against Attacks}  & \textbf{Proven Vulnerabilities}
\\
\midrule
KATAN/KTANTAN~\cite{canniere2009katan} & 2009 & Differential; Linear; Slide; Related-key; Algebraic &  \ac{MITM}~\cite{bogdanov20103}
\\
\hline
Hummingbird \cite{engels2010hummingbird} & 2010 & Linear; Algebraic; Cube; Slide; Related-key & Key-recovery~\cite{saarinen2011cryptanalysis}; Differential~\cite{chen2013cryptanalysis}; Differential fault analysis~\cite{salehani2011differential}
\\ 
\hline
Hummingbird-2 \cite{engels2011hummingbird} & 2011 & Differential; Linear; Algebraic; Cube; Slide & Related-key~\cite{zhang2012real};  Side channel cube~\cite{fan2012security}; Diff Sequence Analysis~\cite{chai2012cryptanalysis}
\\ 
\hline
SEPAR \cite{vahi2020separ} & 2020 & Differential; Linear; Algebraic; Related-key; Key collision\\
\bottomrule
\end{tabular} 
\end{table}
        
        \begin{wrapfigure} {r} {0.5\textwidth}
            \includegraphics[width=\textwidth, height=4.5cm, keepaspectratio]{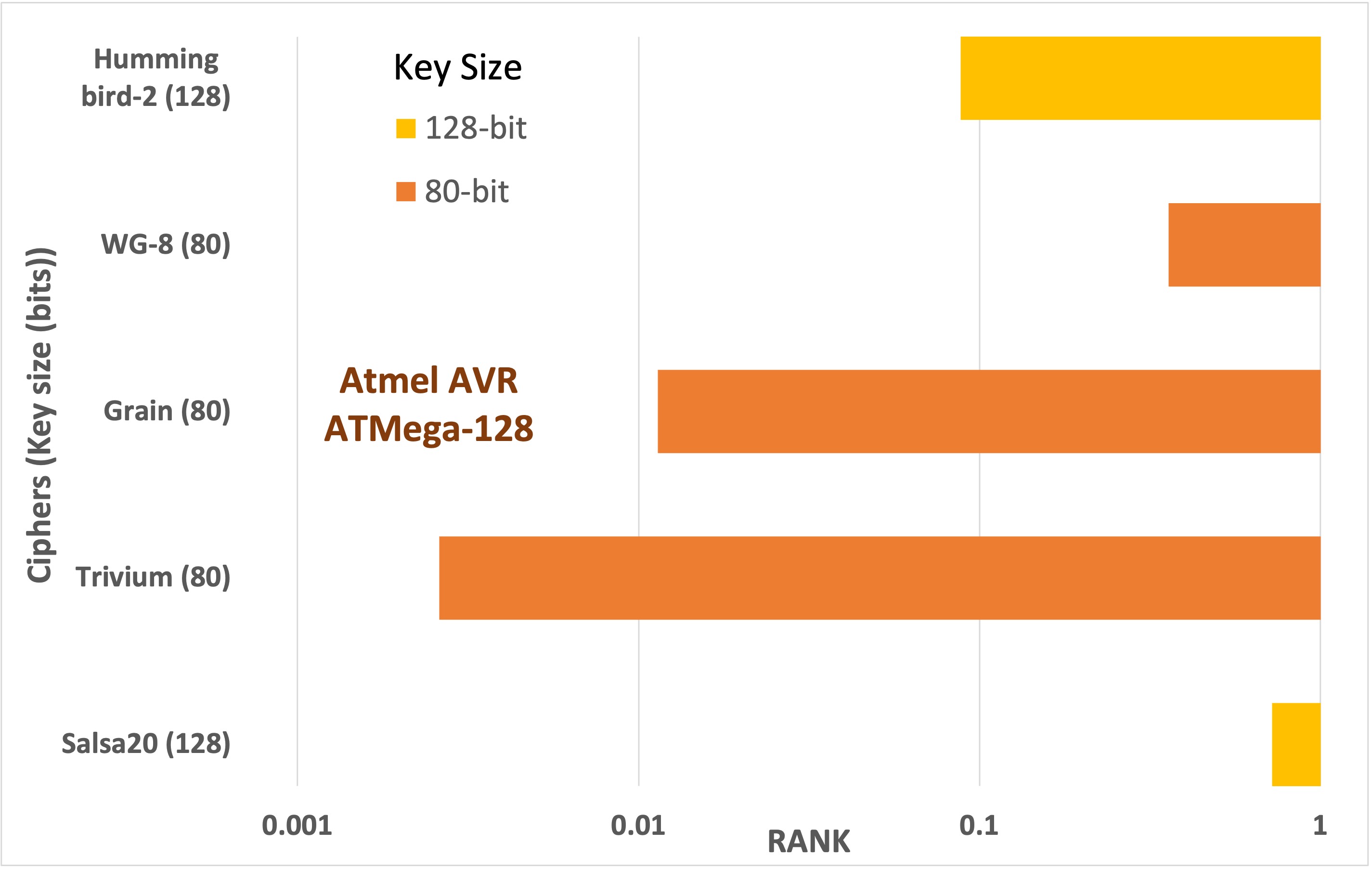}
           ~\caption{RANK of Lightweight Stream Ciphers.}
            \label{fig:RANK_LWSC}
            \Description{RANK of lightweight stream ciphers}
        \end{wrapfigure}
        The comprehensive software implementation was provided for a few \acp{LWSC} on Atmel AVR ATMega-128L~\cite{fan2013wg} as illustrated in Figure~\ref{fig:RANK_LWSC}.  The calculated RANK shows that the Salsa20 outperforms other \acp{LWSC} that were taken into consideration. Salsa20 is followed by WG-8, which is substantially more energy-efficient than Salsa20. Trivium and Grain have the lowest overall software performance as the ciphers are hardware-oriented. Hummingbird is a hybrid cipher with a high code size, but it strikes the optimum balance between RAM, total cycles, and energy usage (see Appendix Table~\ref{tab:SW_imp_OLWSC} for detailed measurements). 
        
    \subsection{Lightweight Hybrid Cipher (LHC)}
        
    \acfp{LHC} combine features of both block and stream ciphers into merged structures that enable efficient software and hardware implementation, as well as improved security against a wide range of vulnerabilities. Such amalgamations are difficult to achieve, and only the following hybrid lightweight ciphers have been developed: KATAN, KTANTAN~\cite{canniere2009katan}, Hummingbird~\cite{engels2010hummingbird, engels2011hummingbird} and SEPAR~\cite{vahi2020separ}. The related strengths and proven vulnerabilities of these ciphers are given in Table~\ref{tab:CompAna_LHC}.
    
    KATAN and KTANTAN are designed for small sensors and devices that are only initialized once, respectively, due to their smaller block sizes. These ciphers draw inspiration from the Trivium stream cipher and adopt its two registers. The compactness is achieved through the use of shift registers, small block size and internal state, and a simple key schedule based on \ac{LFSR}.   
    Hummingbird uses a novel technique of combining block cipher and stream cipher inspired by the rotor mechanism of the Enigma machine~\cite{engels2010hummingbird}. Cryptanalysis performed on Hummingbird found that the cipher is vulnerable to certain attacks~\cite{saarinen2011cryptanalysis, chen2013cryptanalysis} due to the small number of state bits. To tackle those vulnerabilities, the revised Hummingbird-2 cipher~\cite{engels2011hummingbird} introduced MAC capability. Hummingbird-2 can be used to implement security in low-cost, ubiquitous devices such as sensors and \acp{RFID} since it has a very small hardware and software footprint. As \ac{IoT} applications became more popular, SEPAR~\cite{vahi2020separ} was proposed to provide a better balance between security, performance, and cost, providing strong cryptographic features while using less energy. The cipher also sought to handle the vulnerabilities related to Hummingbird. 
    
    The hardware implementation of KATAN in 32-bit, 48-bit, and 64-bit variants exhibit incredibly low \ac{GE} values of 802, 927, and 1054 respectively. Similarly, KTANTAN offers an incredibly low \ac{GE} of 462, 588, and 688. Both implementations have a throughput of 12.5, 18.8, and 25.1 Kbps at 100 KHz. The hardware implementation of SEPAR boasts a throughput of 136.3 Kbps at 12 MHz and an execution time of 117.3 $\mu$s. When compared to Hummingbird-2, which has a throughput of 51 Kbps and an execution time of 316.5 $\mu$s, SEPAR outperforms with faster speed and more efficient execution.
    
    Comparing the software implementations of the \ac{LHC} ciphers such as Hummingbird and SEPAR done on the Atmel AVR ATMega 128L microcontroller~\cite{vahi2020separ} (see Appendix Table~\ref{tab:SW_imp_LHC}) illustrates that Hummingbird-1 has better throughput, lower number of cycles, and code size than SEPAR. An improvement to Hummingbird-1 was made with Hummingbird-2. This indicates that Hummingbirdbird-2 outperforms SEPAR in terms of overall software performance. 
    
    \begin{figure} [t]
        \includegraphics[width=\textwidth, height=6.9cm, keepaspectratio]{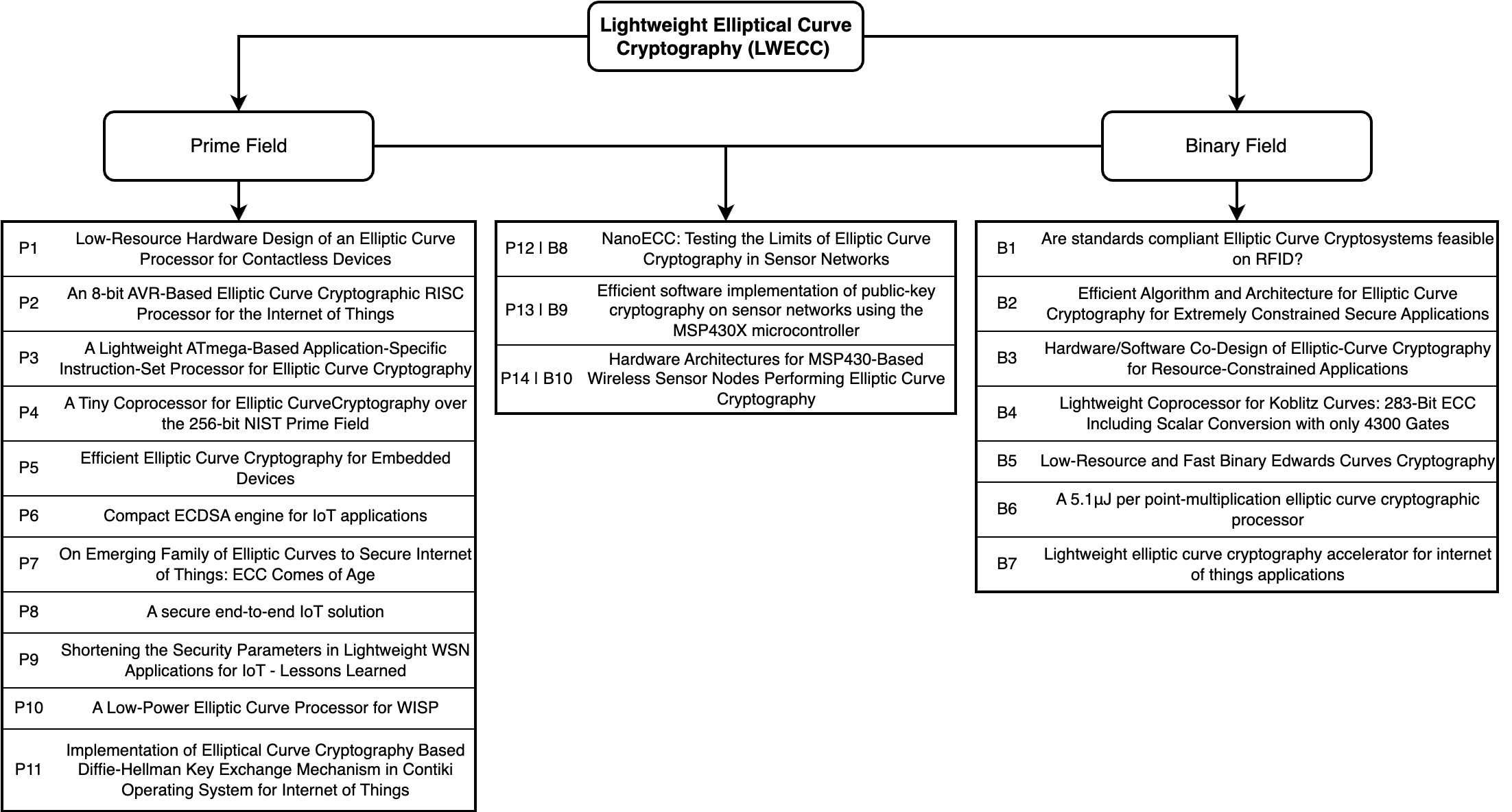}
       ~\caption{Classification of Lightweight Elliptical Curve Implementations.}
        \label{fig:classification_LWECC_cryptography}
        \Description{The Lightweight elliptical curve publications distributed among the prime and binary field.}
    \end{figure}
    
    \section{Lightweight Asymmetric Cipher}
    Public-key cryptography was a game-changing innovation in the field of cryptography. The implementation of algorithms such as RSA in embedded or \ac{IoT} systems is, however, not practical due to the longer key lengths and the delays included in the encryption and decryption process. In response, notable progress towards lightweight asymmetric ciphers has been made by the emergence of \ac{LWECC}.
        
        Elliptic Curve Cryptography (ECC) is a public-key cryptography technique that is based on elliptic curves over finite fields, proposed by Neal Koblitz and Victor Miller in 1985. ECC utilizes shorter keys than RSA for the same level of security and offers very quick key generation, key agreement, and signatures. We categorize the \ac{LWECC} implementation based on their use of either prime or binary field curves, as shown in Figure~\ref{fig:classification_LWECC_cryptography}. This survey only focuses on \ac{LWECC} implementations that are appropriate for resource-limited devices and have detailed implementation measurements available.

        \subsubsection{Prime Field}
  
        A prime field is a finite field over which elliptic curve equations and arithmetic functions are defined. Prime fields are denoted by $F_p$, where $p$ is a prime number. The field $F_p$ comprises integers ranging from 0 to $p-1$, and all mathematical operations such as addition, subtraction, multiplication, and inversion are computed modulo $p$. Table~\ref{tab:CompAna_prime_LWECC} enumerates the publications based on $F_p$ and the associated elliptical curve implementation, as well as the application of each publication. 
        
        \begin{table}[t]
\renewcommand{\arraystretch}{1.5}
\scriptsize
\caption{Comparative Analysis of Prime Field based LWECC implementation.}
\label{tab:CompAna_prime_LWECC}
\begin{tabular}{l c l l l}
\toprule
\textbf{Publication} & \textbf{Year} & \textbf{Curve}  & \textbf{Implementation} & \textbf{Application}
\\
\midrule
P1~\cite{wenger2010low} & 2010 & Weierstrass & Digital Signatures & RFID
\\ 
\hline
P2~\cite{wenger20128} & 2012 & Montgomery, Weierstrass, Edwards & Scalar Multiplication & IoT
\\ 
\hline
P3~\cite{wenger2013lightweight} & 2013 & Weierstrass & Scalar Multiplication & RFID
\\ 
\hline
P4~\cite{bosmans2016tiny} & 2016 & Weierstrass & Scalar Multiplication
\\ 
\hline
P5~\cite{liu2016efficient} & 2016 & Montgomery & Scalar Multiplication & Embedded devices
\\ 
\hline
P6~\cite{yalccin2016compact} & 2016 & Montgomery & Scalar Multiplication & IoT
\\ 
\hline
P7~\cite{liu2016emerging} & 2016  & MoTE & Scalar Multiplication & IoT
\\ 
\hline
P8~\cite{mathur2017secure} & 2017 & Weierstrass & Key Establishment & IoT
 \\ 
\hline
P9~\cite{sojka2017shortening} & 2017 &  & Scalar Multiplication & WSN 
 \\ 
\hline
P10~\cite{cabana2021low} & 2021 & Weierstrass & Scalar Multiplication & RFID
 \\ 
\hline
P11~\cite{thaparimplementation} & 2021 & Koblitz & Key Establishment & IoT
 \\ 
\hline
P12 | B8~\cite{szczechowiak2008nanoecc} & 2008 & Weierstrass & Scalar Multiplication & WSN
\\ 
\hline
P13 | B9~\cite{gouvea2012efficient} & 2012  & Weierstrass & Signatures, key establishment  & WSN
\\ 
\hline
P14 | B10~\cite{wenger2013hardware} & 2013 & Weierstrass & Scalar Multiplication & WSN\\
\bottomrule
\end{tabular} 
\end{table}      
        
        In this section, we will discuss the construction techniques used to provide a lightweight implementation of ECC in the prime field. The most computationally intensive task in ECC is multiplication. Hence, most implementations in this section aim to provide a cost-effective and fast multiplication approach such as P2. P2~\cite{wenger20128} provides an 8-bit Application-Specific Instruction-set Processor (ASIP) design that is implemented on four different families of elliptical curves. All of these curves feature \ac{OPF} that are integrated with a multiply-accumulate unit. The \ac{OPF} improves arithmetic efficiency with a multiplication-based reduction operation, and the multiply-accumulate unit accelerates long integer arithmetic. In case of P9~\cite{sojka2017shortening}, the short ECC algorithm is designed to operate over finite prime fields with small values, typically less than 256 bits. This approach reduces the number of operations required for cryptographic computations, leading to greater efficiency and faster processing times.

        Several \ac{LWECC} implementations, including P1~\cite{wenger2010low}, P3~\cite{wenger2013lightweight}, P4~\cite{bosmans2016tiny}, P8~\cite{mathur2017secure}, and P10~\cite{cabana2021low}, have been applied on the Weierstrass curve. P1 provides an optimized small design for a 16-bit microcontroller executing Elliptic Curve Digital Signature Algorithm (ECDSA) with low resource requirements. This design uses an improved point multiplication technique that involves performing doubling and adding operations in a single function, a Montgomery ladder to calculate multiplication without y-coordinates, and minimal inversion operations. In P3, multi-precision multiplication is accelerated by combining operand-caching multiplication with multiply-accumulate instruction, reducing load, store, and addition operations. P4 implements a memory-mapped coprocessor for a 16-bit microcontroller. It uses a product scanning algorithm for integer multiplication and Mersenne prime for modular reduction, resulting in an efficient and low-cost implementation. P10 uses Montgomery multiplication for modular multiplication and Fermat's theorem for modular multiplicative inverse calculation in programmable RFID tags. 

        The Montgomery curve is used for the implementation in P5~\cite{liu2016efficient} and P6~\cite{yalccin2016compact}. P5 offers \ac{OPF} with performance and security optimizations. Operand caching~\cite{hutter2011fast} and lazy doubling~\cite{seo2013multi} were applied in multiplication and squaring, respectively, while an extended Euclidean algorithm was used for inversion. P6 offers an ECDSA engine that performs point multiplication through Montgomery modular multiplication using two $4\times32$-bit multipliers as adder trees. To minimize area, the modular multiplication unit is applied directly to RAM with only a few registers for operand or temporary storage.

    \begin{figure}
        \includegraphics[width=\linewidth, height=4.4cm, keepaspectratio]{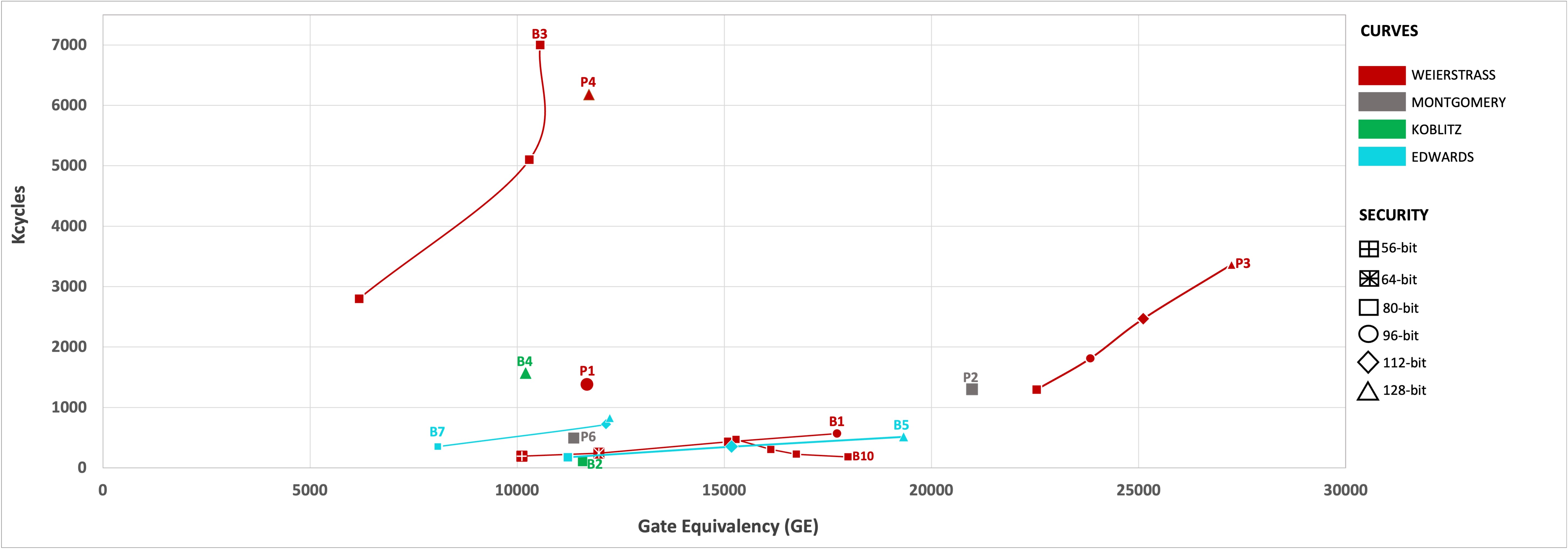}
       ~\caption{Gate Equivalency (GE) vs Cycles of Lightweight Elliptical Curve Cryptographic Implementations.}
        \label{fig:GC_LWECC}
        \Description{GE vs cycles of lightweight Elliptical Curve Cryptographic Implementations}
    \end{figure}
    
        Distinct techniques were employed for implementing P7~\cite{liu2016emerging} on MoTE curves and P11~\cite{thaparimplementation} on Koblitz curves. P7 uses a pseudo-Mersenne prime for fast modular reduction and implements the Reverse Product Scanning (RPS) method~\cite{liu2015reverse} for optimized multiplication and squaring. P11 offers a lightweight and low-power Elliptic-curve Diffie–Hellman (ECDH) implementation by using C language for math operations and the Double and Add Method~\cite{silverman2006introduction} to reduce computational complexity. 
        
        The following paragraph presents publications that offer implementations in both binary and prime fields. P12|B8~\cite{szczechowiak2008nanoecc} utilizes Pairing-Based Cryptography (PBC)~\cite{pairingbasedcryptography} with an optimized MIRACL library~\cite{scott2003miracl} for efficient memory management. The library supports large arithmetic operations and offers complete support for ECC on prime and binary fields. P13|B9~\cite{gouvea2012efficient} implements prime and binary field ECC on the MSP430 microcontroller. In the prime field, the Comba algorithm~\cite{comba1990exponentiation} is used for integer multiplication, and the Montgomery algorithm is used for modular reduction. In the binary field, López–Dahab (LD) algorithm~\cite{lopez2000high} computes polynomial multiplication, and the speed of square operation is increased by using precomputed tables in ROM. P14|B10~\cite{wenger2013hardware} has developed a drop-in module specifically designed to carry out finite-field arithmetic on the MSP430 microcontroller. Since the MSP430 does not have a carry-less multiplier, the module efficiently performs polynomial multiplication by using conditional branching and executing multiply-accumulate operations directly on the memory-mapped multiplier.

        Figure~\ref{fig:GC_LWECC} shows the hardware implementations of publications on the prime field, including P1, P2, P3, P4, and P6. Among the publications based on the Weierstrass curve, P1 offers a better balance of GE and total execution cycles than P3 and P4. P4 consumes slightly more area than P1, but its execution cycles are significantly greater, leading to slower implementation. However, P4 offers a significantly higher level of security, with 128 bits, compared to P1, which provides a relatively lower 96-bit security level. P2 and P6 provide 80-bit security on the Montgomery curve, and P6 provides a better hardware implementation than P2 (see Table~\ref{tab:HW_imp_LWECC} in the appendix for detailed measurements).

        The hardware implementation in P1 and P3 publications was implemented for RFID based on the Weierstrass curve, and P2 and P6 using the Montgomery curve for IoT. P1 was found to be more suitable for RFID due to its balance in GE and cycles, while P6 offers a better low-cost implementation option for IoT.
        
        \begin{figure}
            \includegraphics[width=\textwidth, height=5 cm, keepaspectratio]{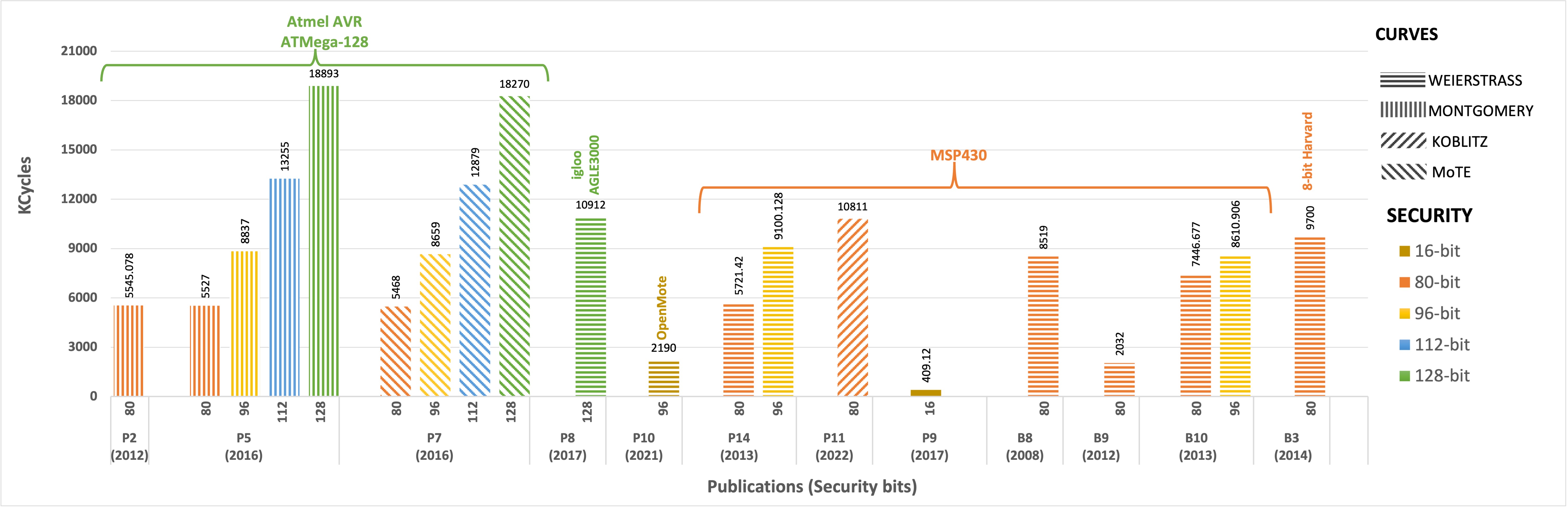}
           ~\caption{Cycles Executed by Lightweight Elliptical Curve Cryptographic Implementations.}
            \label{fig:cycles_LWECC}
            \Description{Cycles of lightweight Elliptical Curve Cryptographic Implementations.}
        \end{figure}
    
        The software implementation presented in Figure~\ref{fig:cycles_LWECC} shows that P2, P5, and P7 were implemented on ATMega-128. The implementation demonstrates that P2 and P5 were implemented on the Montgomery curve, with P5 showing slightly better performance by consuming slightly fewer cycles in terms of 80-bit security. On the other hand, P7 was carried out on the MoTE curve and demonstrated better performance than both P2 and P5. The analysis obtained from the software implementation results of P9, P11, and P14 on MSP430 indicates that P9 consumes the least number of cycles but offers only 16-bit security and is based on a custom shortECC curve. P11 provides 80-bit security carried out on the Koblitz curve, consuming more execution cycles than P9. Meanwhile, P14 offers better performance than P9 carried out on the Weierstrass curve. Taking into consideration all software implementations in Figure~\ref{fig:cycles_LWECC}, P10 offers the best software performance for 96-bit security, while P8 offers the best performance for 128-bit security. For 80-bit and 112-bit security, P7 offers the best software performance (see Table~\ref{tab:SW_imp_LWECC} in the appendix for detailed measurements). 

        Several software implementations were developed for IoT, including P2, P7, P8, and P11. P7 offers the fastest software implementation for IoT with a wide security range. P8 offers 128-bit variant security for IoT and provides the fastest implementation on OpenMote. The publications P9 and P14 provide implementations for WSN. Although P9 offers the fastest implementation, it only provides 16-bit security. In contrast, P14 provides 80 and 96-bit security.
        
        \subsubsection{Binary Field}
        A binary field is a finite field over which arithmetic functions are defined using binary XOR, binary AND, and bitwise XOR. It is represented by $2^m$, where $m$ defines the size of the field. The field is represented by binary polynomials where coefficients are limited to 0 or 1, and the highest degree is $(m-1)$. The following table, presented in Table~\ref{tab:CompAna_binary_LWECC} specifies publications based on $2^m$ and associated elliptic curve implementation, along with their respective applications.
        
        This section focuses on the implementation approaches used to achieve a lightweight implementation of ECC in the binary field. The implementations of B1~\cite{kumar2006standards} and B3~\cite{holler2014hardware} were carried out on the Weierstrass curve. B1 proposes an efficient ASIC implementation of the ECC processor using a modified Montgomery multiplication method and efficient bit-serial multipliers that reduce overall area consumption. B3 utilizes the concept of virtual addressing to develop a hardware accelerator that requires less precomputation, resulting in fast performance and reduced storage demand.

        B2~\cite{azarderakhsh2014efficient}, B4~\cite{sinha2015lightweight}, and B6~\cite{rovzic20175} are implementations that were developed using the Koblitz curve. B2 used a bit-level Gaussian normal basis (GNB) multiplier for point multiplication, and inversions are computed through Fermat's Little Theorem exponentiations~\cite{itoh1988fast}, resulting in a simpler and smaller implementation. B4 provides a design for lightweight and \ac{SCA} resistant coprocessor for 16-bit microcontrollers. The implementation uses lazy reduction, where divisions are implemented with shifts, additions, and subtractions. B6 provides a processor design for ECC with multiple cryptographic protocol support. The register file uses a bit-slice architecture to reduce multiplexers, and computational complexity is reduced with a modular arithmetic logic unit.
        
        B5~\cite{koziel2015low} and B7~\cite{lara2020lightweight} were implemented on the Edwards curve. The addition and doubling formulas for binary Edwards curves enable fast, small, and secure point multiplication. B5 offers a coprocessor design that utilizes GNB to represent field elements and curves. The implementation incorporates Montgomery Ladder, which results in faster processing and minimal register usage. B7 offers an FPGA-based accelerator for ECC operations that reduces hardware requirements with a bit-serial multiplier.  
        
        \begin{table}[t]
\renewcommand{\arraystretch}{1.5}
\scriptsize
\caption{Comparative Analysis of Binary Field based LWECC Implementation.}
\label{tab:CompAna_binary_LWECC}
\begin{tabular}{l c l l l}
\toprule
\textbf{Publication} & \textbf{Year} & \textbf{Curve} & \textbf{Implementation} & \textbf{Application}
\\
\midrule
B1~\cite{kumar2006standards} & 2006 & Weierstrass & Scalar Multiplication & RFID and sensor networks
\\ 
\hline
B2~\cite{azarderakhsh2014efficient} & 2014 & Koblitz & & RFID
\\ 
\hline
B3~\cite{holler2014hardware} & 2014 & Weierstrass & Scalar Multiplication & RFID
\\ 
\hline
B4~\cite{sinha2015lightweight} & 2015 & Koblitz & Scalar Multiplication &
\\ 
\hline
B5~\cite{koziel2015low} & 2015 & Edwards & Scalar Multiplication & RFID
\\ 
\hline
B6~\cite{rovzic20175} & 2016 & Koblitz & Scalar Multiplication & eID and RFID
\\ 
\hline
B7~\cite{lara2020lightweight} & 2019 & Edwards & Scalar Multiplication & IoT\\
\bottomrule
\end{tabular} 
\end{table}
        
        The hardware implementation of the publications presented in Figure~\ref{fig:GC_LWECC} on the binary field shows that among the publications based on the Weierstrass curve, B3 provides the lowest GE but has one of the largest execution cycles. On the other hand, the hardware implementation in B1 and B10 results in fewer cycles but a higher \ac{GE} than B3. Among the different implementations on the Koblitz curve, B4 consumes the least \ac{GE} but requires more computational cycles. B2 requires more area but fewer cycles compared to B4. Additionally, B4 provides 128-bit security, while B2 provides 80-bit security. When comparing the publications carried out on the Edwards curves, B7 is more efficient in terms of GE, while B5 is more efficient in terms of execution cycles. A high-level comparison analysis indicates that hardware implementations on the Edwards curve provide a better balance of GE and total cycles at a lower performance cost.

        The binary curves provide support for efficient and better hardware implementation. B1, B2, B3, B5, and B6 offer secure hardware implementation for RFID technology. Out of these, B3 consumes the least area, but the computational cycles are the highest. In contrast, B2, which is implemented on the Koblitz curve, has the lowest number of cycles while maintaining a better balance of total execution cycles and area consumption. B7 implements IoT with a wide range of security, while B10 provides an 80-bit secure implementation for WSNs.

        Figure~\ref{fig:cycles_LWECC} illustrates the software implementations performed on the Weierstrass curve. B8, B9, and B10 have been implemented on MSP430. B3 has been implemented on an 8-bit Harvard architecture. B9 delivers the highest performance with the least amount of computational clocks required for execution, while B3 is executed with the highest computational clocks, resulting in a comparatively slower performance (see Appendix Table~\ref{tab:SW_imp_LWECC}).

        The software implementation in B8, B9, and B10 was provided for WSN. B9 has the fastest implementation in WSN, while B3, developed for RFID, utilizes relatively higher clock cycles that slow down the performance. 
    
\section{Conclusion}

    This systematic review expounds on the evolution of various lightweight cryptography methods useful for ubiquitous devices or embedded systems. We cover the extensive research done on both symmetric and asymmetric cryptographic algorithms. The best hardware-based lightweight cryptographic algorithms for various key lengths, application regions, and performance metrics are summarized in Table~\ref{tab:HW_imp_conc}. The best software implementation, characterized by the least \ac{CpB} consumption, highest throughput, minimal utilization of RAM and ROM, and peak overall software performance (RANK) for 80-bit and 128-bit key size are summarized in Table~\ref{tab:SW_imp_conc}. 
    \begin{table}[t]
\renewcommand{\arraystretch}{1.1}
\scriptsize
\caption{Best Performing Lightweight Cryptographic Algorithms based on Hardware Metrics (\ac{GE}, Throughput/CpB, and \ac{FoM}) Across Varied Key Sizes (80, 96, 128-bit) and Application Regions.}
\label{tab:HW_imp_conc}
\begin{tabular}{l l l l l l l l l l l }
\toprule

&  & \multicolumn{3}{c}{\textbf{GE}}& \multicolumn{3}{c}{\textbf{Throughput (*CpB for LWECC)}} & \multicolumn{3}{c}{\textbf{FoM}} \\
\cmidrule(lr){3-5}
\cmidrule(lr){6-8}
\cmidrule(lr){9-11}
&  & \textbf{80-bit} & \textbf{96-bit} & \textbf{128-bit}
                    & \textbf{80-bit} & \textbf{96-bit }& \textbf{128-bit} 
                    & \textbf{80-bit} &\textbf{96-bit} & \textbf{128-bit} 
\\ \midrule

\multirow{4}{*}{{\makecell{\textbf{LWBC}}}} &

Region-1 & PRINT & Simeck & CHAM
         & PRINT & Simon & DoT
         & PRINT & Simon & DoT
\\ 
\cline{2-11}
& Region-2 & QTL  & EPCBC & ALLPC 
           & RECTANGLE & SFN & PIPO
           & QTL & SFN & PIPO
\\ 
\cline{2-11}
& Region-3 & KLEIN & SEA & RECTANGLE
           & KLEIN & KLEIN & SKINNY
           & KLEIN & KLEIN & RECTANGLE
\\ 
\cline{2-11}
& Region-4 & & mCrypton & PRINCE 
           & & mCrypton & CLEFIA
           & & mCrypton & PRINCE
\\ 
\hline
\multirow{4}{*}{{\makecell{\textbf{LWSC}}}} &

Region-1 & Sprout & Pandaka & 
         & Sprout &  & 
         & Sprout &  & 
\\ 
\cline{2-11}
& Region-2 & Grain v1  &  & Lizard
           & Grain v1 &  & Grain 128
           & Grain v1 &  & Grain 128
\\ 
\cline{2-11}
& Region-3 & Trivium &  & FAN
           & Trivium &  & 
           & Trivium &  & 
\\ 
\cline{2-11}
& Region-4 & Mickey 1.0 &  & Mickey 2.0 
           & Mickey 1.0 &  & Mickey 2.0 
           & Mickey 1.0 &  & Mickey 2.0 
\\ 

\hline
\multirow{2}{*}{{\makecell{\textbf{LWECC}}}} &

Prime & P6 & P1 & P4
      & P6 & P1 & P3 
\\ 
\cline{2-11}
& Binary  & B3  &  & B4
          & B2 and B10 &  & B5
\\ 
\bottomrule
\end{tabular}
\end{table}
    \begin{table}[t]
\renewcommand{\arraystretch}{1.1}
\scriptsize
\caption{Best Performing Lightweight Cryptographic Algorithms based on Software Metrics (\ac{CpB}, Throughput, RAM, ROM, and RANK) across Varied Key Sizes (80, 128-bit) on different microcontrollers/processors.}
\label{tab:SW_imp_conc}
\begin{tabular}{l l l l l l l l l l l l }
\toprule

&  & \multicolumn{2}{c}{\textbf{CpB}}& \multicolumn{2}{c}{\textbf{Throughput}} & \multicolumn{2}{c}{\textbf{RAM}} & \multicolumn{2}{c}{\textbf{ROM}} & \multicolumn{2}{c}{\textbf{RANK}} \\
\cmidrule(lr){3-4}
\cmidrule(lr){5-6}
\cmidrule(lr){7-8}
\cmidrule(lr){9-10}
\cmidrule(lr){11-12}
&  & \textbf{80-bit}  & \textbf{128-bit}
                    & \textbf{80-bit} & \textbf{128-bit} 
                    & \textbf{80-bit} & \textbf{128-bit} 
                    & \textbf{80-bit} & \textbf{128-bit} 
                    & \textbf{80-bit} & \textbf{128-bit} 
\\ \midrule

\multirow{3}{*}{{\makecell{\textbf{LWBC}}}} &

ATMEGA-128 & RECTANGLE & PIPO
           & RECTANGLE & SPECK
           & LED & PIPO
           & PRESENT & PIPO
           & RECTANGLE & PIPO
\\ 
\cline{2-12}
& ARM-7 (LPC2129) &  & SIMON
                  &  & SIMON
                  &  & 
                  &  & ANU-II
                  &  & SIMON
\\ 
\cline{2-12}
& ATiny 45 & ITUbee & TEA 
           & & 
           & &
           & &
           & ITUbee & TEA 
\\
\hline
\multirow{2}{*}{{\makecell{\textbf{LWSC}}}} &

ATMEGA-128 & WG-8 & Salsa20
           & WG-8 & HB-2
           & WG-8/Grain & SEPAR
           & Trivium & HB-2
           & WG-8 & Salsa20
\\ 
\cline{2-12}
& Pentium 4 & Trivium & Rabbit
            & Trivium & Rabbit
                  &  & 
                  &  &
                  &  &
\\ 
\hline
\multirow{2}{*}{{\makecell{\textbf{LWECC}}}} &

Prime      & P7 & P7
           &  & 
           &  & 
           &  & 
           &  & 
\\ 
\cline{2-12}
& Binary & B9 & 
         &  & 
         & B10 &   
         & B10 &
         &  &
\\ 
\bottomrule
\end{tabular}
\end{table}
    
\bibliographystyle{ACM-Reference-Format}
\bibliography{bibliography}

@String{BIT = "{BIT}" }

@String{Computing = "Computing" }

@String{Computer = "{IEEE} Computer" }

@String{Springer = "Springer-Verlag" }

@inproceedings{papp2015embedded,
  title={Embedded systems security: Threats, vulnerabilities, and attack taxonomy},
  author={Papp, Dorottya and Ma, Zhendong and Buttyan, Levente},
  booktitle={2015 13th Annual Conference on Privacy, Security and Trust (PST)},
  pages={145--152},
  year={2015},
  organization={ieee}
}

@article{beaulieu2015simon,
  title={SIMON and SPECK: Block Ciphers for the Internet of Things},
  author={Beaulieu, Ray and Shors, Douglas and Smith, Jason and Treatman-Clark, Stefan and Weeks, Bryan and Wingers, Louis},
  journal={Cryptology ePrint Archive},
  year={2015}
}

@misc{cryptographic_technology_group_2017,
  author       = {{National Institute of Standards and Technology (NIST)}},
  title        = {Lightweight Cryptography: CSRC},
  howpublished = {\url{https://csrc.nist.gov/projects/lightweight-cryptography}},
  year         = {2017},
  month        = jan,
  note         = {Accessed: 2026-02-16}
}

@article{hatzivasilis2018review,
  title={A review of lightweight block ciphers},
  author={Hatzivasilis, George and Fysarakis, Konstantinos and Papaefstathiou, Ioannis and Manifavas, Charalampos},
  journal={Journal of cryptographic Engineering},
  volume={8},
  number={2},
  pages={141--184},
  year={2018},
  publisher={Springer}
}

@article{sevin2021survey,
  title={A survey on software implementation of lightweight block ciphers for IoT devices},
  author={Sevin, Abdullah and Mohammed, Abdu Ahmed Osman},
  journal={Journal of Ambient Intelligence and Humanized Computing},
  pages={1--15},
  year={2021},
  publisher={Springer}
}

@article{mohd2015survey,
  title={A survey on lightweight block ciphers for low-resource devices: Comparative study and open issues},
  author={Mohd, Bassam J and Hayajneh, Thaier and Vasilakos, Athanasios V},
  journal={Journal of Network and Computer Applications},
  volume={58},
  pages={73--93},
  year={2015},
  publisher={Elsevier}
}

@article{manifavas2016survey,
  title={A survey of lightweight stream ciphers for embedded systems},
  author={Manifavas, Charalampos and Hatzivasilis, George and Fysarakis, Konstantinos and Papaefstathiou, Yannis},
  journal={Security and Communication Networks},
  volume={9},
  number={10},
  pages={1226--1246},
  year={2016},
  publisher={Wiley Online Library}
}

@inproceedings{noura2019lightweight,
  title={Lightweight stream cipher scheme for resource-constrained IoT devices},
  author={Noura, Hassan and Couturier, Rapha{\"e}l and Pham, Congduc and Chehab, Ali},
  booktitle={2019 International Conference on Wireless and Mobile Computing, Networking and Communications (WiMob)},
  pages={1--8},
  year={2019},
  organization={IEEE}
}

@incollection{bokhari2018comparative,
  title={A comparative study on lightweight cryptography},
  author={Bokhari, MU and Hassan, Shabbir},
  booktitle={Cyber Security},
  pages={69--79},
  year={2018},
  publisher={Springer}
}

@inproceedings{sallam2018survey,
  title={A survey on lightweight cryptographic algorithms},
  author={Sallam, Suzan and Beheshti, Babak D},
  booktitle={TENCON 2018-2018 IEEE Region 10 Conference},
  year={2018}
}

@article{dhanda2020lightweight,
  title={Lightweight cryptography: a solution to secure IoT},
  author={Dhanda, Sumit Singh and Singh, Brahmjit and Jindal, Poonam},
  journal={Wireless Personal Communications},
  volume={112},
  number={3},
  pages={1947--1980},
  year={2020},
  publisher={Springer}
}

@incollection{guria2021lightweight,
  title={Lightweight Cryptography in Cloud-Based IoT: An Analytical Approach},
  author={Guria, Payel and Bhattacharyya, Aditya},
  booktitle={Integration and Implementation of the Internet of Things Through Cloud Computing},
  pages={190--216},
  year={2021},
  publisher={IGI Global}
}

@article{shannon1949communication,
  title={Communication theory of secrecy systems},
  author={Shannon, Claude E},
  journal={The Bell system technical journal},
  volume={28},
  number={4},
  pages={656--715},
  year={1949},
  publisher={Nokia Bell Labs}
}

@inproceedings{lim2005mcrypton,
  title={mCrypton--a lightweight block cipher for security of low-cost RFID tags and sensors},
  author={Lim, Chae Hoon and Korkishko, Tymur},
  booktitle={International workshop on information security applications},
  pages={243--258},
  year={2005},
  organization={Springer}
}

@inproceedings{bogdanov2007present,
  title={PRESENT: An ultra-lightweight block cipher},
  author={Bogdanov, Andrey and Knudsen, Lars R and Leander, Gregor and Paar, Christof and Poschmann, Axel and Robshaw, Matthew JB and Seurin, Yannick and Vikkelsoe, Charlotte},
  booktitle={International workshop on cryptographic hardware and embedded systems},
  pages={450--466},
  year={2007},
  organization={Springer}
}

@inproceedings{knudsen2002integral,
  title={Integral cryptanalysis},
  author={Knudsen, Lars and Wagner, David},
  booktitle={International Workshop on Fast Software Encryption},
  pages={112--127},
  year={2002},
  organization={Springer}
}

@article{biham1994new,
  title={New types of cryptanalytic attacks using related keys},
  author={Biham, Eli},
  journal={Journal of Cryptology},
  volume={7},
  number={4},
  pages={229--246},
  year={1994},
  publisher={Springer}
}

@inproceedings{biryukov2000advanced,
  title={Advanced slide attacks},
  author={Biryukov, Alex and Wagner, David},
  booktitle={International conference on theory and applications of cryptographic techniques},
  year={2000}
}

@inproceedings{canniere2009katan,
  title={KATAN and KTANTAN—a family of small and efficient hardware-oriented block ciphers},
  author={Canni{\`e}re, Christophe De and Dunkelman, Orr and Kne{\v{z}}evi{\'c}, Miroslav},
  booktitle={International Workshop on Cryptographic Hardware and Embedded Systems},
  pages={272--288},
  year={2009},
  organization={Springer}
}

@inproceedings{knudsen2010printcipher,
  title={PRINTcipher: a block cipher for IC-printing},
  author={Knudsen, Lars and Leander, Gregor and Poschmann, Axel and Robshaw, Matthew JB},
  booktitle={International Workshop on Cryptographic Hardware and Embedded Systems},
  pages={16--32},
  year={2010},
  organization={Springer}
}

@inproceedings{gong2011klein,
  title={KLEIN: a new family of lightweight block ciphers},
  author={Gong, Zheng and Nikova, Svetla and Law, Yee Wei},
  booktitle={International workshop on radio frequency identification: security and privacy issues},
  pages={1--18},
  year={2011},
  organization={Springer}
}

@inproceedings{guo2011led,
  title={The LED block cipher},
  author={Guo, Jian and Peyrin, Thomas and Poschmann, Axel and Robshaw, Matt},
  booktitle={International workshop on cryptographic hardware and embedded systems},
  pages={326--341},
  year={2011},
  organization={Springer}
}

@inproceedings{borghoff2012prince,
  title={PRINCE--a low-latency block cipher for pervasive computing applications},
  author={Borghoff, Julia and Canteaut, Anne and G{\"u}neysu, Tim and Kavun, Elif and Knezevic, Miroslav and Knudsen, Christian and others},
  booktitle={International conference on the theory and application of cryptology and information security},
  year={2012},
  organization={Springer}
}

@inproceedings{aldabbagh2014olbca,
  title={Olbca: A new lightweight block cipher algorithm},
  author={Aldabbagh, Sufyan Salim Mahmood and Al Shaikhli, Imad Fakhri Taha},
  booktitle={2014 3rd International Conference on Advanced Computer Science Applications and Technologies},
  pages={15--20},
  year={2014},
  organization={IEEE}
}

@inproceedings{banik2015midori,
  title={Midori: A block cipher for low energy},
  author={Banik, Subhadeep and Bogdanov, Andrey and Isobe, Takanori and Shibutani, Kyoji and Hiwatari, Harunaga and Akishita, Toru and Regazzoni, Francesco},
  booktitle={International Conference on the Theory and Application of Cryptology and Information Security},
  pages={411--436},
  year={2015},
  organization={Springer}
}

@article{zhang2015rectangle,
  title={RECTANGLE: a bit-slice lightweight block cipher suitable for multiple platforms},
  author={Zhang, Wentao and Bao, Zhenzhen and Lin, Dongdai and Rijmen, Vincent and Yang, Bohan and Verbauwhede, Ingrid},
  journal={Science China Information Sciences},
  volume={58},
  number={12},
  pages={1--15},
  year={2015},
  publisher={Springer}
}

@article{bansod2016pico,
  title={PICO: An Ultra Lightweight and Low Power Encryption Design for Ubiquitous Computing.},
  author={Bansod, Gaurav and Pisharoty, Narayan and Patil, Abhijit},
  journal={Defence Science Journal},
  volume={66},
  number={3},
  year={2016}
}

@inproceedings{banik2017gift,
  title={GIFT: a small present},
  author={Banik, Subhadeep and Pandey, Sumit Kumar and Peyrin, Thomas and Sasaki, Yu and Sim, Siang Meng and Todo, Yosuke},
  booktitle={International Conference on cryptographic hardware and embedded systems},
  pages={321--345},
  year={2017},
  organization={Springer}
}

@inproceedings{kim2020pipo,
  title={PIPO: A lightweight block cipher with efficient higher-order masking software implementations},
  author={Kim, Hangi and Jeon, Yongjin and Kim, Giyoon and Kim, Jongsung and 
          Sim, Bo-Yeon and Han, Dong-Guk and Seo, Hyeongmin},
  booktitle={International Conference on Information Security and Cryptology},
  pages={99--122},
  year={2020},
  organization={Springer}
}

@inproceedings{beierle2016skinny,
  title={The SKINNY family of block ciphers and its low-latency variant MANTIS},
  author={Beierle, Christof and Jean, J{\'e}r{\'e}my and K{\"o}lbl, Stefan and Leander, Gregor and Moradi, Amir and Peyrin, Thomas and Sasaki, Yu and Sasdrich, Pascal and Sim, Siang Meng},
  booktitle={Annual International Cryptology Conference},
  pages={123--153},
  year={2016},
  organization={Springer}
}

@article{nechvatal2001report,
  title={Report on the development of the Advanced Encryption Standard (AES)},
  author={Nechvatal, James and Barker, Elaine and Bassham, Lawrence and Burr, William and Dworkin, Morris and Foti, James and Roback, Edward},
  journal={Journal of research of the National Institute of Standards and Technology},
  volume={106},
  number={3},
  year={2001},
  publisher={National Institute of Standards and Technology}
}

@inproceedings{moradi2011pushing,
  title={Pushing the limits: A very compact and a threshold implementation of AES},
  author={Moradi, Amir and Poschmann, Axel and Ling, San and Paar, Christof and Wang, Huaxiong},
  booktitle={Annual International Conference on the Theory and Applications of Cryptographic Techniques},
  pages={69--88},
  year={2011},
  organization={Springer}
}

@article{park2009security,
  title={Security analysis of mCrypton proper to low-cost ubiquitous computing devices and applications},
  author={Park, Jong Hyuk},
  journal={International Journal of Communication Systems},
  volume={22},
  number={8},
  pages={959--969},
  year={2009},
  publisher={Wiley Online Library}
}

@inproceedings{hao2015meet,
  title={A meet-in-the-middle attack on round-reduced mCrypton using the differential enumeration technique},
  author={Hao, Yonglin and Bai, Dongxia and Li, Leibo},
  booktitle={International Conference on Network and System Security},
  pages={166--183},
  year={2015},
  organization={Springer}
}

@inproceedings{collard2009statistical,
  title={A statistical saturation attack against the block cipher PRESENT},
  author={Collard, Baudoin and Standaert, F-X},
  booktitle={Cryptographers’ Track at the RSA Conference},
  pages={195--210},
  year={2009},
  organization={Springer}
}

@inproceedings{cho2010linear,
  title={Linear cryptanalysis of reduced-round PRESENT},
  author={Cho, Joo Yeon},
  booktitle={Cryptographers’ Track at the RSA Conference},
  pages={302--317},
  year={2010},
  organization={Springer}
}

@inproceedings{ohkuma2009weak,
  title={Weak keys of reduced-round PRESENT for linear cryptanalysis},
  author={Ohkuma, Kenji},
  booktitle={International Workshop on Selected Areas in Cryptography},
  pages={249--265},
  year={2009},
  organization={Springer}
}

@article{ahmadian2015biclique,
  title={Biclique cryptanalysis of the full-round KLEIN block cipher},
  author={Ahmadian, Zahra and Salmasizadeh, Mahmoud and Aref, Mohammad Reza},
  journal={IET Information Security},
  volume={9},
  number={5},
  pages={294--301},
  year={2015},
  publisher={Wiley Online Library}
}

@inproceedings{cheng2017multidimensional,
  title={Multidimensional Zero-Correlation Linear Cryptanalysis on PRINCE},
  author={Cheng, Lu and Pan, Xiaozhong and Wei, Yuechuan and Lv, Liqun},
  booktitle={International Conference on Emerging Internetworking, Data \& Web Technologies},
  pages={222--232},
  year={2017},
  organization={Springer}
}

@inproceedings{wheeler1994tea,
  title={TEA, a tiny encryption algorithm},
  author={Wheeler, David J and Needham, Roger M},
  booktitle={International workshop on fast software encryption},
  year={1994},
  organization={Springer}
}

@article{needham1997tea,
  title={Tea extensions},
  author={Needham, Roger M and Wheeler, David J},
  journal={Report (Cambridge University, Cambridge, UK, 1997)},
  year={1997}
}

@inproceedings{standaert2006sea,
  title={SEA: A scalable encryption algorithm for small embedded applications},
  author={Standaert, Fran{\c{c}}ois-Xavier and Piret, Gilles and Gershenfeld, Neil and Quisquater, Jean-Jacques},
  booktitle={International Conference on Smart Card Research and Advanced Applications},
  pages={222--236},
  year={2006},
  organization={Springer}
}

@inproceedings{leander2007new,
  title={New lightweight DES variants},
  author={Leander, Gregor and Paar, Christof and Poschmann, Axel and Schramm, Kai},
  booktitle={International workshop on fast software encryption},
  pages={196--210},
  year={2007},
  organization={Springer}
}

@article{davies1995pairs,
  title={Pairs and triplets of DES S-boxes},
  author={Davies, Donald and Murphy, Sean},
  journal={Journal of Cryptology},
  volume={8},
  number={1},
  pages={1--25},
  year={1995},
  publisher={Springer}
}

@inproceedings{izadi2009mibs,
  title={MIBS: a new lightweight block cipher},
  author={Izadi, Maryam and Sadeghiyan, Babak and Sadeghian, Seyed Saeed and Khanooki, Hossein Arabnezhad},
  booktitle={International Conference on Cryptology and Network Security},
  pages={334--348},
  year={2009},
  organization={Springer}
}

@inproceedings{wu2011lblock,
  title={LBlock: a lightweight block cipher},
  author={Wu, Wenling and Zhang, Lei},
  booktitle={International conference on applied cryptography and network security},
  pages={327--344},
  year={2011},
  organization={Springer}
}

@inproceedings{karakocc2013itubee,
  title={ITUbee: a software oriented lightweight block cipher},
  author={Karako{\c{c}}, Ferhat and Demirci, H{\"u}seyin and Harmanc{\i}, A Emre},
  booktitle={International Workshop on Lightweight Cryptography for Security and Privacy},
  pages={16--27},
  year={2013},
  organization={Springer}
}

@misc{cryptoeprint:2013:404,
    author       = {Ray Beaulieu and
		    Douglas Shors and
		    Jason Smith and
		    Stefan Treatman-Clark and
		    Bryan Weeks and
		    Louis Wingers},
    title        = {The SIMON and SPECK Families of Lightweight Block Ciphers},
    howpublished = {Cryptology ePrint Archive, Report 2013/404},
    year         = {2013},
    note         = {\url{https://ia.cr/2013/404}},
}

@inproceedings{baysal2015roadrunner,
  title={Roadrunner: A small and fast bitslice block cipher for low cost 8-bit processors},
  author={Baysal, Adnan and {\c{S}}ahin, S{\"u}hap},
  booktitle={Lightweight Cryptography for Security and Privacy},
  pages={58--76},
  year={2015},
  organization={Springer}
}

@article{aldabbagh2017design,
  title={Design 32-bit lightweight block cipher algorithm (dlbca)},
  author={AlDabbagh, Sufyan Salim Mahmood},
  journal={International Journal of Computer Applications},
  volume={166},
  number={8},
  pages={17--20},
  year={2017},
  publisher={Foundation of Computer Science}
}

@inproceedings{patil2017lici,
  title={LiCi: A new ultra-lightweight block cipher},
  author={Patil, Jagdish and Bansod, Gaurav and Kant, Kumar Shashi},
  booktitle={2017 International Conference on Emerging Trends \& Innovation in ICT (ICEI)},
  pages={40--45},
  year={2017},
  organization={IEEE}
}

@article{aboushosha2020slim,
  title={SLIM: A lightweight block cipher for internet of health things},
  author={Aboushosha, Bassam and Ramadan, Rabie A and Dwivedi, Ashutosh Dhar and El-Sayed, Ayman and Dessouky, Mohamed M},
  journal={IEEE Access},
  volume={8},
  pages={203747--203757},
  year={2020},
  publisher={IEEE}
}

@inproceedings{hong2006hight,
  title={HIGHT: A new block cipher suitable for low-resource device},
  author={Hong, Deukjo and Sung, Jaechul and Hong, Seokhie and 
          Lim, Jongin and Lee, Bon-Seok and Lee, Seog Chung},
  booktitle={International Workshop on Cryptographic Hardware and Embedded Systems},
  pages={46--59},
  year={2006},
  organization={Springer}
}

@inproceedings{shirai2007128,
  title={The 128-bit blockcipher CLEFIA},
  author={Shirai, Taizo and Shibutani, Kyoji and Akishita, Toru and Moriai, Shiho and Iwata, Tetsu},
  booktitle={International workshop on fast software encryption},
  pages={181--195},
  year={2007},
  organization={Springer}
}

@inproceedings{ojha2009twis,
  title={Twis--a lightweight block cipher},
  author={Ojha, Shri Kant and Kumar, Naveen and Jain, Kritika and others},
  booktitle={International Conference on Information Systems Security},
  pages={280--291},
  year={2009},
  organization={Springer}
}

@inproceedings{shibutani2011piccolo,
  title={Piccolo: an ultra-lightweight blockcipher},
  author={Shibutani, Kyoji and Isobe, Takanori and Hiwatari, Harunaga and Mitsuda, Atsushi and Akishita, Toru and Shirai, Taizo},
  booktitle={International workshop on cryptographic hardware and embedded systems},
  pages={342--357},
  year={2011},
  organization={Springer}
}

@inproceedings{suzaki2012textnormal,
  title={TWINE: A Lightweight Block Cipher for Multiple Platforms},
  author={Suzaki, Tomoyasu and Minematsu, Kazuhiko and Morioka, Sumio and Kobayashi, Eita},
  booktitle={International Conference on Selected Areas in Cryptography},
  pages={339--354},
  year={2012},
  organization={Springer}
}

@inproceedings{kolay2014khudra,
  title={Khudra: a new lightweight block cipher for FPGAs},
  author={Kolay, Souvik and Mukhopadhyay, Debdeep},
  booktitle={International Conference on Security, Privacy, and Applied Cryptography Engineering},
  pages={126--145},
  year={2014},
  organization={Springer}
}

@inproceedings{aldabbagh2014hisec,
  title={Hisec: A new lightweight block cipher algorithm},
  author={AlDabbagh, Sufyan Salim Mahmood and Al Shaikhli, Imad Fakhri Taha and Alahmad, Mohammad A},
  booktitle={Proceedings of the 7th International Conference on Security of Information and Networks},
  pages={151--156},
  year={2014}
}

@article{kumar2014few,
  title={FeW: a lightweight block cipher},
  author={Kumar, Manoj and Sk, PAL and Panigrahi, Anupama},
  journal={Turkish Journal of Mathematics and Computer Science},
  volume={11},
  number={2},
  pages={58--73},
  year={2014}
}

@inproceedings{banik2020warp,
  title={WARP: Revisiting GFN for lightweight 128-bit block cipher},
  author={Banik, Subhadeep and Bao, Zhenzhen and Isobe, Takanori and Kubo, Hiroyasu and Liu, Fukang and Minematsu, Kazuhiko and Sakamoto, Kosei and Shibata, Nao and Shigeri, Maki},
  booktitle={International Conference on Selected Areas in Cryptography},
  year={2020},
  organization={Springer}
}

@INPROCEEDINGS{8711697,
  author={Sehrawat, Deepti and Gill, Nasib Singh and Devi, Munisha},
  booktitle={2019 6th International Conference on Signal Processing and Integrated Networks (SPIN)}, 
  title={Comparative Analysis of Lightweight Block Ciphers in IoT-Enabled Smart Environment}, 
  year={2019},
  volume={},
  number={},
  pages={915-920},
  doi={10.1109/SPIN.2019.8711697}
}

@article{li2016qtl,
  title={QTL: a new ultra-lightweight block cipher},
  author={Li, Lang and Liu, Botao and Wang, Hui},
  journal={Microprocessors and Microsystems},
  volume={45},
  pages={45--55},
  year={2016},
  publisher={Elsevier}
}

@article{usman2017sit,
  title={SIT: a lightweight encryption algorithm for secure internet of things},
  author={Usman, Muhammad and Ahmed, Irfan and Aslam, M Imran and Khan, Shujaat and Shah, Usman Ali},
  journal={arXiv preprint arXiv:1704.08688},
  year={2017}
}

@article{li2018sfn,
  title={SFN: A new lightweight block cipher},
  author={Li, Lang and Liu, Botao and Zhou, Yimeng and Zou, Yi},
  journal={Microprocessors and Microsystems},
  volume={60},
  pages={138--150},
  year={2018},
  publisher={Elsevier}
}

@article{biswas2020lrbc,
  title={LRBC: a lightweight block cipher design for resource constrained IoT devices},
  author={Biswas, A and Majumdar, A and Nath, S and Dutta, A and Baishnab, KL},
  journal={Journal of Ambient Intelligence and Humanized Computing},
  pages={1--15},
  year={2020},
  publisher={Springer}
}

@article{roy2021lcb,
  title={LCB: Light Cipher Block An Ultrafast Lightweight Block Cipher For Resource Constrained IOT Security Applications},
  author={Roy, Siddhartha and Roy, Saptarshi and Biswas, Arpita and Baishnab, Krishna Lal},
  journal={KSII Transactions on Internet and Information Systems (TIIS)},
  volume={15},
  number={11},
  pages={4122--4144},
  year={2021},
  publisher={Korean Society for Internet Information}
}

@inproceedings{lu2007cryptanalysis,
  title={Cryptanalysis of reduced versions of the HIGHT block cipher from CHES 2006},
  author={Lu, Jiqiang},
  booktitle={International Conference on Information Security and Cryptology},
  pages={11--26},
  year={2007},
  organization={Springer}
}

@inproceedings{ozen2009lightweight,
  title={Lightweight block ciphers revisited: Cryptanalysis of reduced round PRESENT and HIGHT},
  author={{\"O}zen, Onur and Var{\i}c{\i}, Kerem and Tezcan, Cihangir and Kocair, {\c{C}}elebi},
  booktitle={Australasian Conference on Information Security and Privacy},
  pages={90--107},
  year={2009},
  organization={Springer}
}

@inproceedings{hong2011biclique,
  title={Biclique attack on the full HIGHT},
  author={Hong, Deukjo and Koo, Bonwook and Kwon, Daesung},
  booktitle={International Conference on Information Security and Cryptology},
  pages={365--374},
  year={2011},
  organization={Springer}
}

@inproceedings{tsunoo2008impossible,
  title={Impossible differential cryptanalysis of CLEFIA},
  author={Tsunoo, Yukiyasu and Tsujihara, Etsuko and Shigeri, Maki and Saito, Teruo and Suzaki, Tomoyasu and Kubo, Hiroyasu},
  booktitle={International Workshop on Fast Software Encryption},
  pages={398--411},
  year={2008},
  organization={Springer}
}

@inproceedings{wang2012biclique,
  title={Biclique cryptanalysis of reduced-round piccolo block cipher},
  author={Wang, Yanfeng and Wu, Wenling and Yu, Xiaoli},
  booktitle={International Conference on Information Security Practice and Experience},
  pages={337--352},
  year={2012},
  organization={Springer}
}

@inproceedings{hong2013lea,
  title={LEA: A 128-bit block cipher for fast encryption on common processors},
  author={Hong, Deukjo and Lee, Jung-Keun and Kim, Dong-Chan and Kwon, Daesung and Ryu, Kwon Ho and Lee, Dong-Geon},
  booktitle={international workshop on information security applications},
  pages={3--27},
  year={2013},
  organization={Springer}
}

@inproceedings{yang2015simeck,
  title={The simeck family of lightweight block ciphers},
  author={Yang, Gangqiang and Zhu, Bo and Suder, Valentin and Aagaard, Mark D and Gong, Guang},
  booktitle={International Workshop on Cryptographic Hardware and Embedded Systems},
  pages={307--329},
  year={2015},
  organization={Springer}
}

@inproceedings{dinu2016design,
  title={Design strategies for ARX with provable bounds: Sparx and LAX},
  author={Dinu, Daniel and Perrin, L{\'e}o and Udovenko, Aleksei and Velichkov, Vesselin and Gro{\ss}sch{\"a}dl, Johann and Biryukov, Alex},
  booktitle={International Conference on the Theory and Application of Cryptology and Information Security},
  year={2016},
  organization={Springer}
}

@inproceedings{koo2017cham,
  title={CHAM: A family of lightweight block ciphers for resource-constrained devices},
  author={Koo, Bonwook and Roh, Dongyoung and Kim, Hyeonjin and Jung, Younghoon and Lee, Dong-Geon and Kwon, Daesung},
  booktitle={International conference on information security and cryptology},
  pages={3--25},
  year={2017},
  organization={Springer}
}

@misc{zhang_shi_xu_yao_li_2014, 
title={Sablier v1 }, url={https://competitions.cr.yp.to/round1/sablierv1.pdf}, 
 journal={Cryptographic competitions}, 
 publisher={CAESAR: Competition for Authenticated Encryption: Security, Applicability, and Robustness}, 
 author={Zhang, Bin and Shi, Zhenqing and Xu, Chao and Yao, Yuan and Li, Zhenqi}, year={2014}, 
 month={Mar}
}

@incollection{robshaw2008estream,
  title={The eSTREAM project},
  author={Robshaw, Matthew},
  booktitle={New Stream Cipher Designs},
  pages={1--6},
  year={2008},
  publisher={Springer}
}

@inproceedings{luo2010lightweight,
  title={A lightweight stream cipher WG-7 for RFID encryption and authentication},
  author={Luo, Yiyuan and Chai, Qi and Gong, Guang and Lai, Xuejia},
  booktitle={2010 IEEE Global Telecommunications Conference GLOBECOM 2010},
  pages={1--6},
  year={2010},
  organization={IEEE}
}

@inproceedings{fan2013wg,
  title={Wg-8: A lightweight stream cipher for resource-constrained smart devices},
  author={Fan, Xinxin and Mandal, Kalikinkar and Gong, Guang},
  booktitle={International Conference on Heterogeneous Networking for Quality, Reliability, Security and Robustness},
  pages={617--632},
  year={2013},
  organization={Springer}
}

@article{bernstein2005salsa20,
  title={Salsa20 specification},
  author={Bernstein, Daniel J},
  journal={eSTREAM Project algorithm description, http://www. ecrypt. eu. org/stream/salsa20pf. html},
  year={2005},
  publisher={Citeseer}
}

@inproceedings{bernstein2008chacha,
  title={ChaCha, a variant of Salsa20},
  author={Bernstein, Daniel J and others},
  booktitle={Workshop record of SASC},
  volume={8},
  number={1},
  pages={3--5},
  year={2008}
}

@article{dubrova2017espresso,
  title={Espresso: A stream cipher for 5G wireless communication systems},
  author={Dubrova, Elena and, Martin},
  journal={Cryptography and Communications},
  year={2017},
  publisher={Springer}
}

@article{ghafari2016fruit,
  title={Fruit-v2: ultra-lightweight stream cipher with shorter internal state},
  author={Ghafari, Vahid Amin and Hu, Honggang and Chen, Ying},
  journal={Cryptology ePrint Archive},
  year={2016}
}

@inproceedings{boesgaard2003rabbit,
  title={Rabbit: A new high-performance stream cipher},
  author={Boesgaard, Martin and Vesterager, Mette and Pedersen, Thomas and Christiansen, Jesper and Scavenius, Ove},
  booktitle={International workshop on fast software encryption},
  pages={307--329},
  year={2003},
  organization={Springer}
}

@inproceedings{tahir2008rabbit,
  title={Rabbit-MAC: Lightweight authenticated encryption in wireless sensor networks},
  author={Tahir, Ruhma and Javed, Muhammad Younas and Cheema, Ahmad Raza},
  booktitle={2008 International Conference on Information and Automation},
  pages={573--577},
  year={2008},
  organization={IEEE}
}

@incollection{wu2008stream,
  title={The stream cipher HC-128},
  author={Wu, Hongjun},
  booktitle={New stream cipher designs},
  pages={39--47},
  year={2008},
  publisher={Springer}
}

@incollection{canniere2008trivium,
  title={Trivium},
  author={Canni{\`e}re, Christophe De and Preneel, Bart},
  booktitle={New stream cipher designs},
  pages={244--266},
  year={2008},
  publisher={Springer}
}

@phdthesis{wu_2014, 
 title={ACORN: A Lightweight Authenticated Cipher (v1)}, 
 url={http://competitions.cr.yp.to/round1/acornv1.pdf}, 
 author={Wu, Hongjun}, 
 year={2014}
}

@inproceedings{watanabe2008enocoro,
  title={Enocoro-80: A hardware oriented stream cipher},
  author={Watanabe, Dai and Ideguchi, Kota and Kitahara, Jun and Muto, Kenichiro and Furuichi, Hiroki and Kaneko, Toshinobu},
  booktitle={2008 Third International Conference on Availability, Reliability and Security},
  pages={1294--1300},
  year={2008},
  organization={IEEE}
}

@inproceedings{david2011a2u2,
  title={A2U2: a stream cipher for printed electronics RFID tags},
  author={David, Mathieu and Ranasinghe, Damith C and Larsen, Torben},
  booktitle={2011 IEEE International Conference on RFID},
  pages={176--183},
  year={2011},
  organization={IEEE}
}

@inproceedings{mohandas2020a4,
  title={A4: A lightweight stream cipher},
  author={Mohandas, Nair Arun and Swathi, Adinath and Abhijith, R and Nazar, Ajmal and Sharath, Greeshma},
  booktitle={2020 5th International Conference on Communication and Electronics Systems (ICCES)},
  pages={573--577},
  year={2020},
  organization={IEEE}
}

@inproceedings{kumar2009bean,
  title={BEAN: a lightweight stream cipher},
  author={Kumar, Naveen and Ojha, Shrikant and Jain, Kritika and Lal, Sangeeta},
  booktitle={Proceedings of the 2nd international conference on Security of information and networks},
  pages={168--171},
  year={2009}
}

@inproceedings{chen2010tinystream,
  title={TinyStream: A lightweight and novel stream cipher scheme for wireless sensor networks},
  author={Chen, Tieming and Ge, Liang and Wang, Xiaohao and Cai, Jiamei},
  booktitle={2010 International Conference on Computational Intelligence and Security},
  pages={528--532},
  year={2010},
  organization={IEEE}
}

@article{tian2012quavium,
  title={Quavium-A New Stream Cipher Inspired by Trivium.},
  author={Tian, Yun and Chen, Gongliang and Li, Jianhua},
  journal={J. Comput.},
  volume={7},
  number={5},
  pages={1278--1283},
  year={2012}
}

@article{babbage2005stream,
  title={The stream cipher MICKEY (version 1)},
  author={Babbage, Steve and Dodd, Matthew},
  journal={ECRYPT Stream Cipher Project Report},
  volume={15},
  pages={2005},
  year={2005}
}

@article{diedrich2016comparison,
  title={Comparison of Lightweight Stream Ciphers: MICKEY 2.0, WG-8, Grain and Trivium},
  author={Diedrich, Lennart and Jattke, Patrick and Murati, Lulzim and Senker, Matthias and Wiesmaier, Alexander},
  journal={Unpublished},
  year={2016}
}

@incollection{babbage2008mickey,
  title={The MICKEY stream ciphers},
  author={Babbage, Steve and Dodd, Matthew},
  booktitle={New Stream Cipher Designs},
  pages={191--209},
  year={2008},
  publisher={Springer}
}

@techreport{ISO29192,
type = {Standard},
key = {ISO/IEC 29192-3},
month = mar,
year = {2012},
title = {{Information technology — Security techniques — Lightweight cryptography — Part 3: Stream ciphers}},
volume = {2000},
address = {Geneva, CH},
institution = {International Organization for Standardization}
}

@article{khazaei2005distinguishing,
  title={Distinguishing attack on grain},
  author={Khazaei, Shahram and Hassanzadeh, Mehdi and Kiaei, Mohammad},
  journal={ECRYPT Stream Cipher Project},
  volume={71},
  year={2005}
}

@inproceedings{canniere2008analysis,
  title={Analysis of Grain’s initialization algorithm},
  author={Canni{\`e}re, Christophe De and K{\"u}c{\"u}k, {\"O}zg{\"u}l and Preneel, Bart},
  booktitle={International Conference on Cryptology in Africa},
  pages={276--289},
  year={2008},
  organization={Springer}
}

@inproceedings{armknecht2015lightweight,
  title={On lightweight stream ciphers with shorter internal states},
  author={Armknecht, Frederik and Mikhalev, Vasily},
  booktitle={International Workshop on Fast Software Encryption},
  pages={451--470},
  year={2015},
  organization={Springer}
}

@article{hamann2017lizard,
  title={LIZARD-A lightweight stream cipher for power-constrained devices},
  author={Hamann, Matthias and Krause, Matthias and Meier, Willi},
  journal={IACR Transactions on Symmetric Cryptology},
  volume={2017},
  number={1},
  pages={45--79},
  year={2017},
  publisher={Ruhr-Universit{\"a}t Bochum}
}

@inproceedings{jiao2021fan,
  title={FAN: A Lightweight Authenticated Cryptographic Algorithm},
  author={Jiao, Lin and Feng, Dengguo and Hao, Yonglin and Gong, Xinxin and Du, Shaoyu},
  booktitle={Cryptographers’ Track at the RSA Conference},
  pages={299--325},
  year={2021},
  organization={Springer}
}

@inproceedings{bogdanov2013ale,
  title={ALE: AES-based lightweight authenticated encryption},
  author={Bogdanov, Andrey and Mendel, Florian and Regazzoni, Francesco and Rijmen, Vincent and Tischhauser, Elmar},
  booktitle={International Workshop on Fast Software Encryption},
  pages={447--466},
  year={2013},
  organization={Springer}
}

@inproceedings{jakimoski2011asc,
  title={ASC-1: An authenticated encryption stream cipher},
  author={Jakimoski, Goce and Khajuria, Samant},
  booktitle={International Workshop on Selected Areas in Cryptography},
  pages={356--372},
  year={2011},
  organization={Springer}
}

@incollection{berbain2008sosemanuk,
  title={Sosemanuk, a fast software-oriented stream cipher},
  author={Berbain, C{\^o}me and Billet, Olivier and Canteaut, Anne and Courtois, Nicolas and Gilbert, Henri and Goubin, Louis and Gouget, Aline and Granboulan, Louis and Lauradoux, C{\'e}dric and Minier, Marine and others},
  booktitle={New stream cipher designs},
  pages={98--118},
  year={2008},
  publisher={Springer}
}

@article{das2013car30,
  title={CAR30: A new scalable stream cipher with rule 30},
  author={Das, Sourav and RoyChowdhury, Dipanwita},
  journal={Cryptography and Communications},
  year={2013},
  publisher={Springer}
}

@article{hell2007grain,
  title={Grain: a stream cipher for constrained environments},
  author={Hell, Martin and Johansson, Thomas and Meier, Willi},
  journal={International journal of wireless and mobile computing},
  volume={2},
  number={1},
  pages={86--93},
  year={2007},
  publisher={Inderscience Publishers}
}

@inproceedings{hell2006stream,
  title={A stream cipher proposal: Grain-128},
  author={Hell, Martin and Johansson, Thomas and Maximov, Alexander and Meier, Willi},
  booktitle={2006 IEEE International Symposium on Information Theory},
  pages={1614--1618},
  year={2006},
  organization={IEEE}
}

@article{gren2011grain,
  title={Grain-128a: a new version of Grain-128 with optional authentication},
  author={Ågren, Martin and Hell, Martin and Johansson, Thomas and Meier, Willi},
  journal={International Journal of Wireless and Mobile Computing},
  volume={5},
  number={1},
  pages={48--59},
  year={2011},
  publisher={Inderscience Publishers}
}

@inproceedings{engels2010hummingbird,
  title={Hummingbird: ultra-lightweight cryptography for resource-constrained devices},
  author={Engels, Daniel and Fan, Xinxin and Gong, Guang and Hu, Honggang and Smith, Eric M},
  booktitle={International conference on financial cryptography and data security},
  pages={3--18},
  year={2010},
  organization={Springer}
}

@inproceedings{engels2011hummingbird,
  title={The Hummingbird-2 lightweight authenticated encryption algorithm},
  author={Engels, Daniel and Saarinen, Markku-Juhani O and Schweitzer, Peter and Smith, Eric M},
  booktitle={International workshop on radio frequency identification: Security and privacy issues},
  pages={19--31},
  year={2011},
  organization={Springer}
}

@inproceedings{bonnerji2018cozmo,
  title={COZMO-A new lightweight stream cipher},
  author={Bonnerji, Rhea and Sarkar, Simanta and Rarhi, Krishnendu and Bhattacharya, Abhishek},
  booktitle={2018 Second International Conference on Green Computing and Internet of Things (ICGCIoT)},
  pages={565--568},
  year={2018},
  organization={IEEE}
}

@inproceedings{isobe2012security,
  title={Security analysis of the lightweight block ciphers XTEA, LED and Piccolo},
  author={Isobe, Takanori and Shibutani, Kyoji},
  booktitle={Australasian Conference on Information Security and Privacy},
  pages={71--86},
  year={2012},
  organization={Springer}
}

@inproceedings{kelsey1997related,
  title={Related-key cryptanalysis of 3-way, biham-des, cast, des-x, newdes, rc2, and tea},
  author={Kelsey, John and Schneier, Bruce and Wagner, David},
  booktitle={International Conference on Information and Communications Security},
  pages={233--246},
  year={1997},
  organization={Springer}
}

@article{ding2013cryptanalysis,
  title={Cryptanalysis of MICKEY family of stream ciphers},
  author={Ding, Lin and Guan, Jie},
  journal={Security and Communication Networks},
  volume={6},
  number={8},
  pages={936--941},
  year={2013},
  publisher={Wiley Online Library}
}

@article{zhang2018survey,
  title={Survey of design and security evaluation of authenticated encryption algorithms in the CAESAR competition},
  author={Zhang, Fan and Liang, Zi-yuan and Yang, Bo-lin and Zhao, Xin-jie and Guo, Shi-ze and Ren, Kui},
  journal={Frontiers of Information Technology \& Electronic Engineering},
  volume={19},
  number={12},
  pages={1475--1499},
  year={2018},
  publisher={Springer}
}

@article{bansod2016anu,
  title={ANU: an ultra lightweight cipher design for security in IoT},
  author={Bansod, Gaurav and Patil, Abhijit and Sutar, Swapnil and Pisharoty, Narayan},
  journal={Security and Communication Networks},
  volume={9},
  number={18},
  pages={5238--5251},
  year={2016},
  publisher={Wiley Online Library}
}

@inproceedings{dahiphale2017anu,
  title={ANU-II: A fast and efficient lightweight encryption design for security in IoT},
  author={Dahiphale, Vijay and Bansod, Gaurav and Patil, Jagdish},
  booktitle={2017 International Conference on Big Data, IoT and Data Science (BID)},
  pages={130--137},
  year={2017},
  organization={IEEE}
}

@inproceedings{chen2014pandaka,
  title={Pandaka: A lightweight cipher for RFID systems},
  author={Chen, Min and Chen, Shigang and Xiao, Qingjun},
  booktitle={IEEE INFOCOM},
  pages={172--180},
  year={2014}
}

@article{bansod2017boron,
  title={BORON: an ultra-lightweight and low power encryption design for pervasive computing},
  author={Bansod, Gaurav and Pisharoty, Narayan and Patil, Abhijit},
  journal={Frontiers of Information Technology \& Electronic Engineering},
  volume={18},
  number={3},
  pages={317--331},
  year={2017},
  publisher={Springer}
}

@article{guo2021shadow,
  title={Shadow: A lightweight block cipher for IoT nodes},
  author={Guo, Ying and Li, Lang and Liu, Botao},
  journal={IEEE Internet of Things Journal},
  volume={8},
  number={16},
  pages={13014--13023},
  year={2021},
  publisher={IEEE}
}

@inproceedings{patil2019dot,
  title={Dot: A new ultra-lightweight sp network encryption design for resource-constrained environment},
  author={Patil, Jagdish and Bansod, Gaurav and Kant, Kumar Shashi},
  booktitle={Proceedings of the 2nd International Conference on Data Engineering and Communication Technology},
  pages={249--257},
  year={2019},
  organization={Springer}
}

@article{feng2022scenery,
  title={SCENERY: a lightweight block cipher based on Feistel structure},
  author={Feng, Jingya and Li, Lang},
  journal={Frontiers of Computer Science},
  volume={16},
  number={3},
  pages={1--10},
  year={2022},
  publisher={Springer}
}

@inproceedings{shantha2018sat_jo,
  title={Sat\_Jo: an enhanced lightweight block cipher for the internet of things},
  author={Shantha, Mary Joshitta R and Arockiam, L},
  booktitle={2018 Second International Conference on Intelligent Computing and Control Systems (ICICCS)},
  pages={1146--1150},
  year={2018},
  organization={IEEE}
}

@inproceedings{joshitta2019security,
  title={Security analysis of SAT\_Jo lightweight block cipher for data security in healthcare IoT},
  author={Joshitta, R Shantha Mary and Arockiam, Lawrence and Malarchelvi, PD Sheba Kezia},
  booktitle={Proceedings of the 2019 3rd International Conference on Cloud and Big Data Computing},
  pages={111--116},
  year={2019}
}

@inproceedings{cheng2021allpc,
  title={ALLPC: A Lightweight Block Cipher Based on Generalized Feistel Networks for IoT},
  author={Cheng, Junhua and Guo, Songtao and He, Jing},
  booktitle={2021 IEEE International Performance, Computing, and Communications Conference (IPCCC)},
  pages={1--8},
  year={2021},
  organization={IEEE}
}

@article{vahi2020separ,
  title={SEPAR: A new lightweight hybrid encryption algorithm with a novel design approach for IoT},
  author={Vahi, Arsalan and Jafarali Jassbi, Somaye},
  journal={Wireless Personal Communications},
  volume={114},
  number={3},
  pages={2283--2314},
  year={2020},
  publisher={Springer}
}

@article{sehrawat2020ultra,
  title={Ultra BRIGHT: a tiny and fast ultra lightweight block cipher for IoT},
  author={Sehrawat, Deepti and Gill, Nasib},
  journal={Int J Sci Technol Res},
  volume={9},
  pages={1063},
  year={2020}
}

@inproceedings{kocher1999differential,
  title={Differential power analysis},
  author={Kocher, Paul and Jaffe, Joshua and Jun, Benjamin},
  booktitle={Annual international cryptology conference},
  pages={388--397},
  year={1999},
  organization={Springer}
}

@inproceedings{bogdanov20103,
  title={A 3-subset meet-in-the-middle attack: cryptanalysis of the lightweight block cipher KTANTAN},
  author={Bogdanov, Andrey and Rechberger, Christian},
  booktitle={International Workshop on Selected Areas in Cryptography},
  pages={229--240},
  year={2010},
  organization={Springer}
}

@inproceedings{leander2011cryptanalysis,
  title={A cryptanalysis of PRINTcipher: the invariant subspace attack},
  author={Leander, Gregor and Abdelraheem, Mohamed Ahmed and AlKhzaimi, Hoda and Zenner, Erik},
  booktitle={Annual Cryptology Conference},
  pages={206--221},
  year={2011},
  organization={Springer}
}

@article{qiu2021integral,
  title={Integral Distinguishers of the Full-Round Lightweight Block Cipher SAT\_Jo},
  author={Qiu, Xueying and Wei, Yongzhuang and Hodzic, Samir and Pasalic, Enes},
  journal={Security and Communication Networks},
  volume={2021},
  year={2021},
  publisher={Hindawi}
}

@article{noura2023lesca,
  title={LESCA: LightwEight Stream Cipher Algorithm for emerging systems},
  author={Noura, Hassan and Salman, Ola and Couturier, Rapha{\"e}l and Chehab, Ali},
  journal={Ad Hoc Networks},
  volume={138},
  pages={102999},
  year={2023},
  publisher={Elsevier}
}

@inproceedings{lallemand2015cryptanalysis,
  title={Cryptanalysis of KLEIN},
  author={Lallemand, Virginie and Naya-Plasencia, Mar{\'\i}a},
  booktitle={Fast Software Encryption: 21st International Workshop, FSE 2014, London, UK, March 3-5, 2014. Revised Selected Papers},
  pages={451--470},
  year={2015},
  organization={Springer}
}

@article{jeong2012biclique,
  title={Biclique cryptanalysis of lightweight block ciphers PRESENT, Piccolo and LED},
  author={Jeong, Kitae and Kang, HyungChul and Lee, Changhoon and Sung, Jaechul and Hong, Seokhie},
  journal={Cryptology ePrint Archive},
  year={2012}
}

@inproceedings{walter2013optimizing,
  title={Optimizing guessing strategies for algebraic cryptanalysis with applications to EPCBC},
  author={Walter, Michael and Bulygin, Stanislav and Buchmann, Johannes},
  booktitle={Information Security and Cryptology: 8th International Conference, Inscrypt 2012, Beijing, China},
  year={2013},
  organization={Springer}
}

@article{zhao2015truncated,
  title={Truncated differential cryptanalysis of PRINCE},
  author={Zhao, Guangyao and Sun, Bing and Li, Chao and Su, Jinshu},
  journal={Security and Communication Networks},
  volume={8},
  number={16},
  pages={2875--2887},
  year={2015},
  publisher={Wiley Online Library}
}

@inproceedings{gerault2016related,
  title={Related-key cryptanalysis of midori},
  author={G{\'e}rault, David and Lafourcade, Pascal},
  booktitle={Progress in Cryptology--INDOCRYPT 2016: 17th International Conference on Cryptology in India, Kolkata, India, December 11-14, 2016, Proceedings},
  pages={287--304},
  year={2016},
  organization={Springer}
}

@inproceedings{chen2017impossible,
  title={Impossible differential cryptanalysis of midori},
  author={Chen, Zhan and Wang, XY},
  booktitle={Mechatronics and Automation Engineering: Proceedings of the International Conference on Mechatronics and Automation Engineering (ICMAE2016)},
  pages={221--229},
  year={2017},
  organization={World Scientific}
}

@inproceedings{hong2004differential,
  title={Differential Cryptanalysis of TEA and XTEA},
  author={Hong, Seokhie and Hong, Deukjo and Ko, Youngdai and Chang, Donghoon and Lee, Wonil and Lee, Sangjin},
  booktitle={Information Security and Cryptology-ICISC 2003: 6th International Conference, Seoul, Korea, November 27-28, 2003. Revised Papers 6},
  year={2004},
  organization={Springer}
}

@inproceedings{bogdanov2012zero,
  title={Zero correlation linear cryptanalysis with reduced data complexity},
  author={Bogdanov, Andrey and Wang, Meiqin},
  booktitle={Fast Software Encryption: 19th International Workshop, FSE 2012, Washington, DC, USA, March 19-21, 2012. Revised Selected Papers},
  pages={29--48},
  year={2012},
  organization={Springer}
}

@article{faghihi2016biclique,
  title={Biclique cryptanalysis of MIBS-80 and PRESENT-80 block ciphers},
  author={Faghihi Sereshgi, Mohammad Hossein and Dakhilalian, Mohammad and Shakiba, Mohsen},
  journal={Security and Communication Networks},
  volume={9},
  number={1},
  pages={27--33},
  year={2016},
  publisher={Wiley Online Library}
}

@article{fu2016differential,
  title={Differential fault attack on ITUbee block cipher},
  author={Fu, Shan and Xu, Guoai and Pan, Juan and Wang, Zongyue and Wang, An},
  journal={ACM Transactions on Embedded Computing Systems (TECS)},
  volume={16},
  number={2},
  pages={1--10},
  year={2016},
  publisher={ACM New York, NY, USA}
}

@inproceedings{wang2012security,
  title={Security on LBlock against biclique cryptanalysis},
  author={Wang, Yanfeng and Wu, Wenling and Yu, Xiaoli and Zhang, Lei},
  booktitle={Information Security Applications: 13th International Workshop, WISA 2012, Jeju Island, Korea, August 16-18, 2012, Revised Selected Papers 13},
  pages={1--14},
  year={2012},
  organization={Springer}
}

@inproceedings{li2013cube,
  title={Cube cryptanalysis of LBlock with noisy leakage},
  author={Li, Zhenqi and Zhang, Bin and Yao, Yuan and Lin, Dongdai},
  booktitle={Information Security and Cryptology--ICISC 2012: 15th International Conference, Seoul, Korea, November 28-30, 2012, Revised Selected Papers 15},
  pages={141--155},
  year={2013},
  organization={Springer}
}

@article{fan2022differential,
  title={Differential cryptanalysis of full-round ANU-II ultra-lightweight block cipher},
  author={Fan, Ting and Li, Lingchen and Wei, Yongzhuang and Pasalic, Enes},
  journal={International Journal of Distributed Sensor Networks},
  volume={18},
  number={9},
  pages={15501329221119398},
  year={2022},
  publisher={SAGE Publications Sage UK: London, England}
}

@inproceedings{bogdanov2014zero,
  title={Zero-correlation linear cryptanalysis with FFT and improved attacks on ISO standards Camellia and CLEFIA},
  author={Bogdanov, Andrey and Geng, Huizheng and Wang, Meiqin and Wen, Long and Collard, Baudoin},
  booktitle={Selected Areas in Cryptography -- SAC 2013: 20th International Conference},
  year={2014},
  organization={Springer}
}

@article{zheng2019security,
  title={Security of Khudra Against Meet-in-the-Middle-Type Cryptanalysis},
  author={ZHENG, Yafei and WU, Wenling},
  journal={Chinese Journal of Electronics},
  volume={28},
  number={3},
  pages={482--488},
  year={2019},
  publisher={Wiley Online Library}
}

@article{teh2022differential,
  title={Differential cryptanalysis of WARP},
  author={Teh, Je Sen and Biryukov, Alex},
  journal={Journal of Information Security and Applications},
  volume={70},
  pages={103316},
  year={2022},
  publisher={Elsevier}
}

@inproceedings{ccoban2017cryptanalysis,
  title={Cryptanalysis of QTL block cipher},
  author={{\c{C}}oban, Mustafa and Karako{\c{c}}, Ferhat and {\"O}zen, Mehmet},
  booktitle={Lightweight Cryptography for Security and Privacy: 5th International Workshop, LightSec 2016, Aksaray, Turkey, September 21-22, 2016, Revised Selected Papers},
  pages={60--68},
  year={2017},
  organization={Springer}
}

@article{sadeghi2018cryptanalysis,
  title={Cryptanalysis of SFN Block Cipher},
  author={Sadeghi, Sadegh and Bagheri, Nasour},
  journal={Cryptology ePrint Archive},
  year={2018}
}

@inproceedings{chan2023differential,
  title={Differential Cryptanalysis of Lightweight Block Ciphers SLIM and LCB},
  author={Chan, Yen Yee and Khor, Cher-Yin and Teh, Je Sen and Teng, Wei Jian and Jamil, Norziana},
  booktitle={Emerging Information Security and Applications: Third International Conference, EISA 2022},
  pages={55--67},
  year={2023},
  organization={Springer}
}

@inproceedings{lu2008cryptanalysis,
  title={Cryptanalysis of Rabbit.},
  author={Lu, Yi and Wang, Huaxiong and Ling, San},
  booktitle={ISC},
  volume={8},
  pages={204--214},
  year={2008},
  organization={Springer}
}

@inproceedings{lee2008cryptanalysis,
  title={Cryptanalysis of SOSEMANUK and SNOW 2.0 using linear masks},
  author={Lee, Jung-Keun and Lee, Dong Hoon and Park, Sangwoo},
  booktitle={Advances in Cryptology-ASIACRYPT 2008: 14th International Conference on the Theory and Application of Cryptology and Information Security},
  pages={524--538},
  year={2008},
  organization={Springer}
}

@inproceedings{hojsik2008differential,
  title={Differential fault analysis of Trivium},
  author={Hojs{\'\i}k, Michal and Rudolf, Bohuslav},
  booktitle={Fast Software Encryption: 15th International Workshop, FSE 2008, Lausanne, Switzerland, February 10-13, 2008, Revised Selected Papers 15},
  pages={158--172},
  year={2008},
  organization={Springer}
}

@article{kazmi2017algebraic,
  title={Algebraic side channel attack on trivium and grain ciphers},
  author={Kazmi, Asif Raza and Afzal, Mehreen and Amjad, Muhammad Faisal and Abbas, Haider and Yang, Xiaodong},
  journal={IEEE Access},
  volume={5},
  pages={23958--23968},
  year={2017},
  publisher={IEEE}
}

@article{shan2014related,
  title={Related-key differential attack on round reduced RECTANGLE-80},
  author={Shan, Jinyong and Hu, Lei and Song, Ling and Sun, Siwei and Ma, Xiaoshuang},
  journal={Cryptology ePrint Archive},
  year={2014}
}

@article{kumar2020optimal,
  title={Optimal differential trails in lightweight block ciphers ANU and PICO},
  author={Kumar, Manoj and Suresh, TS and Pal, Saibal K and Panigrahi, Anupama},
  journal={Cryptologia},
  volume={44},
  number={1},
  pages={68--78},
  year={2020},
  publisher={Taylor \& Francis}
}

@article{liu2020integral,
  title={Integral attack on PICO algorithm based on division property},
  author={LIU, Zongfu and YUAN, Zheng and ZHAO, Chenxi and ZHU, Liang},
  journal={Journal of Computer Applications},
  volume={40},
  number={10},
  pages={2967},
  year={2020}
}

@article{han2019unbalanced,
  title={Unbalanced biclique cryptanalysis of full-round GIFT},
  author={Han, Guoyong and Zhao, Hongluan and Zhao, Chunquan},
  journal={IEEE Access},
  volume={7},
  year={2019},
  publisher={IEEE}
}

@article{cao2019related,
  title={Related-key differential cryptanalysis of the reduced-round block cipher GIFT},
  author={Cao, Meichun and Zhang, Wenying},
  journal={IEEE Access},
  volume={7},
  year={2019},
  publisher={IEEE}
}

@article{kumar2022full,
  title={Full-round differential attack on DoT block cipher},
  author={Kumar, Manoj},
  journal={Journal of Discrete Mathematical Sciences and Cryptography},
  pages={1--13},
  year={2022},
  publisher={Taylor \& Francis}
}

@article{kim2022integral,
  title={Integral Cryptanalysis of Lightweight Block Cipher PIPO},
  author={Kim, Sunyeop and Kim, Jeseong and Kim, Seonggyeom and Hong, Deukjo and Sung, Jaechul and Hong, Seokhie},
  journal={IEEE Access},
  volume={10},
  pages={110195--110204},
  year={2022},
  publisher={IEEE}
}

@inproceedings{lim2022differential,
  title={Differential fault attack on lightweight block cipher PIPO},
  author={Lim, Seonghyuck and Han, Jaeseung and Lee, Tae-Ho and Han, Dong-Guk},
  booktitle={Information Security and Cryptology--ICISC 2021: 24th International Conference, Seoul, South Korea, December 1--3, 2021, Revised Selected Papers},
  pages={296--307},
  year={2022},
  organization={Springer}
}

@article{yang2016truncated,
  title={Truncated Differential Analysis of Round-Reduced RoadRunneR Block Cipher},
  author={Yang, Qianqian and Hu, Lei and Sun, Siwei and Song, Ling},
  journal={Cryptology ePrint Archive},
  year={2016}
}

@inproceedings{sasaki2018related,
  title={Related-key boomerang attacks on full ANU lightweight block cipher},
  author={Sasaki, Yu},
  booktitle={Applied Cryptography and Network Security: 16th International Conference, ACNS 2018, Leuven, Belgium, July 2-4, 2018, Proceedings 16},
  pages={421--439},
  year={2018},
  organization={Springer}
}

@inproceedings{koccak2012cryptanalysis,
  title={Cryptanalysis of TWIS block cipher},
  author={Ko{\c{c}}ak, Onur and {\"O}ztop, Ne{\c{s}}e},
  booktitle={Research in Cryptology: 4th Western European Workshop, WEWoRC 2011, Weimar, Germany, July 20-22, 2011, Revised Selected Papers 4},
  pages={109--121},
  year={2012},
  organization={Springer}
}

@inproceedings{ccoban2012biclique,
  title={Biclique cryptanalysis of TWINE},
  author={{\c{C}}oban, Mustafa and Karako{\c{c}}, Ferhat and Bozta{\c{s}}, {\"O}zkan},
  booktitle={Cryptology and Network Security: 11th International Conference, CANS 2012, Darmstadt, Germany, December 12-14, 2012. Proceedings 11},
  pages={43--55},
  year={2012},
  organization={Springer}
}

@article{wei2019related,
  title={Related-key impossible differential cryptanalysis on lightweight cipher TWINE},
  author={Wei, Yuechuan and Xu, Peng and Rong, Yisheng},
  journal={Journal of Ambient Intelligence and Humanized Computing},
  volume={10},
  number={2},
  pages={509--517},
  year={2019},
  publisher={Springer}
}

@article{abed2013cryptanalysis,
  title={Cryptanalysis of the speck family of block ciphers},
  author={Abed, Farzaneh and List, Eik and Lucks, Stefan and Wenzel, Jakob},
  journal={Cryptology Archive},
  year= {2013}
}

@article{zhang2016zero,
  title={Zero Correlation Linear Cryptanalysis on LEA Family Ciphers.},
  author={Zhang, Kai and Guan, Jie and Hu, Bin},
  journal={J. Commun.},
  volume={11},
  number={7},
  pages={677--685},
  year={2016}
}

@article{dwivedi2018differential,
  title={Differential cryptanalysis of round-reduced LEA},
  author={Dwivedi, Ashutosh Dhar and Srivastava, Gautam},
  journal={IEEE Access},
  volume={6},
  pages={79105--79113},
  year={2018},
  publisher={IEEE}
}

@article{sadeghi2018improved,
  title={Improved zero-correlation and impossible differential cryptanalysis of reduced-round SIMECK block cipher},
  author={Sadeghi, Sadegh and Bagheri, Nasour},
  journal={IET Information Security},
  volume={12},
  number={4},
  pages={314--325},
  year={2018},
  publisher={Wiley Online Library}
}

@inproceedings{ramesh2019side,
  title={Side channel analysis of sparx-64/128: Cryptanalysis and countermeasures},
  author={Ramesh, Sumesh Manjunath and AlKhzaimi, Hoda},
  booktitle={Progress in Cryptology--AFRICACRYPT 2019: 11th International Conference on Cryptology in Africa, Rabat, Morocco, July 9--11, 2019, Proceedings 11},
  year={2019},
  organization={Springer}
}

@inproceedings{ryabko2019gradient,
  title={Gradient Cryptanalysis of Block Cipher CHAM 64/128},
  author={Ryabko, Boris and Soskov, Alexander and Fionov, Andrey},
  booktitle={2019 XVI International Symposium" Problems of Redundancy in Information and Control Systems"(REDUNDANCY)},
  pages={211--215},
  year={2019},
  organization={IEEE}
}

@article{crowley2005truncated,
  title={Truncated differential cryptanalysis of five rounds of Salsa20},
  author={Crowley, Paul},
  journal={Cryptology ePrint Archive},
  year={2005}
}

@inproceedings{shao2012related,
  title={Related-cipher attack on Salsa20},
  author={Shao, Zeng-yu and Ding, Lin},
  booktitle={2012 Fourth International Conference on Computational and Information Sciences},
  pages={1182--1185},
  year={2012},
  organization={IEEE}
}

@phdthesis{raizada2015some,
  title={Some results on analysis and implementation of HC-128 stream cipher},
  author={Raizada, Shashwat},
  year={2015},
  school={Indian Statistical Institute-Kolkata}
}

@article{tsunoo2006evaluation,
  title={Evaluation of SOSEMANUK with regard to guess-and-determine attacks},
  author={Tsunoo, Yukiyasu and Saito, Teruo and Shigeri, Maki and Suzaki, Tomoyasu and Ahmadi, Hadi and Eghlidos, Taraneh and Khazaei, Shahram},
  journal={SASC 2006 Stream Ciphers Revisited},
  pages={25},
  year={2006},
  publisher={Citeseer}
}

@article{orumiehchiha2012cryptanalysis,
  title={Cryptanalysis of WG-7: a lightweight stream cipher},
  author={Orumiehchiha, Mohammad Ali and Pieprzyk, Josef and Steinfeld, Ron},
  journal={Cryptography and Communications},
  volume={4},
  pages={277--285},
  year={2012},
  publisher={Springer}
}

@inproceedings{banik2012differential,
  title={A differential fault attack on the grain family of stream ciphers},
  author={Banik, Subhadeep and Maitra, Subhamoy and Sarkar, Santanu},
  booktitle={Cryptographic Hardware and Embedded Systems--CHES 2012: 14th International Workshop, Leuven, Belgium, September 9-12, 2012. Proceedings 14},
  year={2012},
  organization={Springer}
}

@article{ding2013related,
  title={Related key chosen IV attack on Grain-128a stream cipher},
  author={Ding, Lin and Guan, Jie},
  journal={IEEE Transactions on Information Forensics},
  year={2013},
  publisher={IEEE}
}

@inproceedings{feng2014cryptanalysis,
  title={Cryptanalysis on the authenticated cipher sablier},
  author={Feng, Xiutao and Zhang, Fan},
  booktitle={Network and System Security: 8th International Conference, NSS 2014, Xi’an, China, October 15-17, 2014, Proceedings 8},
  pages={198--208},
  year={2014},
  organization={Springer}
}

@inproceedings{yang2019cube,
  title={Cube cryptanalysis of round-reduced ACORN},
  author={Yang, Jingchun and Liu, Meicheng and Lin, Dongdai},
  booktitle={Information Security: 22nd International Conference, ISC 2019, New York City, NY, USA, September 16--18, 2019, Proceedings},
  pages={44--64},
  year={2019},
  organization={Springer}
}

@article{barbero2022rotational,
  title={Rotational Cryptanalysis on ChaCha Stream Cipher},
  author={Barbero, Stefano and Bazzanella, Danilo and Bellini, Emanuele},
  journal={Symmetry},
  volume={14},
  number={6},
  pages={1087},
  year={2022},
  publisher={MDPI}
}

@article{maitra2016chosen,
  title={Chosen IV cryptanalysis on reduced round ChaCha and Salsa},
  author={Maitra, Subhamoy},
  journal={Discrete Applied Mathematics},
  volume={208},
  pages={88--97},
  year={2016},
  publisher={Elsevier}
}

@inproceedings{watanabe2010update,
  title={Update on enocoro stream cipher},
  author={Watanabe, Dai and Owada, Toru and Okamoto, Kazuto and Igarashi, Yasutaka and Kaneko, Toshinobu},
  booktitle={2010 International Symposium On Information Theory \& Its Applications},
  pages={778--783},
  year={2010},
  organization={IEEE}
}

@article{ding2015slide,
  title={Slide attack on standard stream cipher Enocoro-80 in the related-key chosen IV setting},
  author={Ding, Lin and Jin, Chenhui and Guan, Jie},
  journal={Pervasive and Mobile Computing},
  volume={24},
  pages={224--230},
  year={2015},
  publisher={Elsevier}
}

@inproceedings{aagren2011cryptanalysis,
  title={Cryptanalysis of the stream cipher BEAN},
  author={{\AA}gren, Martin and Hell, Martin},
  booktitle={Proceedings of the 4th international conference on Security of information and networks},
  pages={21--28},
  year={2011}
}

@inproceedings{abdelraheem2011cryptanalysis,
  title={Cryptanalysis of the light-weight cipher A2U2},
  author={Abdelraheem, Mohamed Ahmed and Borghoff, Julia and Zenner, Erik and David, Mathieu},
  booktitle={Cryptography and Coding: 13th IMA International Conference, IMACC 2011, Oxford, UK, December 12-15, 2011. Proceedings 13},
  pages={375--390},
  year={2011},
  organization={Springer}
}

@article{ding2014cryptanalysis,
  title={Cryptanalysis of lightweight WG-8 stream cipher},
  author={Ding, Lin and Jin, Chenhui and Guan, Jie and Wang, Qiuyan},
  journal={IEEE Transactions on Information Forensics and Security},
  volume={9},
  number={4},
  pages={645--652},
  year={2014},
  publisher={IEEE}
}

@article{rostami2019cryptanalysis,
  title={Cryptanalysis of WG-8 and WG-16 stream ciphers},
  author={Rostami, Saeed and Shakour, Elham and Orumiehchiha, Mohammad Ali and Pieprzyk, Josef},
  journal={Cryptography and Communications},
  volume={11},
  pages={351--362},
  year={2019},
  publisher={Springer}
}

@inproceedings{yarom2015evaluation,
  title={Evaluation and cryptanalysis of the Pandaka lightweight cipher},
  author={Yarom, Yuval and Li, Gefei and Ranasinghe, Damith C},
  booktitle={Applied Cryptography and Network Security: 13th International Conference, ACNS 2015, New York, NY, USA, June 2-5, 2015, Revised Selected Papers 13},
  pages={370--385},
  year={2015},
  organization={Springer}
}

@inproceedings{esgin2016practical,
  title={Practical cryptanalysis of full Sprout with TMD tradeoff attacks},
  author={Esgin, Muhammed F and Kara, Orhun},
  booktitle={Selected Areas in Cryptography--SAC 2015: 22nd International Conference, Sackville, NB, Canada, August 12--14, 2015, Revised Selected Papers 22},
  pages={67--85},
  year={2016},
  organization={Springer}
}

@article{wang2019fast,
  title={Fast correlation attacks on grain-like small state stream ciphers and cryptanalysis of plantlet, fruit-v2 and fruit-80},
  author={Wang, Shichang and Liu, Meicheng and Lin, Dongdai and Ma, Li},
  journal={Cryptology ePrint Archive},
  year={2019}
}

@article{yao2021cryptanalysis,
  title={Cryptanalysis of the class of maximum period galois NLFSR-based stream ciphers},
  author={Yao, Ge and Parampalli, Udaya},
  journal={Cryptography and Communications},
  volume={13},
  pages={847--864},
  year={2021},
  publisher={Springer}
}

@inproceedings{siddhanti2017differential,
  title={Differential fault attack on grain v1, ACORN v3 and lizard},
  author={Siddhanti, Akhilesh and Sarkar, Santanu and Maitra, Subhamoy and Chattopadhyay, Anupam},
  booktitle={Security, Privacy, and Applied Cryptography Engineering: 7th International Conference, SPACE 2017, Goa, India, December 13-17, 2017, Proceedings 7},
  pages={247--263},
  year={2017},
  organization={Springer}
}

@article{maitra2017tmdto,
  title={A TMDTO attack against Lizard},
  author={Maitra, Subhamoy and Sinha, Nishant and Siddhanti, Akhilesh and Anand, Ravi and Gangopadhyay, Sugata},
  journal={IEEE Transactions on Computers},
  volume={67},
  number={5},
  pages={733--739},
  year={2017},
  publisher={IEEE}
}

@article{banik2017some,
  title={Some cryptanalytic results on Lizard},
  author={Banik, Subhadeep and Isobe, Takanori},
  journal={Cryptology ePrint Archive},
  year={2017}
}

@inproceedings{chen2013cryptanalysis,
  title={Cryptanalysis of the lightweight block cipher hummingbird-1},
  author={Chen, Xunjun and Zhu, Yuelong and Gong, Zheng and Luo, Yiyuan},
  booktitle={2013 Fourth International Conference on Emerging Intelligent Data and Web Technologies},
  pages={515--518},
  year={2013},
  organization={IEEE}
}

@inproceedings{salehani2011differential,
  title={Differential fault analysis of Hummingbird},
  author={Salehani, Yaser Esmaeili and Youssef, Amr},
  booktitle={Proceedings of the International Conference on Security and Cryptography},
  pages={357--361},
  year={2011},
  organization={IEEE}
}

@inproceedings{saarinen2011cryptanalysis,
  title={Cryptanalysis of hummingbird-1},
  author={Saarinen, Markku-Juhani O},
  booktitle={Fast Software Encryption: 18th International Workshop, FSE 2011, Lyngby, Denmark, February 13-16, 2011, Revised Selected Papers 18},
  pages={328--341},
  year={2011},
  organization={Springer}
}

@article{zhang2012real,
  title={Real time related key attack on Hummingbird-2},
  author={Zhang, Kai and Ding, Lin and Li, Junzhi and Guan, Jie},
  journal={KSII Transactions on Internet and Information Systems (TIIS)},
  volume={6},
  number={8},
  pages={1946--1963},
  year={2012},
  publisher={Korean Society for Internet Information}
}

@inproceedings{fan2012security,
  title={On the security of hummingbird-2 against side channel cube attacks},
  author={Fan, Xinxin and Gong, Guang},
  booktitle={Research in Cryptology: 4th Western European Workshop, WEWoRC 2011, Weimar, Germany, July 20-22, 2011, Revised Selected Papers 4},
  pages={18--29},
  year={2012},
  organization={Springer}
}

@article{chai2012cryptanalysis,
  title={A cryptanalysis of HummingBird-2: The differential sequence analysis},
  author={Chai, Qi and Gong, Guang},
  journal={Cryptology ePrint Archive},
  year={2012}
}

@inproceedings{raddum2015algebraic,
  title={Algebraic analysis of the simon block cipher family},
  author={Raddum, H{\aa}vard},
  booktitle={Progress in Cryptology--LATINCRYPT 2015: 4th International Conference on Cryptology and Information Security in Latin America, Guadalajara, Mexico, August 23-26, 2015, Proceedings 4},
  pages={157--169},
  year={2015},
  organization={Springer}
}

@inproceedings{yoshikawa2016power,
  title={Power analysis attack and its countermeasure for a lightweight block cipher Simon},
  author={Yoshikawa, Masaya and Nozaki, Yusuke},
  booktitle={Information Technology: New Generations: 13th International Conference on Information Technology},
  pages={151--160},
  year={2016},
  organization={Springer}
}

@inproceedings{biryukov2015differential,
  title={Differential analysis of block ciphers SIMON and SPECK},
  author={Biryukov, Alex and Roy, Arnab and Velichkov, Vesselin},
  booktitle={Fast Software Encryption: 21st International Workshop, FSE 2014, London, UK, March 3-5, 2014. Revised Selected Papers 21},
  pages={546--570},
  year={2015},
  organization={Springer}
}

@inproceedings{su2011full,
  title={Full-round differential attack on TWIS block cipher},
  author={Su, Bozhan and Wu, Wenling and Zhang, Lei and Li, Yanjun},
  booktitle={Information Security Applications: 11th International Workshop, WISA 2010, Jeju Island, Korea, August 24-26, 2010, Revised Selected Papers 11},
  pages={234--242},
  year={2011},
  organization={Springer}
}

@inproceedings{tupsamudre2014differential,
  title={Differential fault analysis on the families of SIMON and SPECK ciphers},
  author={Tupsamudre, Harshal and Bisht, Shikha and Mukhopadhyay, Debdeep},
  booktitle={2014 Workshop on Fault Diagnosis and Tolerance in Cryptography},
  pages={40--48},
  year={2014},
  organization={IEEE}
}

@article{lim1998crypton,
  title={CRYPTON: A new 128-bit block cipher},
  author={Lim, Chae Hoon},
  journal={NIsT AEs Proposal},
  year={1998},
  publisher={Citeseer}
}

@article{THABIT2023100759,
title = {Cryptography Algorithms for Enhancing IoT Security},
journal = {Internet of Things},
volume = {22},
pages = {100759},
year = {2023},
issn = {2542-6605},
doi = {https://doi.org/10.1016/j.iot.2023.100759},
url = {https://www.sciencedirect.com/science/article/pii/S2542660523000823},
author = {Fursan Thabit and Ozgu Can and Asia Othman Aljahdali and Ghaleb H. Al-Gaphari and Hoda A. Alkhzaimi},
}

@article{badel2010armadillo,
    title = {ARMADILLO: A Multi-purpose Cryptographic Primitive Dedicated to Hardware},
    author = {Stéphane Badel and Nilay Dağtekin and Jorge Nakahara and Khaled Ouafi and Nicolas Reffé and Pouyan Sepehrdad and Petr Sušil and Serge Vaudenay},
    pages = {398-412},
    year = {2010},
    doi = {10.1007/978-3-642-15031-9_27}
}

@inproceedings{jassim2021survey,
  title={A survey on stream ciphers for constrained environments},
  author={Jassim, Sameeh A and Farhan, Alaa K},
  booktitle={2021 1st Babylon International Conference on Information Technology and Science (BICITS)},
  pages={228--233},
  year={2021},
  organization={IEEE}
}

@article{singh2019lightweight,
  title={Lightweight Ciphers for Internet of things: A Survey},
  author={Singh, Amarpreet and Singh, Sandeep and Singh, Gurpreet},
  journal={International Journal of Innovative Technology and Exploring Engineering},
  volume={8},
  number={7},
  pages={1973--1981},
  year={2019}
}

@article{tawalbeh2017security,
  title={Security in Wireless Sensor Networks Using Lightweight Cryptography.},
  author={Tawalbeh, Hala and Hashish, Sonia and Tawalbeh, Loai and Aldairi, Anwar},
  journal={Journal of Information Assurance \& Security},
  volume={12},
  number={4},
  year={2017}
}

@inproceedings{kaps2008chai,
  title={Chai-tea, cryptographic hardware implementations of xtea},
  author={Kaps, Jens-Peter},
  booktitle={Progress in Cryptology-INDOCRYPT 2008: 9th International Conference on Cryptology in India, Kharagpur, India, December 14-17, 2008. Proceedings 9},
  pages={363--375},
  year={2008},
  organization={Springer}
}

@article{kitsos2012comparative,
  title={A comparative study of hardware architectures for lightweight block ciphers},
  author={Kitsos, Paris and Sklavos, Nicolas and Parousi, Maria and Skodras, Athanassios N},
  journal={Computers \& Electrical Engineering},
  volume={38},
  number={1},
  pages={148--160},
  year={2012},
  publisher={Elsevier}
}

@inproceedings{liu2016efficient,
  title={Efficient hardware implementation of roadrunner for lightweight application},
  author={Liu, Juhua and Bai, Guoqiang and Wu, Xingjun},
  booktitle={2016 IEEE Trustcom/BigDataSE/ISPA},
  pages={224--227},
  year={2016},
  organization={IEEE}
}

@book{david2012lightweight,
  title={Lightweight Cryptography for Passive RFID Tags},
  author={David, Mathieu},
  year={2012}
}

@article{lim2009implementation,
  title={Implementation of HIGHT cryptic circuit for RFID tag},
  author={Lim, Young-Il and Lee, Je-Hoon and You, Younggap and Cho, Kyoung-Rok},
  journal={IEICE Electronics Express},
  volume={6},
  number={4},
  pages={180--186},
  year={2009},
  publisher={The Institute of Electronics, Information and Communication Engineers}
}

@article{sulaiman2018modified,
  title={Modified 128-EEA2 Algorithm by Using HISEC Lightweight Block CipherAlgorithm with Improving the Security and Cost Factors},
  author={Sulaiman, Alyaa Ghanim and AlDabbagh, Sufyan Salim Mahmood},
  journal={Indonesian Journal of Electrical Engineering and Computer Science},
  volume={10},
  number={1},
  year={2018}
}

@article{rashidi2020efficient,
  title={Efficient and flexible hardware structures of the 128 bit CLEFIA block cipher},
  author={Rashidi, Bahram},
  journal={IET Computers \& Digital Techniques},
  volume={14},
  number={2},
  pages={69--79},
  year={2020},
  publisher={Wiley Online Library}
}

@article{jattke2016comparison,
  title={Comparison of two Lightweight Stream Ciphers: MICKEY 2.0 \& WG-8},
  author={Jattke, Patrick and Senker, Matthias and Wiesmaier, Alexander},
  year={2016}
}

@inproceedings{bilgin2013fides,
  title={Fides: Lightweight authenticated cipher with side-channel resistance for constrained hardware},
  author={Bilgin, Beg{\"u}l and Bogdanov, Andrey and Kne{\v{z}}evi{\'c}, Miroslav and Mendel, Florian and Wang, Qingju},
  booktitle={Cryptographic Hardware and Embedded Systems-CHES 2013},
  pages={142--158},
  year={2013},
  organization={Springer}
}

@inproceedings{canright2005very,
  title={A very compact S-box for AES},
  author={Canright, David},
  booktitle={International Workshop on Cryptographic Hardware and Embedded Systems},
  year={2005},
  organization={Springer}
}

@inproceedings{wenger2010low,
  title={Low-resource hardware design of an elliptic curve processor for contactless devices},
  author={Wenger, Erich and Feldhofer, Martin and Felber, Norbert},
  booktitle={International Workshop on Information Security Applications},
  pages={92--106},
  year={2010},
  organization={Springer}
}

@inproceedings{wenger20128,
  title={An 8-bit AVR-based elliptic curve cryptographic RISC processor for the internet of things},
  author={Wenger, Erich and Grossschadl, Johann},
  booktitle={2012 45th Annual IEEE/ACM International Symposium on Microarchitecture Workshops},
  pages={39--46},
  year={2012},
  organization={IEEE}
}

@inproceedings{wenger2013lightweight,
  title={A lightweight ATmega-based application-specific instruction-set processor for elliptic curve cryptography},
  author={Wenger, Erich},
  booktitle={Lightweight Cryptography for Security and Privacy: Second International Workshop, LightSec 2013},
  year={2013},
  organization={Springer}
}

@inproceedings{bosmans2016tiny,
  title={A tiny coprocessor for elliptic curve cryptography over the 256-bit NIST prime field},
  author={Bosmans, Jeroen and Roy, Sujoy Sinha and Jarvinen, Kimmo and Verbauwhede, Ingrid},
  booktitle={29th International Conference on VLSI Design and 15th International Conference on Embedded Systems},
  pages={523--528},
  year={2016},
  organization={IEEE}
}

@article{yalccin2016compact,
  title={Compact ECDSA engine for IoT applications},
  author={Yal{\c{c}}in, T},
  journal={Electronics Letters},
  volume={52},
  number={15},
  pages={1310--1312},
  year={2016},
  publisher={Wiley Online Library}
}

@article{liu2016emerging,
  title={On emerging family of elliptic curves to secure internet of things: ECC comes of age},
  author={Liu, Zhe and Huang, Xinyi and Hu, Zhi and Khan, Muhammad Khurram and Seo, Hwajeong and Zhou, Lu},
  journal={IEEE Transactions on Dependable and Secure Computing},
  volume={14},
  number={3},
  pages={237--248},
  year={2016},
  publisher={IEEE}
}

@article{mathur2017secure,
  title={A secure end-to-end IoT solution},
  author={Mathur, Avijit and Newe, Thomas and Elgenaidi, Walid and Rao, Muzaffar and Dooly, Gerard and Toal, Daniel},
  journal={Sensors and Actuators A: Physical},
  volume={263},
  pages={291--299},
  year={2017},
  publisher={Elsevier}
}

@inproceedings{sojka2017shortening,
  title={Shortening the security parameters in lightweight WSN applications for IoT-lessons learned},
  author={Sojka-Piotrowska, Anna and Langendoerfer, Peter},
  booktitle={2017 IEEE International Conference on Pervasive Computing and Communications Workshops (PerCom Workshops)},
  pages={636--641},
  year={2017},
  organization={IEEE}
}

@inproceedings{cabana2021low,
  title={A Low-Power Elliptic Curve Processor for WISP},
  author={Cabana, Igor Mendez and Silva-C{\'a}rdenas, Carlos},
  booktitle={2021 19th IEEE International New Circuits and Systems Conference (NEWCAS)},
  pages={1--4},
  year={2021},
  organization={IEEE}
}

@article{thaparimplementation,
  title={Implementation of Elliptical Curve Cryptography Based Diffie-Hellman Key Exchange Mechanism in Contiki Operating System for Internet of Things},
  author={Thapar, Prateek and Batra, Usha},
  journal={International Journal of Electrical and Electronics Research (IJEER)},
  volume={10},
  year={2022},
  pages={335--340}
}

@inproceedings{kumar2006standards,
  title={Are standards compliant elliptic curve cryptosystems feasible on RFID},
  author={Kumar, Sandeep and Paar, Christof},
  booktitle={Workshop on RFID security},
  year={2006}
}

@article{azarderakhsh2014efficient,
  title={Efficient algorithm and architecture for elliptic curve cryptography for extremely constrained secure applications},
  author={Azarderakhsh, Reza and J{\"a}rvinen, Kimmo U and Mozaffari-Kermani, Mehran},
  journal={IEEE Transactions on Circuits and Systems I: Regular Papers},
  volume={61},
  number={4},
  pages={1144--1155},
  year={2014},
  publisher={IEEE}
}

@inproceedings{holler2014hardware,
  title={Hardware/software co-design of elliptic-curve cryptography for resource-constrained applications},
  author={H{\"o}ller, Andrea and Druml, Norbert and Kreiner, Christian and Steger, Christian and Felicijan, Tomaz},
  booktitle={Proceedings of the 51st Annual Design Automation Conference},
  pages={1--6},
  year={2014}
}

@inproceedings{sinha2015lightweight,
  title={Lightweight coprocessor for Koblitz curves: 283-bit ECC including scalar conversion with only 4300 gates},
  author={Sinha Roy, Sujoy and J{\"a}rvinen, Kimmo and Verbauwhede, Ingrid},
  booktitle={International workshop on cryptographic hardware and embedded systems},
  pages={102--122},
  year={2015},
  organization={Springer}
}

@inproceedings{koziel2015low,
  title={Low-resource and fast binary edwards curves cryptography},
  author={Koziel, Brian and Azarderakhsh, Reza and Mozaffari-Kermani, Mehran},
  booktitle={Progress in Cryptology--INDOCRYPT 2015: 16th International Conference on Cryptology in India, Bangalore, India, December 6-9, 2015, Proceedings 16},
  year={2015},
  organization={Springer}
}

@article{rovzic20175,
  title={A 5.1 $\mu$J per point-multiplication elliptic curve cryptographic processor},
  author={Ro{\v{z}}i{\'c}, Vladimir and Reparaz, Oscar and Verbauwhede, Ingrid},
  journal={International Journal of Circuit Theory and Applications},
  volume={45},
  number={2},
  pages={170--187},
  year={2017},
  publisher={Wiley Online Library}
}

@article{lara2020lightweight,
  title={Lightweight elliptic curve cryptography accelerator for internet of things applications},
  author={Lara-Nino, Carlos Andres and Diaz-Perez, Arturo and Morales-Sandoval, Miguel},
  journal={Ad Hoc Networks},
  volume={103},
  pages={102159},
  year={2020},
  publisher={Elsevier}
}

@inproceedings{szczechowiak2008nanoecc,
  title={NanoECC: Testing the limits of elliptic curve cryptography in sensor networks},
  author={Szczechowiak, Piotr and Oliveira, Leonardo B and Scott, Michael and Collier, Martin and Dahab, Ricardo},
  booktitle={Wireless Sensor Networks: 5th European Conference, EWSN 2008, Bologna, Italy. Proceedings},
  pages={305--320},
  year={2008},
  organization={Springer}
}

@article{gouvea2012efficient,
  title={Efficient software implementation of public-key cryptography on sensor networks using the MSP430X microcontroller},
  author={Gouv{\^e}a, Conrado PL and Oliveira, Leonardo B and L{\'o}pez, Julio},
  journal={Journal of Cryptographic Engineering},
  volume={2},
  pages={19--29},
  year={2012},
  publisher={Springer}
}

@inproceedings{wenger2013hardware,
  title={Hardware architectures for MSP430-based wireless sensor nodes performing elliptic curve cryptography},
  author={Wenger, Erich},
  booktitle={Applied Cryptography and Network Security: 11th International Conference, ACNS 2013, Banff, AB, Canada, June 25-28, 2013. Proceedings 11},
  year={2013},
  organization={Springer}
}

@inproceedings{hutter2011fast,
  title={Fast multi-precision multiplication for public-key cryptography on embedded microprocessors},
  author={Hutter, Michael and Wenger, Erich},
  booktitle={Cryptographic Hardware and Embedded Systems--CHES 2011: 13th International Workshop. Proceedings 13},
  pages={459--474},
  year={2011},
  organization={Springer}
}

@inproceedings{seo2013multi,
  title={Multi-precision squaring for public-key cryptography on embedded microprocessors},
  author={Seo, Hwajeong and Liu, Zhe and Choi, Jongseok and Kim, Howon},
  booktitle={International Conference on Cryptology in India},
  pages={227--243},
  year={2013},
  organization={Springer}
}

@inproceedings{liu2015reverse,
  title={Reverse product-scanning multiplication and squaring on 8-bit AVR processors},
  author={Liu, Zhe and Seo, Hwajeong and Gro{\ss}sch{\"a}dl, Johann and Kim, Howon},
  booktitle={Information and Communications Security: 16th International Conference},
  pages={158--175},
  year={2015},
  organization={Springer}
}

@article{silverman2006introduction,
  title={An introduction to the theory of elliptic curves},
  author={Silverman, Joseph H},
  journal={Brown University. June},
  volume={19},
  pages={2006},
  year={2006}
}

@article{itoh1988fast,
  title={A fast algorithm for computing multiplicative inverses in GF (2m) using normal bases},
  author={Itoh, Toshiya and Tsujii, Shigeo},
  journal={Information and computation},
  volume={78},
  number={3},
  pages={171--177},
  year={1988},
  publisher={Elsevier}
}

@article{pairingbasedcryptography,
author = {Oliveira, Leonardo and Dahab, Ricardo},
year = {2006},
month = {01},
pages = {},
title = {Pairing-Based Cryptography for Sensor Networks}
}

@article{scott2003miracl,
  title={MIRACL-A multiprecision integer and rational arithmetic C/C++ library},
  author={Scott, Michael},
  journal={http://www. shamus. ie},
  year={2003},
  publisher={Shamus Software Ltd}
}

@article{comba1990exponentiation,
  title={Exponentiation cryptosystems on the IBM PC},
  author={Comba, Paul G.},
  journal={IBM systems journal},
  volume={29},
  number={4},
  pages={526--538},
  year={1990},
  publisher={IBM}
}

@inproceedings{lopez2000high,
  title={High-speed software multiplication in F2m},
  author={L{\'o}pez, Julio and Dahab, Ricardo},
  booktitle={International Conference on Cryptology in India},
  year={2000},
  organization={Springer}
}

@inproceedings{good2008hardware,
  title={Hardware performance of eStream phase-III stream cipher candidates},
  author={Good, Tim and Benaissa, Mohammed},
  booktitle={Proc. of Workshop on the State of the Art of Stream Ciphers (SACS’08)},
  year={2008}
}

@inproceedings{daemen1998fast,
  title={Fast hashing and stream encryption with PANAMA},
  author={Daemen, Joan and Clapp, Craig},
  booktitle={International Workshop on FSE},
  pages={60--74},
  year={1998},
  organization={Springer}
}

\appendix

\section{Lightweight Block Ciphers}
The results of the software and hardware implementation obtained from the respective research publications are tabulated in the appendix, accompanied by corresponding bar graphs. 

    \subsection{Substitution Permutation Network (SPN)}

        \begin{table} [H]
\scriptsize
\caption{Hardware implementation of SPN Block Ciphers.}
\label{tab:HW_imp_SPN}
\begin{tabular*}{\textwidth}{@{\extracolsep{\fill}} l c c c c c c c c c c}
\toprule
\multicolumn{2}{l}{\textbf{Cipher}} &
  \textbf{Year} &
  \begin{tabular}[c]{@{}c@{}}\textbf{Block}\\  \textbf{Length}\\ \textbf{(bits})\end{tabular} &
  \begin{tabular}[c]{@{}c@{}}\textbf{Key}\\  \textbf{Size}\\ \textbf{(bits)}\end{tabular} &
  \textbf{Rounds} &
  \textbf{GE} &
  \begin{tabular}[c]{@{}c@{}}\textbf{Throughput}\\  \textbf{(@ 100 KHz)}\end{tabular} &
  \begin{tabular}[c]{@{}c@{}}\textbf{Power} \\  \textbf{Consumption}\\  \textbf{($\mu$W)}\end{tabular} &
  \begin{tabular}[c]{@{}c@{}}\textbf{Figure of Merit}\\ \textbf{(FOM)}\end{tabular} &
  \begin{tabular}[c]{@{}c@{}}\textbf{CMOS} \\ \textbf{Tech}\\ \textbf{($\mu$m)}\end{tabular} \\                     
\hline
mCrypton~\cite{lim2005mcrypton}     & mCrypton-64    & 2005 & 64  & 64       & 12  & 3473   & 492.3  &       & 408.15  & 0.13 \\
              & mCrypton-96    &      &     & 96       &     & 3789   &        &       & 342.91  &      \\
              & mCrypton-128   &      &     & 128      &     & 4108   &        &       & 291.72  &      \\ \hline
              
PRESENT~\cite{bogdanov2007present}    & PRESENT-80     & 2007 & 64  & 80       & 31  & 1570   & 200    & 5     & 811.39  & 0.18 \\
              & PRESENT-128    &      &     & 128      &     & 1884   &        &       & 563.47  &      \\ \hline
              
PRINT~\cite{knudsen2010printcipher}     & PRINTcipher-48 & 2010 & 48  & 80       & 47  & 503    & 6.25   & 2.6   & 247.03  & 0.18 \\
              & PRINTcipher-96 &      & 96  & 160      & 95  & 967    & 3.13   &       & 33.47   &      \\ \hline
              
AES (compact)~\cite{moradi2011pushing}  & AES-128        & 2011 & 128 & 128      & 10  & 2400   & 57     & 6.66  & 98.96   & 0.18 \\ \hline

KLEIN~\cite{gong2011klein}             & KLEIN-64       & 2011 & 64  & 64       & 12  & 2475   & 207    &       & 337.92  & 0.18 \\
              & KLEIN-80       &      &     & 80       & 16  & 2629   &        &       & 299.49  &      \\
              & KLEIN-96       &      &     & 90       & 20  & 2769   &        &       & 269.98  &      \\ \hline
              
LED~\cite{guo2011led}         & LED-64         & 2011 & 64  & 64       & 8   & 966    & 5.1    & 1.67  & 54.65   & 0.18 \\
              & LED-80         &      &     & 80       & 12  & 1040   & 3.4    &       & 31.43   &      \\
              & LED-96         &      &     & 96       & 12  & 1116   & 3.4    &       & 27.3    &      \\
              & LED-128        &      &     & 128      & 12  & 1265   & 3.4    & 2.2   & 21.25   &      \\ \hline
              
EPCBC~\cite{singh2019lightweight}       & EPCBC-48       & 2011 & 48  & 96       & 32  & 1008   & 12.12  & 2.21  & 119.28  & 0.18 \\
              & EPCBC-96       &      & 96  &          &     & 1333   &        & 3.63  & 68.21   &      \\ \hline
              
PRINCE~\cite{tawalbeh2017security}      &                & 2012 & 64  & 128      & 12  & 3491   & 529.9  & 111.3*& 434.8  & 0.09 \\ \hline

OLBCA~\cite{aldabbagh2014olbca}       &                & 2014 & 64  & 80       & 22  & 1521   &        &       &        &      \\ \hline

Midori~\cite{banik2015midori}      &                & 2015 & 64  & 64       & 16  & 1542   &        & 60.6* &        & 0.09 \\ \hline

RECTANGLE~\cite{zhang2015rectangle}     & RECTANGLE-80   & 2015 & 64  & 80       & 25  & 1599.5 & 246    & 74.31*& 960.94 & 0.13 \\
              & RECTANGLE-128  &      &     & 128      &     & 2063.5 &        & 72.15*& 577.45 &      \\ \hline
              
SKINNY~\cite{beierle2016skinny}       & SKINNY-64      & 2016 & 64  & 64       & 32  & 1223   & 200    &       & 1337.14 & 0.18 \\
              &                &      &     & 128      & 36  & 1696   & 177.78 &       & 618.04  &      \\
              &                &      &     & 192      & 40  & 2183   & 160    &       & 335.75  &      \\ \cline{2-10}
              
              & SKINNY-128     &      & 128 & 128      & 40  & 2391   & 320    &       & 559.75  &      \\
              &                &      &     & 256      & 48  & 3312   & 266.67 &       & 243.1   &      \\
              &                &      &     & 384      & 56  & 4268   & 228.57 &       & 125.48  &      \\ \hline
              
PICO~\cite{bansod2016pico}          &                & 2016 & 64  & 128      & 32  & 1877   & 129    &       & 366.15  & 0.18 \\ \hline

BORON~\cite{bansod2017boron}        & BORON-80       & 2017 & 64  & 80       & 25  & 1626   &        &       &         & 0.18 \\
              & BORON-128      &      &     & 128      &     & 1939   &        &       &         &      \\ \hline
              
GIFT~\cite{banik2017gift}         & GIFT-64        & 2017 & 64  & 128      & 28  & 1345   &        &       &         &       \\
              & GIFT-128       &      & 128 &          & 40  & 1997   &        &       &         &        \\ \hline
              
SAT\_Jo~\cite{joshitta2019security}       &                & 2018 & 64  & 80       & 31  & 1167   &        &      &         & 0.09 \\ \hline

DoT~\cite{patil2019dot}        &                & 2019 & 64  & 128      & 31  & 993    & 53.77  &       & 545.31  &        \\ \hline

PIPO~\cite{kim2020pipo}          & PIPO-128       & 2021 & 64  & 128      & 13  & 1446   & 492    &       & 2353.04 & 0.13 \\
              & PIPO-256       &      &     & 256      & 17  & 1602   & 376    &       & 1465.08 &      \\ \bottomrule
\multicolumn{11} {r@{}}{* represents measurement calculated at 10 MHz}
\end{tabular*}
\end{table}
        \begin{table}
\scriptsize
\caption{Software Implementation of SPN Block Ciphers.}
\label{tab:SW_imp_SPN}
\begin{tabular*}{\textwidth}{@{\extracolsep{\fill}} l c c c c c c c c c c c }
\toprule
\multicolumn{2}{l}{\textbf{Cipher}} &
  \begin{tabular}[c]{@{}c@{}}\textbf{Block}\\  \textbf{Length}\\ \textbf{(bits)}\end{tabular} &
  \begin{tabular}[c]{@{}c@{}}\textbf{Key}\\  \textbf{Size}\\ \textbf{(bits)}\end{tabular} &
  \begin{tabular}[c]{@{}c@{}}\textbf{Code}\\  \textbf{Size}\\ \textbf{(bytes)}\end{tabular} &
  \begin{tabular}[c]{@{}c@{}} \textbf{RAM} \\ \textbf{(bytes)}\end{tabular} &
  \begin{tabular}[c]{@{}c@{}} \textbf{Encryption} \\ \textbf{(cycles)}\end{tabular} &
  \begin{tabular}[c]{@{}c@{}}\textbf{Throughput}\\  \textbf{(Kbps)} \end{tabular} &
  \begin{tabular}[c]{@{}c@{}}\textbf{Energy} \\  \textbf{($\mu$J/bit)}\end{tabular} &
  \begin{tabular}[c]{@{}c@{}} \textbf{Cycles/byte} \\ \textbf{(CpB)} \end{tabular} &
  \begin{tabular}[c]{@{}c@{}} \textbf{RANK} \end{tabular} &
  \begin{tabular}[c]{@{}c@{}}\textbf{Hardware} \\ \textbf{Config} \end{tabular} \\              
\midrule
mCrypton      & &64    & 64  & 2146   & 288  & 305408   & 3.33  & 19.26   & 38176  & 0.00962 & \multirow{7}{*}{\rotatebox{90}{\makecell{Atmel AVR \\ ATMega-128 \\\cite{sevin2021survey}}}} \\ \cline{1-11}
              
PRESENT       & &64    & 80  & 1254   & 418  & 1538064   & 0.66  & 96.89  & 192258 & 0.00249 &  \\       \cline{1-11}

KLEIN         & &64    & 64  & 2224   & 160  & 200368    & 3.86  & 16.59  & 25046  & 0.01569 & \\       \cline{1-11}
              
LED           & &64    & 80  & 2000   & 202  & 2613152   & 0.39  & 163.48 & 326644 & 0.00127 &     \\       \cline{1-11}
              
PRINCE        & &64    & 128 & 2130   & 192  & 171472    & 5.95  & 10.77  & 21434  & 0.018565 &  \\       \cline{1-11}

RECTANGLE     & &64    & 80  & 1420   & 370  & 103424    & 9.12  & 7.03   & 12928  & 0.03581 & \\       \cline{1-11}

MANTIS        & &64    & 128 & 714    & 220  & 190016    & 5.38  & 11.92  & 23752  & 0.03648 & \\ 

\hline

PRESENT       & &64    & 128  & 660   & 280  & 10784     &       &        & 1349   & 0.60761& \multirow{3}{*}{\makecell{Atmel AVR \\ ATMega-128 \\~\cite{kim2020pipo}}} \\ \cline{1-11}

RECTANGLE     & &64    & 128  & 466   & 204  & 3224      &        &       & 403   & 2.83912 & \\       \cline{1-11}

PIPO          & &64    & 128 & 320    & 31   & 1576      &       &        & 197    & 13.2883 & \\

\hline
PRESENT          & &64    & 128 & 2756   & 1384  & 31784 & 24.16  &        & 3973 & 0.03215 & \multirow{3}{*}{\makecell{ARM7-LPC2129\\~\cite{bansod2016pico, bansod2017boron, patil2019dot}}} \\       \cline{1-11}

PICO          & &64    & 128 & 2504   & 1256 & 49611     & 15.48 &        & 6201.38& 0.03215 & \\       \cline{1-11}

BORON         & &64    & 128 & 2408   & 1256 & 7998      & 96.02 &        & 999.75 & 0.2033 &  \\       \cline{1-11}

DoT           & &64    & 128 & 2464   & 1256 & 14283     & 53.77 &        & 1738.38& 0.11256 &  \\       \bottomrule
\end{tabular*}
\end{table}

    \begin{figure}
        \includegraphics[width=\textwidth, height=10 cm, keepaspectratio]{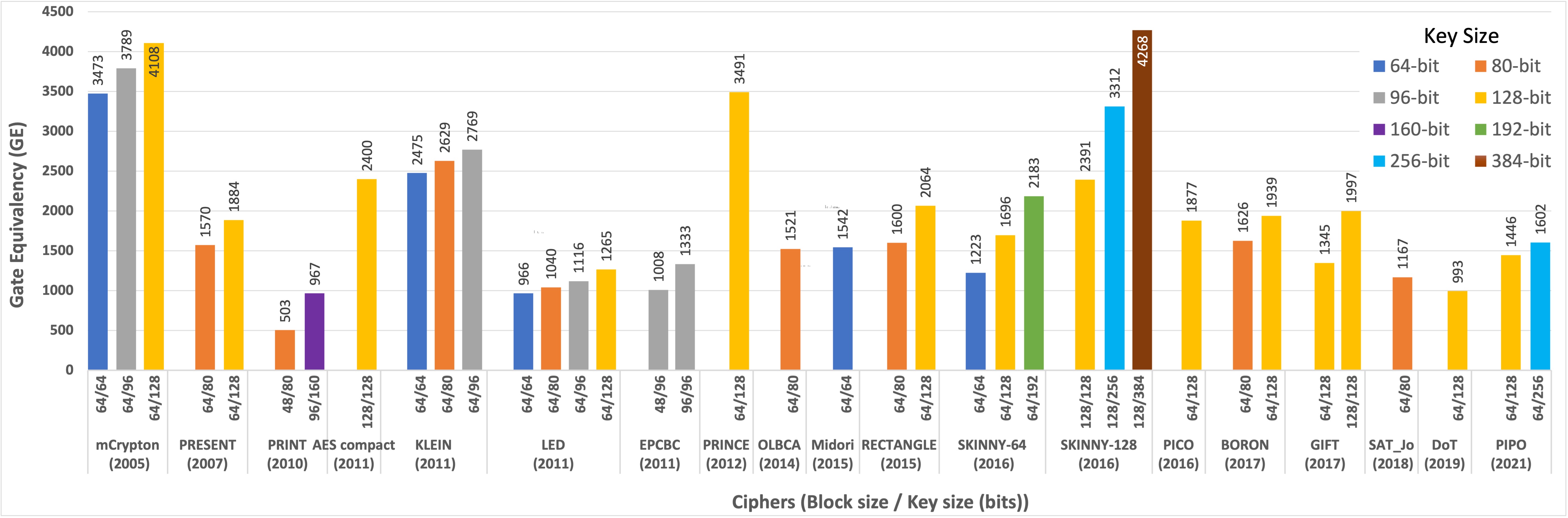}
       ~\caption{Gate Equivalency (GE) of SPN Block Ciphers.}
        \label{fig:GE_SPN}
        \Description{The area consumption of SPN block ciphers}
    \end{figure}

    \begin{figure}
        \includegraphics[width=\textwidth, height=10 cm, keepaspectratio]{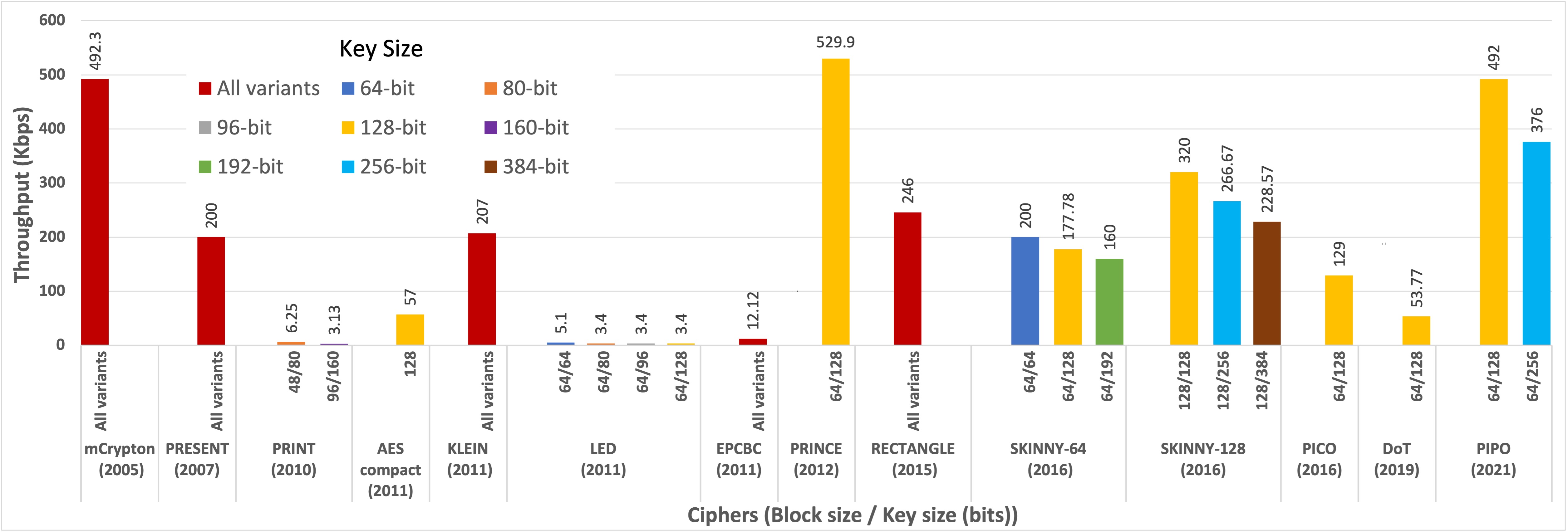}
       ~\caption{Throughput of SPN block ciphers}
        \label{fig:Throughput_SPN}
        \Description{Throughput of SPN block ciphers}
    \end{figure}

    \subsection{Feistel Network (FN)}

    \begin{table} [H]
\scriptsize
\caption{Hardware Implementation of FN Block Ciphers}
\label{tab:HW_imp_FN}
\begin{tabular*}{\textwidth}{@{\extracolsep{\fill}} l c c c c c c c c c}
\toprule
\multicolumn{2}{l}{\textbf{Cipher}} &
  \textbf{Year} &
  \begin{tabular}[c]{@{}c@{}}\textbf{Block}\\  \textbf{Length}\\ \textbf{(bits)}\end{tabular} &
  \begin{tabular}[c]{@{}c@{}}\textbf{Key}\\  \textbf{Size}\\ \textbf{(bits)}\end{tabular} &
  \textbf{Rounds} &
  \textbf{GE} &
  \begin{tabular}[c]{@{}c@{}}\textbf{Throughput}\\  \textbf{(@ 100 KHz)}\end{tabular} &
  \begin{tabular}[c]{@{}c@{}}\textbf{Figure of Merit}\\ \textbf{(FOM)}\end{tabular} &
  \begin{tabular}[c]{@{}c@{}}\textbf{CMOS} \\ \textbf{Tech}\\ \textbf{($\mu$m)}\end{tabular}                       
\\ 
\midrule
TEA~\cite{david2012lightweight}         &           & 1994 & 64  & 128   & 32 & 1984 & 22     & 55.89   & 0.35 \\ \hline

XTEA~\cite{kaps2008chai}        &           & 1997 & 64  & 128   & 64 & 2521 & 57.1   & 89.84   & 0.13 \\ \hline

SEA~\cite{bogdanov2007present}    &           & 2006 & 96   & 96    & 32 & 2280 & 103    & 198.14  & 0.13 \\ \hline

DESL~\cite{leander2007new}        &           & 2007 & 64  & 56    & 16 & 1848 & 44.4   & 130.01  & 0.18 \\ \hline

DESXL~\cite{kitsos2012comparative} &           & 2007 & 64  & 184   &    & 2168 & 44.4   & 94.46   & 0.18 \\ \hline

MIBS~\cite{izadi2009mibs}        &           & 2009 & 64  & 64/80 & 32 & 1400 & 200    & 1020.41 & 0.18 \\ \hline

LBLOCK~\cite{wu2011lblock}      &           & 2011 & 64  & 80    & 32 & 1320 & 200    & 1147.84 & 0.18 \\ \hline

SIMON~\cite{cryptoeprint:2013:404}       & SIMON-48  & 2013 & 48  & 96    & 36 & 763  & 15     & 257.66  & 0.13 \\ \cline{4-9}
            & SIMON-64  &      & 64  & 96    & 42 & 838  & 17.8   & 253.47  &      \\
            &           &      &     & 128   & 44 & 1000 & 16.7   & 167     &      \\ \cline{4-9}
            & SIMON-96  &      & 96  & 96    & 52 & 984  & 14.8   & 152.85  &      \\
            & SIMON-128 &      & 128 & 128   & 68 & 1317 & 22.9   & 132.03  &      \\ \hline
            
RoadRunneR~\cite{liu2016efficient} &           & 2015 & 64  & 80    & 10 & 1388 & 156     & 809.74  & 0.18 \\ \hline

ANU         & ANU-I~\cite{bansod2016anu}     & 2016 & 64  & 128   & 25 & 1015 &        &         & 0.13 \\ \cline{2-8}\\
            & ANU-II~\cite{dahiphale2017anu}    & 2017 & 64  & 128   & 25 & 1010 &        &         &      \\ \hline
            
DLBCA~\cite{aldabbagh2017design}      &           & 2017 & 32  & 80    & 15 & 1116 &        &         &      \\ \hline

LiCi~\cite{patil2017lici}       &           & 2017 & 64  & 128   & 31 & 1153 & 305    & 2294.25 &     \\ \hline

SLIM~\cite{aboushosha2020slim}        &           & 2020 & 32  & 80    & 32 & 553  &        &         & 0.13 \\ \hline
SCENERY~\cite{feng2022scenery}    &           & 2022 & 64  & 80    & 28 & 1438 & 228.57 & 1105.35 & 0.18 \\ \bottomrule
\end{tabular*}
\end{table}

    In the software implementation performed on Atmel AVR ATtiny45, the code size is specified, but not the RAM. The authors of these works claim that all internal variables were saved in CPU registers, and SRAM was only used to hold plaintext, ciphertext, and the master key. Thus, we omit RAM usage when calculating RANK on this platform.

    \begin{table}[H]
\scriptsize
\caption{Software Implementation of FN Block Ciphers.}
\label{tab:SW_imp_FN}
\begin{tabular*}{\textwidth}{@{\extracolsep{\fill}} l c c c c c c c c c c c }
\toprule
\multicolumn{2}{l}{\textbf{Cipher}} &
  \begin{tabular}[c]{@{}c@{}}\textbf{Block}\\  \textbf{Length}\\ \textbf{(bits)}\end{tabular} &
  \begin{tabular}[c]{@{}c@{}}\textbf{Key}\\  \textbf{Size}\\ \textbf{(bits)}\end{tabular} &
  \begin{tabular}[c]{@{}c@{}}\textbf{Code}\\  \textbf{Size}\\ \textbf{(bytes)}\end{tabular} &
  \begin{tabular}[c]{@{}c@{}} \textbf{RAM} \\ \textbf{(bytes)}\end{tabular} &
  \begin{tabular}[c]{@{}c@{}} \textbf{Encryption} \\ \textbf{(cycles)}\end{tabular} &
  \begin{tabular}[c]{@{}c@{}}\textbf{Throughput}\\  \textbf{(Kbps)} \end{tabular} &
  \begin{tabular}[c]{@{}c@{}}\textbf{Energy} \\  \textbf{($\mu$J/bit)}\end{tabular} &
  \begin{tabular}[c]{@{}c@{}} \textbf{Cycles/byte} \end{tabular} &
  \begin{tabular}[c]{@{}c@{}} \textbf{RANK} \end{tabular} &
  \begin{tabular}[c]{@{}c@{}}\textbf{Hardware} \\ \textbf{Config} \end{tabular}                       
\\ 
\midrule
XTEA          & &64    & 128 & 1672   & 184  & 175042    & 5.84  & 10.97   & 21880.25 & 0.0224 & \multirow{7}{*}{\rotatebox{90}{\makecell{Atmel AVR \\ ATMega-128\\~\cite{sevin2021survey}}}}   \\       \cline{1-11}
              
SEA           & &96    & 96  & 1338   & 788  & 186186    & 5.01  & 8.53    & 15515.5  & 0.02212 \\       \cline{1-11}

DESXL         & &64    & 64  & 11444  & 480  & 921200    & 1.05  & 61.06   & 115150   & 0.0007  \\       \cline{1-11}
              
LBLOCK        & &64    & 80  & 1692   & 290  & 149488    & 6.62  & 9.67    & 18686   & 0.02355  \\       \cline{1-11}

SIMON         & &64    & 128 & 1464   & 344  & 33360     & 27.58 & 2.32    & 4170    & 0.11144  \\       \cline{1-11}

RoadRunneR    & &64    & 128 & 1356   & 316  & 133200    & 7.62  & 8.40    & 16650   & 0.03021  \\       \hline

TEA           & &64    & 128  & 672   &      & 7408      &       &         & 926     & 1.60701 & \multirow{4}{*}{\makecell{Atmel AVR \\ ATtiny45\\~\cite{karakocc2013itubee}}}   \\       \cline{1-11}

SEA           & &96    & 96  & 450    &      & 41604     &       &         & 3467    & 0.64096  \\       \cline{1-11}

DESXL         & &64    & 184  & 868   &      & 84602     &       &         & 10575   & 0.10894  \\       \cline{1-11}

LBlock        & &64    & 80  &        &      & 3955      &       &         &  494    & 5.3573  \\  \cline{1-11}

ITUbee        & &80    & 80  &  716   &      & 2607      &       &         &  261    & 5.3573  \\       \hline

SIMON         & &64    & 128 & 2324   & 1256 & 1268     & 605   &         & 158.5 & 1.30462    & \multirow{4}{*}{\makecell{ARM7-LPC2129\\~\cite{patil2017lici, dahiphale2017anu}}}    \\       \cline{1-11}

ANU-I         & &64    & 128 & 2316   & 1304 & 20502     & 37.46 &         & 2562.75 & 0.07925 &  \\       \cline{1-11}

ANU-II        & &64    & 128 & 1852   & 1272 & 1756      & 431.31&        & 219.5   & 1.03635  &  \\       \cline{1-11}

LiCi          & &64    & 128 & 1944   & 1256 & 2518     & 305    &        & 314.75  & 0.713   &  \\       \cline{1-11}

SCENERY       & &64    & 80  &        &      & 1516     &       &        &  189.5   &    \\       \bottomrule
\end{tabular*}
\end{table}

    \begin{figure} [H]
        \includegraphics[width=\textwidth, height=11 cm, keepaspectratio]{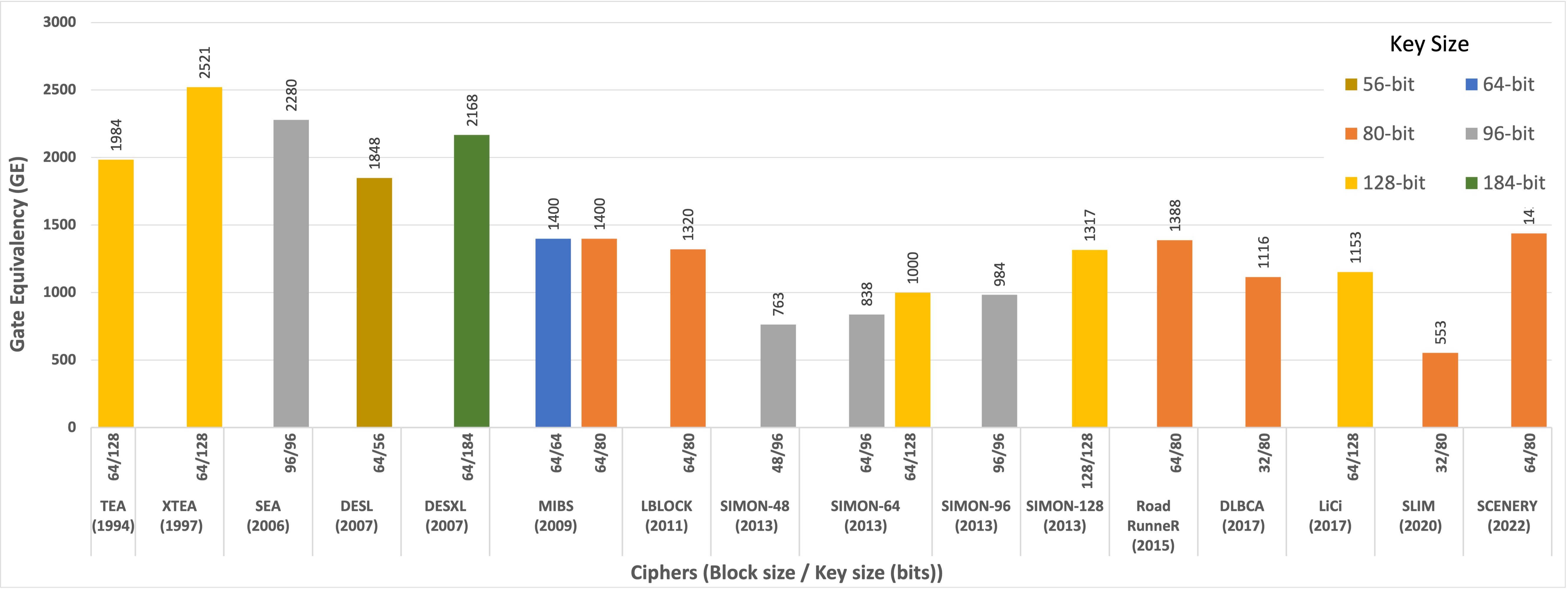}
       ~\caption{Gate Equivalency (GE) of FN Block Ciphers.}
        \label{fig:GE_FN}
        \Description{The area consumption of FN block ciphers}
    \end{figure}
    
    \begin{figure} [H]
        \includegraphics[width=\textwidth, height=11 cm, keepaspectratio]{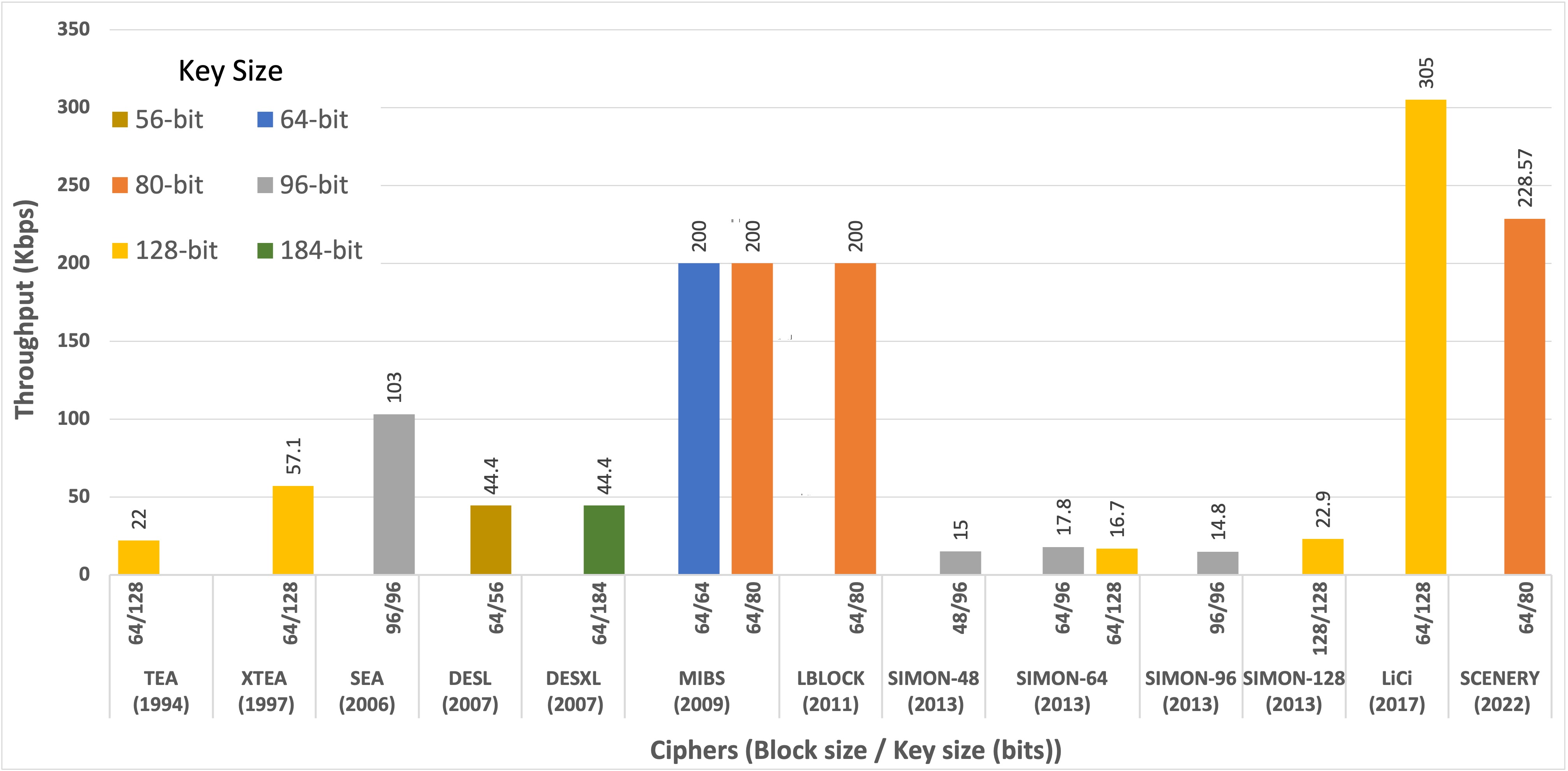}
       ~\caption{Throughput of FN Block Ciphers.}
        \label{fig:Throughput_FN}
        \Description{Throughput of FN block ciphers}
    \end{figure}
    
\newpage
    \subsection{Generalized Feistel Network (GFN) and Lai-Massey Designs (LMD)}

    \begin{table}[H]
\scriptsize
\caption{Hardware Implementation of GFN and LMD Block Ciphers.}
\label{tab:HW_imp_GFN_LMD}
\begin{tabular*}{\textwidth}{@{\extracolsep{\fill}} l c c c c c c c c c c}
\toprule
\multicolumn{2}{l}{\textbf{\textbf{Cipher}}} &
  \textbf{Year} &
  \begin{tabular}[c]{@{}c@{}}\textbf{Block}\\  \textbf{Length}\\ \textbf{(bits)}\end{tabular} &
  \begin{tabular}[c]{@{}c@{}}Key\\  \textbf{Size}\\ \textbf{(bits)}\end{tabular} &
  \textbf{Rounds} &
  \textbf{GE} &
  \begin{tabular}[c]{@{}c@{}}\textbf{Throughput}\\  \textbf{(@ 100 KHz)}\end{tabular} &
  \begin{tabular}[c]{@{}c@{}}\textbf{Power} \\  \textbf{Consumption}\\  \textbf{($\mu$W)}\end{tabular} &
  \begin{tabular}[c]{@{}c@{}}\textbf{Figure of Merit}\\ \textbf{(FOM)}\end{tabular} &
  \begin{tabular}[c]{@{}c@{}}\textbf{CMOS} \\ \textbf{Tech}\\ \textbf{($\mu$m)}\end{tabular} \\
  \midrule
\multicolumn{11}{c}{\textbf{Generalized Feistel Network (GFN)}}                         
\\ 
\hline
HIGHT~\cite{lim2009implementation}  &             & 2006 & 64  & 128 & 32    & 2608    & 188.2  &      & 276.7  & 0.25  \\ \hline

CLEFIA~\cite{rashidi2020efficient}  & CLEFIA-128  & 2007 & 64  & 128 & 18    & 5979    & 711.11 &      & 198.92  & 0.09  \\
        & CLEFIA-192  &      & 192 &     & 22    & 8536    & 581.8  &      & 79.85   &       \\
        & CLEFIA-256  &      & 256 &     & 26    & 8482    & 492.3  &      & 68.43   &       \\ \hline

Piccolo~\cite{shibutani2011piccolo}  & piccolo-80  & 2011 & 64  & 80  & 26/32 & 1274    & 237.04 &      & 1460.44 & 0.13  \\
        & piccolo-128 &      &     & 128 & 32    & 1362    &        &      & 1277.81 &       \\ \hline
        
TWINE~\cite{suzaki2012textnormal}   & Twine-80    & 2012 & 64  & 80  & 36    & 1799    & 178    &      & 549.99  & 0.09  \\
        & Twine-128   &      &     & 128 &       & 2285    &        &      & 341.21  &       \\ \hline
        
Khudra  &             & 2014 & 64  & 80  & 18    & 1939.4  & 177.8  &      & 472.91  & 0.18  \\ \hline

HISEC~\cite{sulaiman2018modified}   &             & 2014 & 64  & 80  & 15    & 1694.08 &        &      &         &       \\ \hline

FeW~\cite{kumar2014few}     & FeW-80      & 2019 & 64  & 80  & 32    &         &        &      &         &        \\
        & FeW-128     &      &     & 128 &       &         &        &      &         &        \\ \hline
        
WARP~\cite{banik2020warp}    &             & 2021 & 128 & 128 & 26    & 925     &        & 34.6*&        &         \\ \hline

ALLPC~\cite{cheng2021allpc}   &             & 2021 & 64  & 128 & 24    & 1023    &        &      &        & 0.13  \\ \hline

\multicolumn{11}{c}{\textbf{Lai-Massey Designs (LMD)}}                       \\ \hline

QTL~\cite{li2016qtl}     & QTL-80      & 2016 & 64  & 80  & 25    & 1025    & 200    & 1.55 & 1903.63 & 0.18    \\
        & QTL-128     &      &     & 128 & 31    & 1206    &        & 2.16 & 1375    &         \\ \hline
        
SIT~\cite{usman2017sit}     &             & 2017 & 64  & 64  & 5     &         &        &      &         &          \\ \hline

SFN~\cite{li2018sfn}     &             & 2018 & 64  & 96  & 32    & 1876.04 & 200    & 1.97 & 568.28  & 0.18     \\ \hline
LRBC~\cite{biswas2020lrbc}    &             & 2020 & 16  & 16  & 24    & 258.9   &        & 11.4 &         & 0.065    \\ \hline
LCB~\cite{roy2021lcb}    &             & 2021 & 32  & 64  &       & 224     &        &      &         & 0.028    \\ \bottomrule
\multicolumn{11} {r@{}}{* represents measurement calculated at 10 MHz}
\end{tabular*}
\end{table}

    \begin{table}[H]
\scriptsize
\caption{Software Implementation of GFN and LMD Block Ciphers.}
\label{tab:SW_imp_GFN_LMD}
\begin{tabular*}{\textwidth}{@{\extracolsep{\fill}} l c c c c c c c c c c c }
\toprule
\multicolumn{2}{l}{\textbf{Cipher}} &
  \begin{tabular}[c]{@{}c@{}}\textbf{Block}\\  \textbf{Length}\\ \textbf{(bits)}\end{tabular} &
  \begin{tabular}[c]{@{}c@{}}\textbf{Key}\\  \textbf{Size}\\ \textbf{(bits)}\end{tabular} &
  \begin{tabular}[c]{@{}c@{}}\textbf{Code}\\  \textbf{Size}\\ \textbf{(bytes)}\end{tabular} &
  \begin{tabular}[c]{@{}c@{}} \textbf{RAM} \\ \textbf{(bytes)}\end{tabular} &
  \begin{tabular}[c]{@{}c@{}} \textbf{Encryption} \\ \textbf{(cycles)}\end{tabular} &
  \begin{tabular}[c]{@{}c@{}}\textbf{Throughput}\\  \textbf{(Kbps)} \end{tabular} &
  \begin{tabular}[c]{@{}c@{}}\textbf{Energy} \\  \textbf{($\mu$J/bit)}\end{tabular} &
  \begin{tabular}[c]{@{}c@{}} \textbf{Cycles/byte} \end{tabular} &
  \begin{tabular}[c]{@{}c@{}} \textbf{RANK} \end{tabular} &
  \begin{tabular}[c]{@{}c@{}}\textbf{Hardware} \\ \textbf{Config} \end{tabular}  \\ \midrule
\multicolumn{12}{c}{\textbf{Generalized Feistel Network (GFN)}}                         
\\ \hline
HIGHT         & &64    & 128 & 1346   & 440  & 116080    & 8.62  & 7.43    & 14510 & 0.03096  & \multirow{4}{*}{\rotatebox{90}{\makecell{Atmel AVR \\ ATMega-128 \\~\cite{sevin2021survey}}}}   \\       \cline{1-11}
              
CLEFIA        & &128   & 256 & 3198   & 432  & 180384    & 5.04  & 6.36    & 11274 & 0.02184  & \\       \cline{1-11}

Piccolo       & &64    & 80  & 1858   & 270  & 295696    & 3.44  & 18.61   & 36962 & 0.01128  &   \\       \cline{1-11}
              
WARP          & &128   & 128 & 1330   & 256  & 68698     &       &         & 4293.6& 0.12644  &   \\       \hline
              
\multicolumn{12}{c}{\textbf{Lai-Massey Designs (LMD)}}   \\ \hline

SFN           & &64    & 96  &        &      & 4736      &       &         & 592   &  & \multirow{2}{*}{\makecell{ATMega-328 \\~\cite{usman2017sit}}}  \\       \cline{1-11}

SIT           & &64    & 64 & 826     & 22   & 3006      &       &         & 375.75 & 3.05902   \\       \bottomrule

\end{tabular*}
\end{table}

    \begin{figure} [H]
        \includegraphics[width=\textwidth, height=12 cm, keepaspectratio]{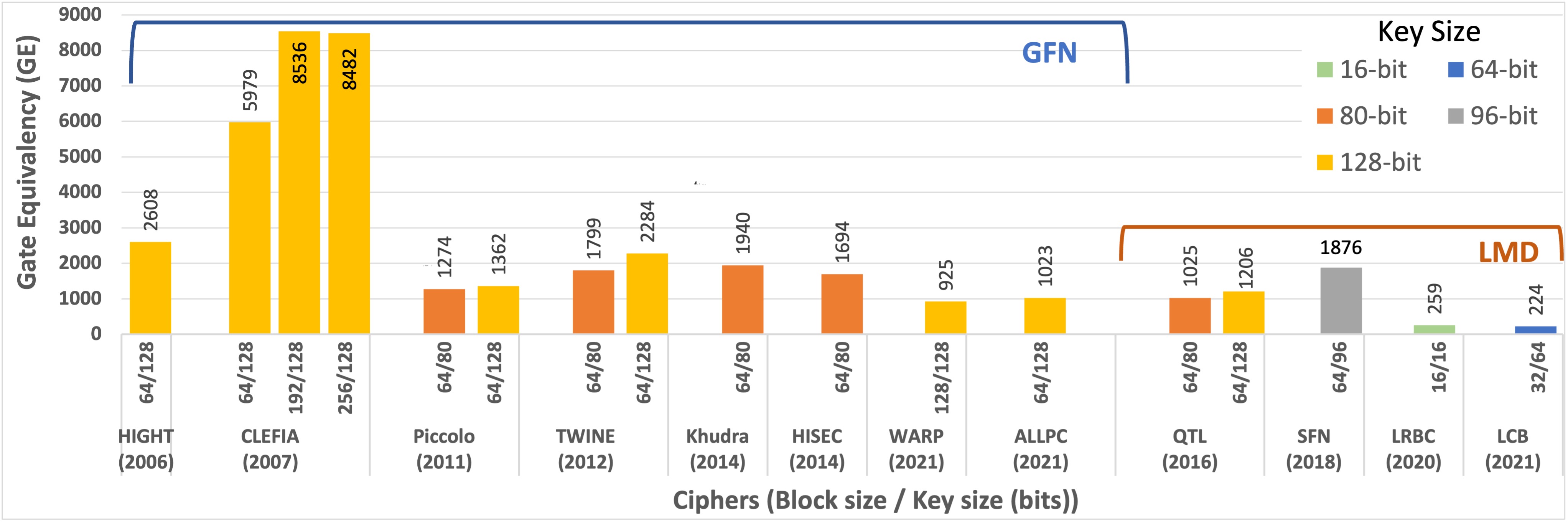}
       ~\caption{Gate Equivalency of GFN and LMD Block Ciphers.}
        \label{fig:GE_GFN_LMD}
        \Description{The area consumption of GFN and LMD block ciphers}
    \end{figure}

    \begin{figure} [H]
        \includegraphics[width=\textwidth, height=11 cm, keepaspectratio]{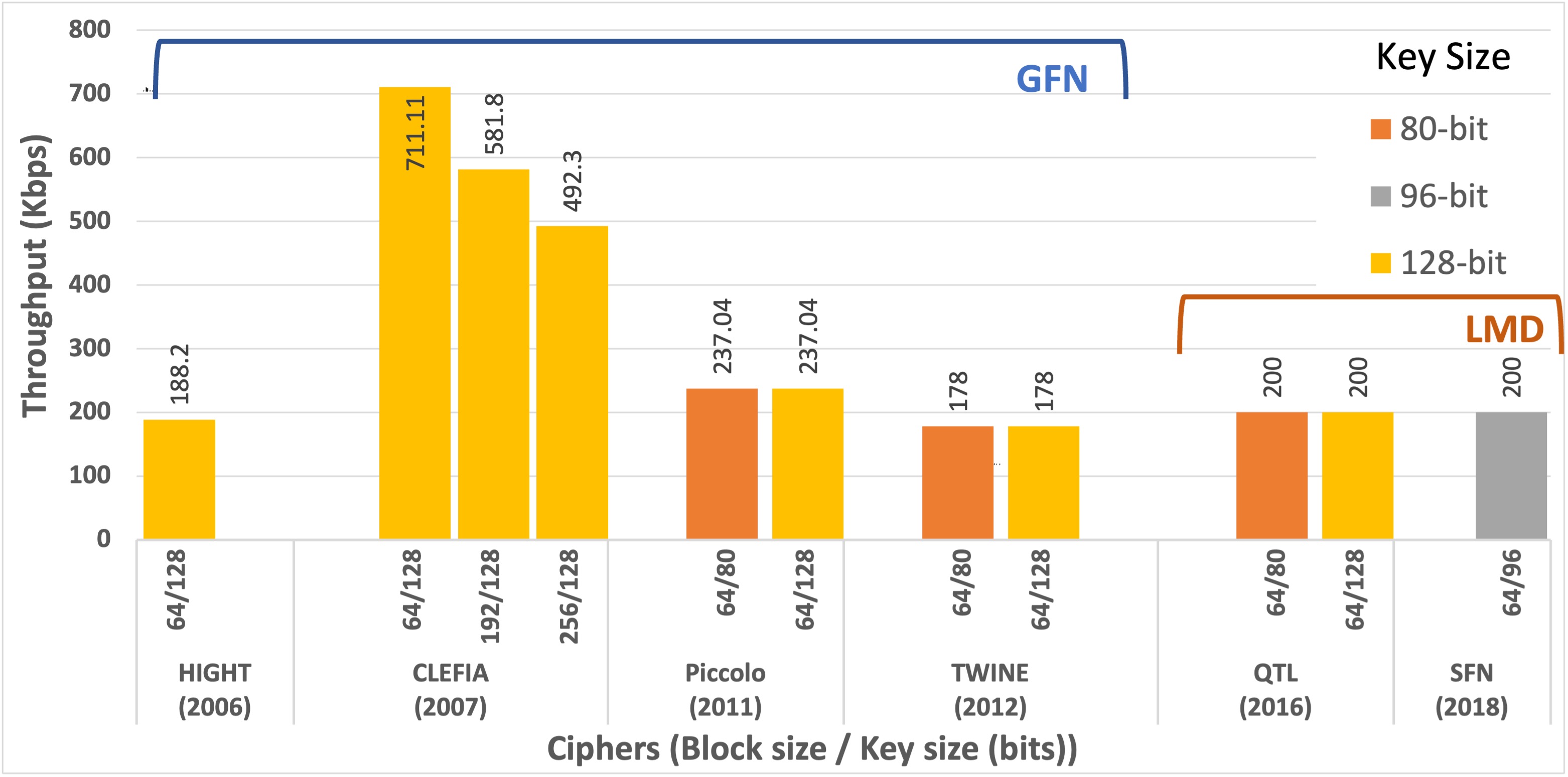}
       ~\caption{Throughput of GFN and LMD Block Ciphers}
        \label{fig:Throughput_GFN_LMD}
        \Description{Throughput of GFN and LMD block ciphers}
    \end{figure}
    
\newpage
    \subsection{Addition Rotation eXclusive-or- (ARX)}

    \begin{table} [H]
\scriptsize
\caption{Hardware Implementation of ARX Block Ciphers.}
\label{tab:HW_imp_ARX}
\begin{tabular*}{\textwidth}{@{\extracolsep{\fill}} l c c c c c c c c c}
\toprule
\multicolumn{2}{l}{\textbf{Cipher}} &
  \textbf{Year} &
  \begin{tabular}[c]{@{}c@{}}\textbf{Block}\\  \textbf{Length}\\ \textbf{(bits)}\end{tabular} &
  \begin{tabular}[c]{@{}c@{}}\textbf{Key}\\  \textbf{Size}\\ \textbf{(bits)}\end{tabular} &
  \textbf{Rounds} &
  \textbf{GE} &
  \begin{tabular}[c]{@{}c@{}}\textbf{Throughput}\\  \textbf{(@ 100 KHz)}\end{tabular} &
  \begin{tabular}[c]{@{}c@{}}\textbf{Figure of Merit}\\ \textbf{(FOM)}\end{tabular} &
  \begin{tabular}[c]{@{}c@{}}\textbf{CMOS} \\ \textbf{Tech}\\ \textbf{($\mu$m)}\end{tabular}                
\\ 
\midrule
SPECK~\cite{cryptoeprint:2013:404}  & SPECK-48  & 2013 & 48     & 96  & 36 & 884  & 12    & 153.56 & 0.13 \\ \cline{4-9}
       & SPECK-64  &      & 64     & 96  & 42 & 984  & 14.5  & 149.75 &      \\
       &           &      &        & 128 & 44 & 1127 & 13.8  & 108.65 &      \\ \cline{4-9}
       & SPECK 96  &      & 96     & 96  & 52 & 1134 & 13.8  & 107.31 &      \\
       & SPECK-128 &      & 128    & 128 & 68 & 1396 & 12.1  & 62.09  &      \\ \hline
       
LEA~\cite{hong2013lea}    &           & 2013 & 128    & 128 & 24 & 3826 & 76.19 & 52.05  & 0.13 \\ \hline

Simeck~\cite{yang2015simeck} & Simeck-32 & 2015 & 32     & 64  & 32 & 505  & 5.6   & 219.59 & 0.13 \\
       & Simeck-48 &      & 48     & 96  & 36 & 715  & 5     & 97.8   &      \\
       & Simeck 64 &      & 64     & 128 & 44 & 924  & 4.2   & 49.19  &      \\ \hline

CHAM~\cite{koo2017cham}   & CHAM-64   & 2017 & 64     & 128 & 80 & 665  & 5     & 113.06 & 0.13 \\ \cline{4-9}
       & CHAM-128  &      & 128    & 128 & 80 & 1057 & 5     & 44.75  &      \\
       &           &      &        & 256 & 96 & 1180 & 4.2   & 30.16  &      \\ \hline
       
Shadow~\cite{guo2021shadow} & Shadow-32 & 2021 & 32     & 64  & 16 & 836  &       &        & 0.18 \\
       & Shadow-64 &      & 64     & 128 & 32 & 1689 &       &        &      \\ \bottomrule
\end{tabular*}
\end{table}

    \begin{table}[H]
\scriptsize
\caption{Software Implementation of ARX Block Ciphers.}
\label{tab:SW_imp_ARX}
\begin{tabular*}{\textwidth}{@{\extracolsep{\fill}} l c c c c c c c c c c c }
\toprule
\multicolumn{2}{l}{\textbf{Cipher}} &
  \begin{tabular}[c]{@{}c@{}}\textbf{Block}\\  \textbf{Length}\\ \textbf{(bits)}\end{tabular} &
  \begin{tabular}[c]{@{}c@{}}\textbf{Key}\\  \textbf{Size}\\ \textbf{(bits)}\end{tabular} &
  \begin{tabular}[c]{@{}c@{}}\textbf{Code}\\  \textbf{Size}\\ \textbf{(bytes)}\end{tabular} &
  \begin{tabular}[c]{@{}c@{}} \textbf{RAM} \\ \textbf{(bytes)}\end{tabular} &
  \begin{tabular}[c]{@{}c@{}} \textbf{Encryption} \\ \textbf{(cycles)}\end{tabular} &
  \begin{tabular}[c]{@{}c@{}}\textbf{Throughput}\\  \textbf{(Kbps)} \end{tabular} &
  \begin{tabular}[c]{@{}c@{}}\textbf{Energy} \\  \textbf{($\mu$J/bit)}\end{tabular} &
  \begin{tabular}[c]{@{}c@{}} \textbf{Cycles/byte} \end{tabular} &
  \begin{tabular}[c]{@{}c@{}} \textbf{RANK} \end{tabular} &
  \begin{tabular}[c]{@{}c@{}}\textbf{Hardware} \\ \textbf{Config} \end{tabular}                       
\\ 
\midrule
SPECK         & &64    & 128 & 2232   & 276  & 20560     & 45.88  & 1.4    & 2570  & 0.13976  & \multirow{4}{*}{\makecell{Atmel AVR \\ ATMega-128 \\~\cite{sevin2021survey}}}  \\       \cline{1-11}
              
LEA           & &128   & 128 & 3010   & 576  & 25272     & 34.09  & 0.94   & 1579.5& 0.15212  &    \\       \cline{1-11}

SPARX         & &64    & 128 & 1904   & 368  & 89664     & 11.06  & 5.79   & 11208 & 0.0338  &    \\       \cline{1-11}
              
CHAM          & &64    & 128 & 1108   & 208  & 334688    & 3.05   & 20.97  & 41836 & 0.01568 &    \\       \bottomrule

\end{tabular*}
\end{table}

    \begin{figure} [H]
        \includegraphics[width=14.3 cm, height=8 cm, keepaspectratio]{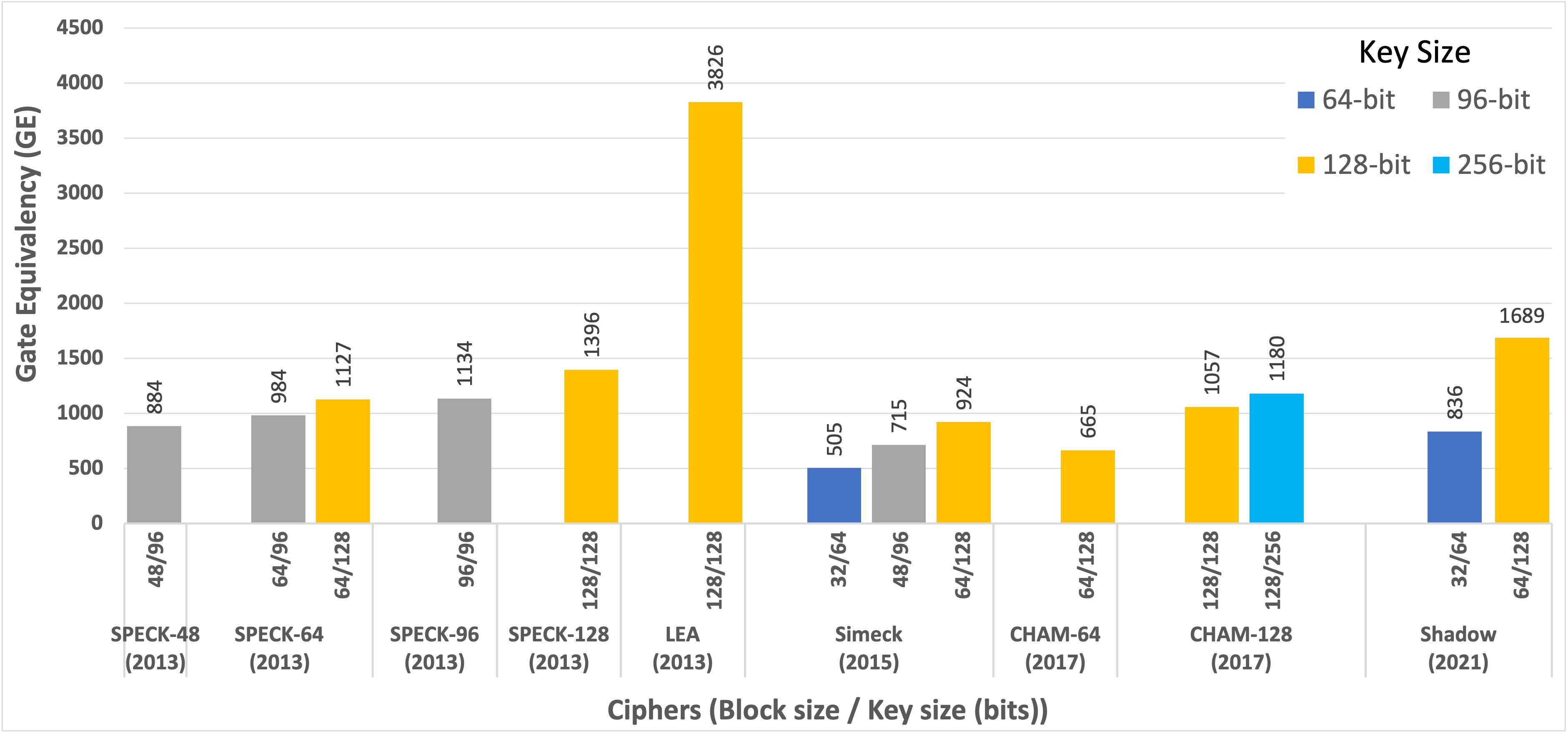}
       ~\caption{Gate Equivalency (GE) of ARX Block Ciphers.}
        \label{fig:GE_ARX}
        \Description{The area consumption of ARX block ciphers}
    \end{figure}

    \begin{figure} [H]
        \includegraphics[width=\textwidth, height=10 cm, keepaspectratio]{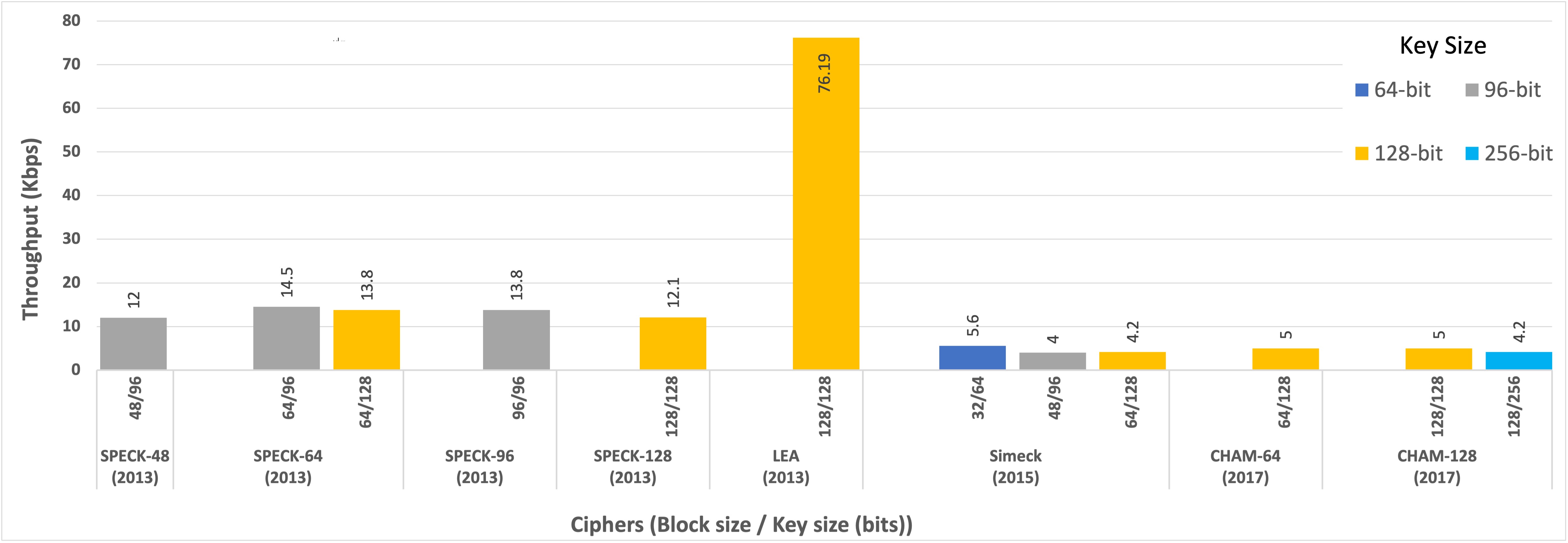}
       ~\caption{Throughput of ARX Block Ciphers.}
        \label{fig:Throughput_ARX}
        \Description{Throughput of GFN block ciphers}
    \end{figure}
    
\section{Lightweight Stream Ciphers}

    \subsection{eSTREAM ciphers}

    \begin{table} [H]
\scriptsize
\caption{Hardware Implementation of eSTREAM Ciphers.}
\begin{tabular*}{\textwidth}{@{\extracolsep{\fill}}l c c c c c c c c c}
\toprule
\multicolumn{2}{l}{\textbf{Cipher}} &
  \textbf{Year} &
  \begin{tabular}[c]{@{}c@{}}\textbf{IV}\\\textbf{(bits)}\end{tabular} &
  \begin{tabular}[c]{@{}c@{}}\textbf{Key}\\  \textbf{Size}\\ \textbf{(bits)}\end{tabular} &
  \textbf{Architecture} &
  \textbf{GE} &
  \begin{tabular}[c]{@{}c@{}}\textbf{Throughput}\\  \textbf{(@ 10 MHz)} \\ \textbf{(Mbps)}\end{tabular} &
  \begin{tabular}[c]{@{}c@{}}\textbf{Figure of Merit} \\ \textbf{(FOM)}\end{tabular} &
  \begin{tabular}[c]{@{}c@{}}\textbf{CMOS} \\ \textbf{Tech}\\ \textbf{($\mu$m)}\end{tabular} \\\midrule
\multicolumn{9}{l}{\textbf{eSTREAM Project: Profile 1 (Software Oriented)}}                                                               \\ \hline
Rabbit~\cite{manifavas2016survey}    &            & 2003 & 64    & 128 & \begin{tabular}[c]{@{}c@{}}Chaotic Map\\  Table\end{tabular} & 3800 &     &   & 0.18 \\ \hline

Salsa20/r~\cite{bernstein2005salsa20} & Salsa20/20 & 2005 &       &     & ARX                                                          &      &     &   & 0.13 \\
          & Salsa20/12 &      &       &     &                                                              &      &     &   &      \\
          & Salsa20/8  &      &       &     &                                                              &      &     &   &      \\ \hline
HC-128~\cite{wu2008stream}    &            & 2008 & 128   & 128 & \begin{tabular}[c]{@{}c@{}}NLFSR + \\ S-Box\end{tabular}     &      &     &   & 0.13 \\ \hline

Sosemanuk~\cite{berbain2008sosemanuk} &            & 2008 & 128 & \begin{tabular}[c]{@{}c@{}}128/\\ 256\end{tabular} & \begin{tabular}[c]{@{}c@{}}FSM + \\ LFSR\end{tabular} & 18819 & & &0.09 \\ \hline
\multicolumn{9}{l}{\textbf{eSTREAM Project: Profile 2 (Hardware Oriented)}}                                                                \\ \hline
Trivium~\cite{good2008hardware}   &            & 2006 & 80    & 80  & 3SHR                                                         & 2580 & 327.9 & 4926 & 0.13 \\ \hline
Grain~\cite{good2008hardware}     & Grain      & 2005 & 64    & 80  & \begin{tabular}[c]{@{}c@{}}LFSR + \\ NLFSR +\end{tabular}    & 1294 & 724.6  & 43274.2 &0.13 \\
          & Grain 128  & 2006 & 96    & 128 & Boolean filter                                               & 1857 & 925.9 & 26849     \\ \hline
Mickey~\cite{good2008hardware}     & Mickey 1.0 & 2005 & 0-80  & 80  & Galois LFSR +                                                & 3188 & 454.5 & 4471.9 & 0.13 \\
          & Mickey 2.0 & 2008 & 0-128 & 128 & NLFSR                                                        & 5039 & 413.2     & 1627.3    \\ \bottomrule
\end{tabular*}
\label{tab:HW_imp_eSTREAM}
\end{table}
    \begin{table}[H]
\scriptsize
\caption{Software Implementation of eSTREAM Ciphers.}
\label{tab:SW_imp_eSTREAM}
\begin{tabular*}{\textwidth}{@{\extracolsep{\fill}} l c c c c c c c }
\toprule
\multicolumn{2}{l}{\textbf{Cipher}} &
  \begin{tabular}[c]{@{}c@{}}\textbf{IV} \\ \textbf{(bits)}\end{tabular} &
  \begin{tabular}[c]{@{}c@{}}\textbf{Key}\\  \textbf{Size}\\ \textbf{(bits)}\end{tabular} &
  \begin{tabular}[c]{@{}c@{}}\textbf{Total}\\ \textbf{Cycles} \end{tabular} &
  \begin{tabular}[c]{@{}c@{}} \textbf{Cycles/byte} \end{tabular} &
  \begin{tabular}[c]{@{}c@{}} \textbf{Encryption Speed} \\ \textbf{(Mbps)}\end{tabular} &
  \begin{tabular}[c]{@{}c@{}}\textbf{Hardware} \\ \textbf{Config} \end{tabular} \\
\multicolumn{8}{l}{\textbf{eSTREAM  Project: Profile 1 (Software Oriented)}}                         
\\ 
\midrule
Rabbit    & &64    & 128 & 726330  & 9.97  & 2409.13 & \multirow{10}{*}{\rotatebox{90}{\makecell{Intel(R), Pentium(R) 4 \\ CPU 3.00 GHz \\cache size 1024 KB\\~\cite{jassim2021survey}}}}   \\       \cline{1-7}
              
Salsa20   & &64    & 128 & 737715  & 16.27 & 1475.78    \\       
          & &64    & 256 & 738157  & 16.01 & 1499.84    \\        \cline{1-7}

HC-128    & &128   & 128 & 672150  & 287.56& 83.51      \\        \cline{1-7}
              
Sosemanuk & &64    & 128 & 738135  & 14.63 & 1641.40     \\       \cline{1-7}

\multicolumn{8}{l}{\textbf{eSTREAM  Project: Profile 2  (Hardware Oriented)}}  \\ \cline{1-7}

Trivium    & &80    & 80  & 736222  & 14.19  & 1692.39    \\       \cline{1-7}

Grain      & &64    & 80  & 16732072& 4340.44& 5.53    \\       \cline{1-7}

Mickey 1.0 & &80    & 80  & 4846455 & 1272.40& 18.87   \\       

Mickey 2.0 & &128   & 128 & 91223775& 22427.07& 1.07   
\\\bottomrule

\end{tabular*}
\end{table}

    \subsection{LWAE ciphers}

    \begin{table} [H]
\scriptsize
\caption{Hardware implementation of LWAE Stream Ciphers.}
\begin{tabular*}{\textwidth}{@{\extracolsep{\fill}} l c c c c c c c}
\toprule
\textbf{Cipher} &
  \begin{tabular}[c]{@{}c@{}}\textbf{Year} \end{tabular} &
  \begin{tabular}[c]{@{}c@{}}\textbf{IV}\\ \textbf{(bits)}\end{tabular} &
  \begin{tabular}[c]{@{}c@{}}\textbf{Key}\\  \textbf{Size}\\ \textbf{(bits)}\end{tabular} &
  \begin{tabular}[c]{@{}c@{}}\textbf{Authentication} \\ \textbf{Tag Size}\\ \textbf{(bits)}\end{tabular} &
  \textbf{Architecture} &
  \textbf{GE} &
  \begin{tabular}[c]{@{}c@{}}\textbf{CMOS} \\ \textbf{Tech}\\ \textbf{($\mu$m)}\end{tabular}                                                                  \\ \midrule
Rabbit-MAC~\cite{tahir2008rabbit}    & 2008 & 64  & 128 &     & Chaotic Map Table                                              &      &      \\ \hline
WG-7~\cite{luo2010lightweight}          & 2010 & 81  & 80  & 80  & LFSR +WG                                                       &      & 0.09 \\ \hline
Hummingbird-1~\cite{engels2010hummingbird} & 2010 & 64  & 256 &     & Hybrid                                                         &      & 0.13 \\
Hummingbird-2~\cite{engels2011hummingbird} & 2011 & 64  & 128 &     &                                                                & 2332 &      \\ \hline
ASC-1~\cite{bilgin2013fides}         & 2011 & 56  & 128 &     & SPN                                                            & 4964 & 0.65 \\ \hline
Grain-128a~\cite{diedrich2016comparison}   & 2011 & 96  & 128 &     & NLFSR + LFSR                                                   & 1857 & 0.65 \\ \hline
Sablier~\cite{zhang_shi_xu_yao_li_2014}    & 2014 & 80  & 80  & 32  & ARX                                                            & 1925 & 0.18 \\ \hline
ACORN~\cite{wu_2014}      & 2014 & 128 & 128 & 128 & 6 LFSR                                                         &      & 0.13 \\ \hline
ALE~\cite{bilgin2013fides}          & 2014 & 128 & 128 & 128 & SPN                                                            & 2700 & 0.65 \\ \hline
FAN~\cite{jiao2021fan}         & 2021 & 64  & 128 & 72  & \begin{tabular}[c]{@{}c@{}}NLFSR + LFSR +\\ S-Box\end{tabular} & 2327 & 0.09 \\ \bottomrule
\end{tabular*}
\label{tab:HW_imp_LWAE}
\end{table}

    \begin{figure} [H]
        \includegraphics[width=\textwidth, height=12 cm, keepaspectratio]{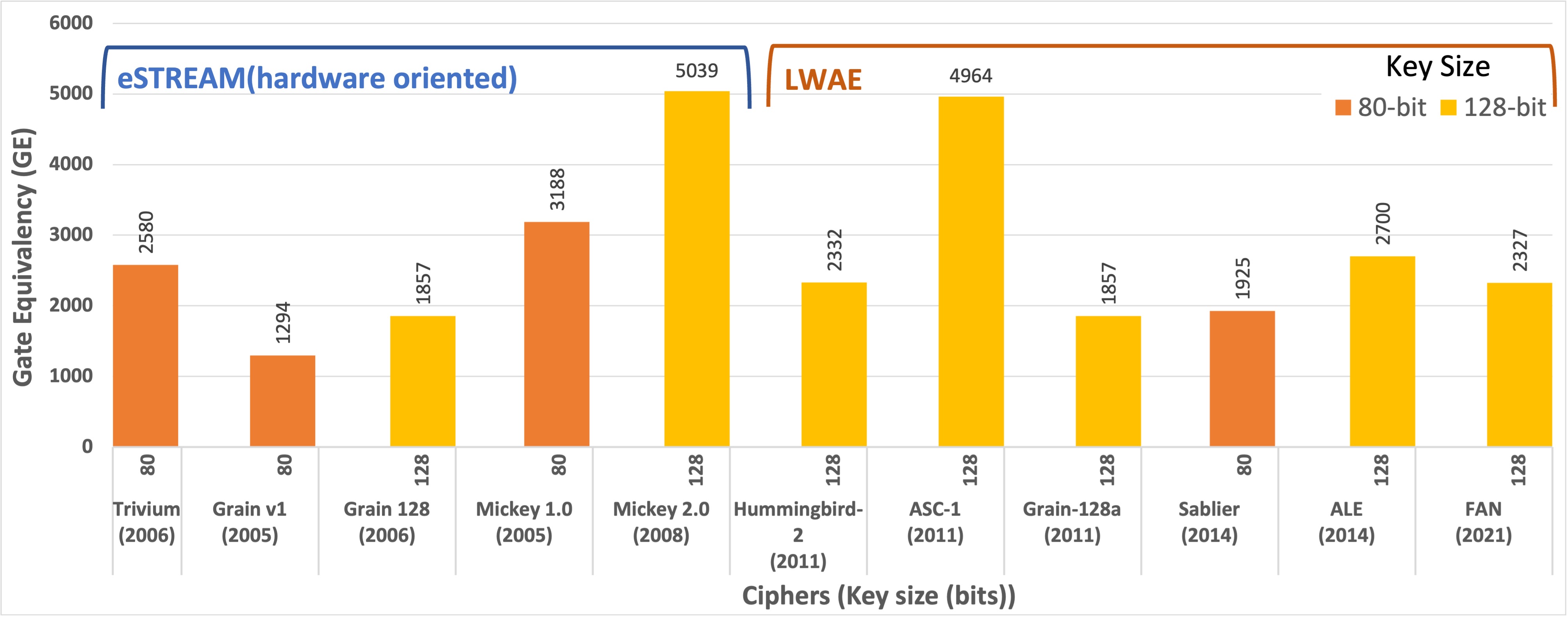}
       ~\caption{Gate Equivalency (GE) of eSTREAM and LWAE Stream Ciphers.}
        \label{fig:GE_eSTREAM_LWAE}
        \Description{The area consumption of eSTREAM and LWAE block ciphers}
    \end{figure}

    \begin{figure} [H]
        \includegraphics[width=\textwidth, height=6 cm, keepaspectratio]{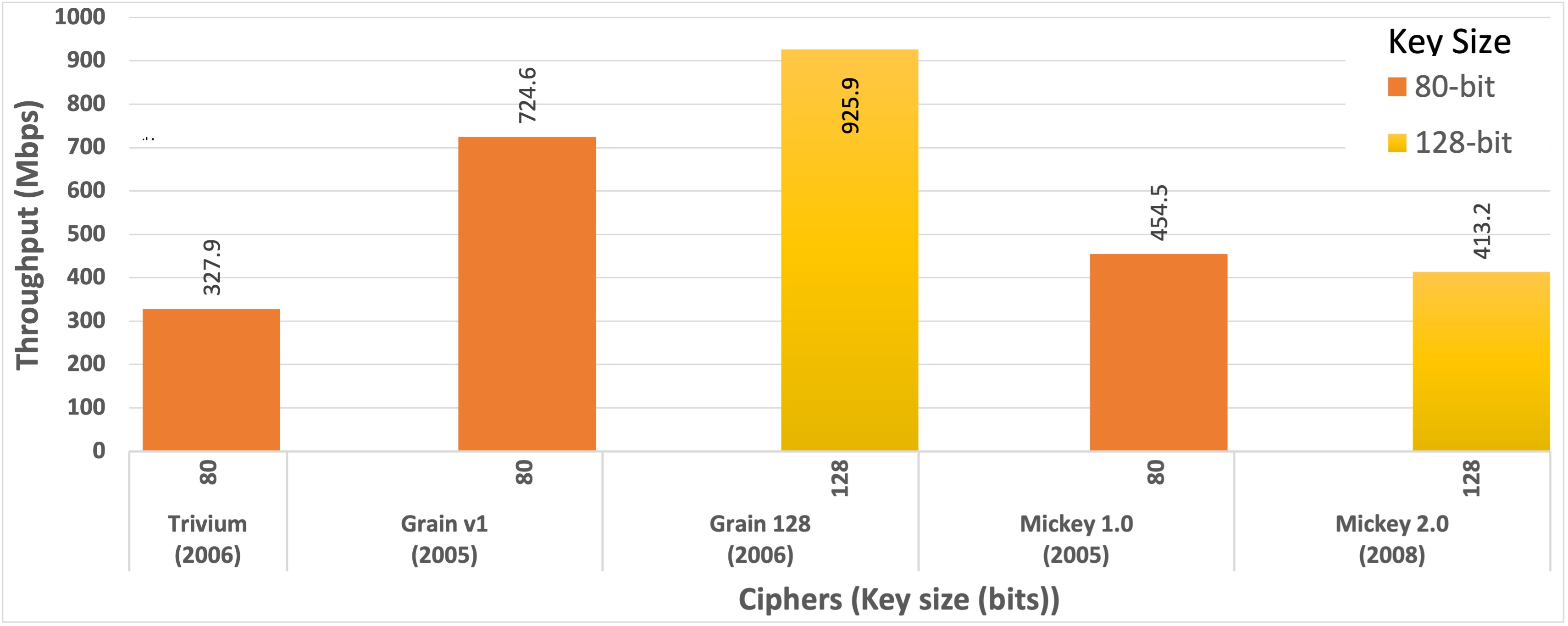}
       ~\caption{Throughput of eSTREAM Ciphers.}
        \label{fig:Throughput_eSTREAM}
        \Description{The throughput of eSTREAM ciphers}
    \end{figure}
    
    \subsection{The remaining lightweight stream ciphers}
    \begin{table} [H]
\scriptsize
\caption{Hardware Implementation of Other Lightweight Cryptographic Stream Ciphers.}
\begin{tabular*}{\textwidth}{@{\extracolsep{\fill}}l c c c c c c c c}
\toprule
\textbf{Cipher} &
  \textbf{Year} &
  \textbf{IV} &
  \begin{tabular}[c]{@{}c@{}}\textbf{Key}\\  \textbf{Size}\end{tabular} &
  \textbf{Architecture} &
  \textbf{GE} &
  \begin{tabular}[c]{@{}c@{}}\textbf{Throughput}\\  (@ 100 KHz)\end{tabular} &
  \begin{tabular}[c]{@{}c@{}}\textbf{Figure of Merit}\\  \textbf{(FoM)}\end{tabular} &
  \begin{tabular}[c]{@{}c@{}}\textbf{CMOS} \\ \textbf{Tech}\\ \textbf{($\mu$m)}\end{tabular} \\\midrule
ChaCha~\cite{hamann2017lizard}    & 2008 & 128 & 256   & ARX                  & 750  &    &   & 0.09 \\ \hline

Enocoro-80~\cite{watanabe2008enocoro} & 2008 & 64  & 80    & PRNG                 &      &    &   & 0.09 \\ \hline

BEAN~\cite{kumar2009bean}       & 2009 & 80  &       & 2FCSR+S-Box          &      &    &   &      \\ \hline

TinyStream~\cite{chen2010tinystream} & 2010 &     &       & TPM                  &      &    &   &      \\ \hline

A2U2~\cite{david2011a2u2}       & 2011 &     & 56    & 2 NFSR + LFSR        & 500  & 50 & 2000 & 0.13 \\ \hline

Quavium~\cite{tian2012quavium}    & 2012 & 80  & 80    & 4 LFSR               & 3496 &    &   & 0.09 \\ \hline

CAR30~\cite{das2013car30}      & 2013 & 120 & 128   & Cellular Automata    &      &    &   &      \\ \hline

WG-8~\cite{jattke2016comparison}       & 2013 & 81  & 80    & WG + LFSR            & 1786 & 100& 313.5  & 0.13 \\ \hline

Pandaka-16~\cite{chen2014pandaka} & 2014 &     & 96    & Shift Registers + FF & 760  &    &   &      \\
Pandaka-32~\cite{chen2014pandaka} &      &     & 192   &                      & 1520 &    &   &      \\ \hline

Sprout~\cite{armknecht2015lightweight}     & 2015 & 80  & 80    & LFSR + NLFSR         & 813  & 100& 1512.93 & 0.18 \\ \hline

Fruit-v2~\cite{ghafari2016fruit}   & 2016 & 64  & 64/80 & LFSR + NLFSR         & 990  & 100& 1020.3  & 0.09 \\ \hline

Espresso~\cite{dubrova2017espresso}   & 2017 & 96  & 128   & NLFSR                & 1497 &    &   & 0.09 \\ \hline

Lizard~\cite{hamann2017lizard}     & 2017 & 64  & 128   & NLFSR                & 1161 &    &   & 0.18 \\ \bottomrule
\end{tabular*}
\label{tab:HW_imp_OLWSC}
\end{table}
    \begin{table}[H]
\scriptsize
\caption{Software Implementation of Other Lightweight Stream Ciphers.}
\label{tab:SW_imp_OLWSC}
\begin{tabular*}{\textwidth}{@{\extracolsep{\fill}} l c c c c c c c c c c }
\toprule
\multicolumn{2}{l}{\textbf{Cipher}} &
  \begin{tabular}[c]{@{}c@{}}\textbf{IV} \\ \textbf{(bits)}\end{tabular} &
  \begin{tabular}[c]{@{}c@{}}\textbf{Key}\\  \textbf{Size}\\ \textbf{(bits)}\end{tabular} &
  \begin{tabular}[c]{@{}c@{}}\textbf{Code}\\  \textbf{Size}\\ \textbf{(bytes)}\end{tabular} &
  \begin{tabular}[c]{@{}c@{}} \textbf{SRAM} \\ \textbf{(bytes)}\end{tabular} &
  \begin{tabular}[c]{@{}c@{}} \textbf{Encryption} \\ \textbf{(cycles)}\end{tabular} &
  \begin{tabular}[c]{@{}c@{}}\textbf{Throughput}\\  \textbf{(Kbps)} \end{tabular} &
  \begin{tabular}[c]{@{}c@{}}\textbf{Energy} \\  \textbf{($\mu$J/bit)}\end{tabular} &
  \begin{tabular}[c]{@{}c@{}} \textbf{RANK} \end{tabular} &
  \begin{tabular}[c]{@{}c@{}}\textbf{Hardware} \\ \textbf{Config} \end{tabular}
                      
\\ 
\midrule
Salsa20        & &64    & 128 & 3842   & 256  & 318      & 83.7   & 101.56  & 0.72224 &  \multirow{8}{*}{\rotatebox{90}{\makecell{Atmel AVR \\ ATMega-128L \\~\cite{fan2013wg}}}}  \\       \cline{1-10}
              
Trivium       & &80    & 80  & 424    & 36   & 775726    & 12     & 7.07   & 0.0026 &   \\       \cline{1-10}

Grain         & &64    & 80  & 778    & 20   & 107336    & 12.9   & 6.57   & 0.01139 &   \\       \cline{1-10}
              
WG-7          & &81    & 80 & 938     &      & 20917     & 34     & 2.5    &         &  \\       \cline{1-10}

WG-8          & &81    & 80 & 1984    & 20   & 1379      & 185.5  & 0.458  & 0.35828 &  \\       \cline{1-10}

Hummingbird-1 & &      & 256& 1308    &      & 14735     & 34.9   & 2.433  &         &   \\       \cline{1-10}

Hummingbird-2 & &      & 128& 3600    & 114  & 2970      & 171.8  & 0.495  & 0.08796    &   \\       
\bottomrule
\end{tabular*}
\end{table}
    \begin{figure} [H]
        \includegraphics[width=14cm, height=6 cm, keepaspectratio]{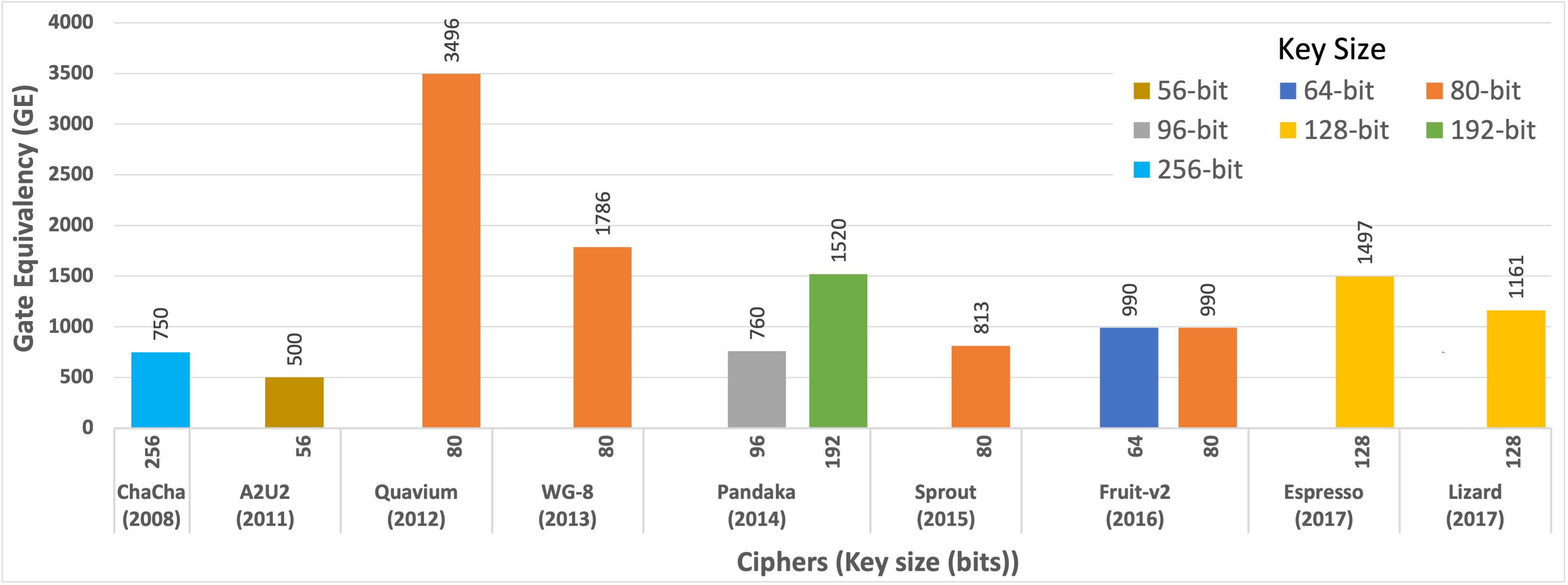}
       ~\caption{Gate Equivalency (GE) of Other Lightweight Stream Ciphers.}
        \label{fig:GE_OLWSC}
        \Description{The area consumption of other lightweight stream ciphers}
    \end{figure}
    \begin{figure} [H]
        \includegraphics[width=\textwidth, height=6 cm, keepaspectratio]{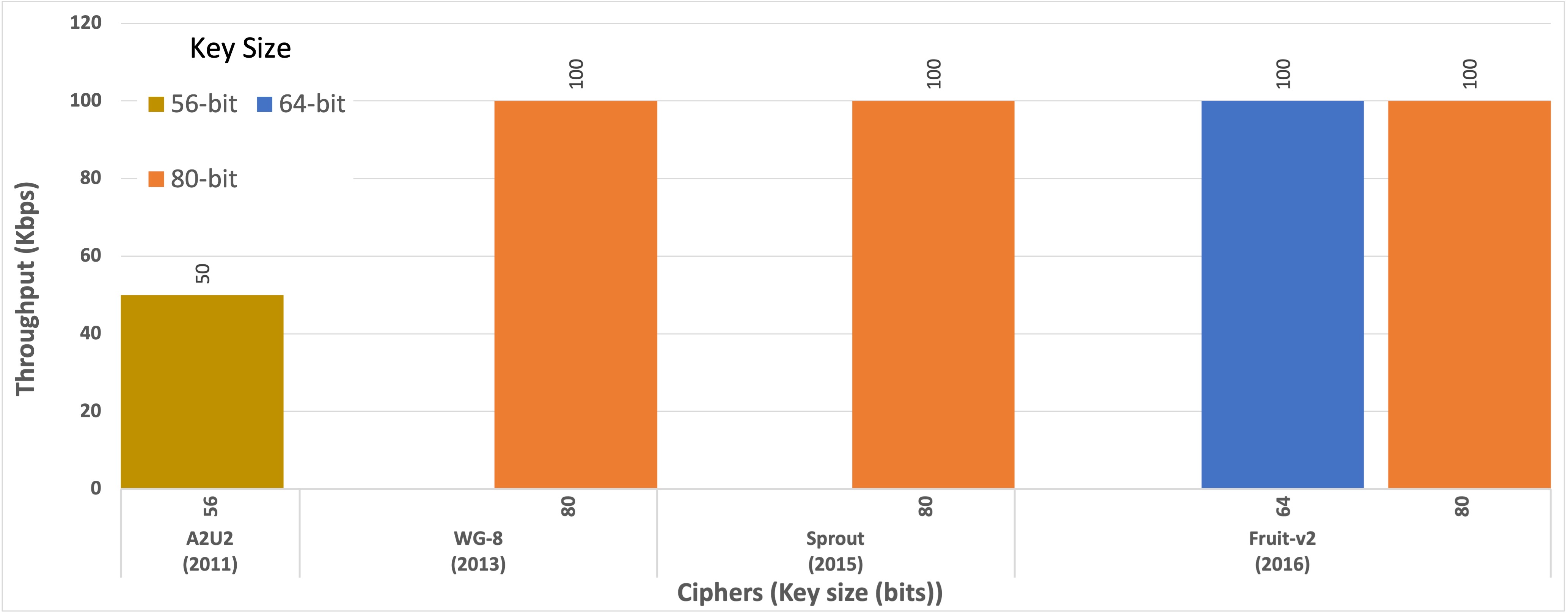}
       ~\caption{Throughput of Other Lightweight Stream Ciphers.}
        \label{fig:Throughput_OLWSC}
        \Description{The throughput of other lightweight stream ciphers}
    \end{figure}
\section{Lightweight Hybrid Ciphers}
    
    \begin{table} [H]
\scriptsize
\caption{Software Implementation of Lightweight Hybrid Ciphers (LHC).}
\label{tab:SW_imp_LHC}
\begin{tabular*}{\textwidth}{@{\extracolsep{\fill}} l c c c c c c c c }
\toprule
\multicolumn{2}{l}{\textbf{Cipher}} &
  \begin{tabular}[c]{@{}c@{}}\textbf{Block}\\  \textbf{Length}\\ \textbf{(bits)}\end{tabular} &
  \begin{tabular}[c]{@{}c@{}}\textbf{Key}\\  \textbf{Size}\\ \textbf{(bits)}\end{tabular} &
  \begin{tabular}[c]{@{}c@{}}\textbf{Code}\\  \textbf{Size}\\ \textbf{(bytes)}\end{tabular} &
  \begin{tabular}[c]{@{}c@{}} \textbf{RAM} \\ \textbf{(bytes)}\end{tabular} &
  \begin{tabular}[c]{@{}c@{}} \textbf{Encryption} \\ \textbf{(cycles)}\end{tabular} &
  \begin{tabular}[c]{@{}c@{}}\textbf{Throughput}\\  \textbf{(Kbps)} \end{tabular} &
  \begin{tabular}[c]{@{}c@{}}\textbf{Hardware} \\ \textbf{Config} \end{tabular}                       
\\ 
\midrule
Hummingbird-1  & &16 & 256 & 3680 &    & 3.664  & 17.5 &  \multirow{2}{*}{\makecell{Atmel AVR \\ ATMega-128~\cite{vahi2020separ}}}  \\ \cline{1-8}

SEPAR         & &16 & 256  & 3860 & 32 & 9.761  & 1.3   \\       

\bottomrule
\end{tabular*}
\end{table}


\section{PRIME and BINARY Field Implementations}
    \begin{table} [H]
\scriptsize
\caption{Hardware implementation of LWECC Algorithms.}
\begin{tabular*}{\textwidth}{@{\extracolsep{\fill}}l l c c c c c c c c c}
\toprule
   & \multicolumn{2}{l}{\textbf{Publication}}  &
  \textbf{Year} &
  \begin{tabular}[c]{@{}c@{}}\textbf{Curve Type}\end{tabular} &
  \begin{tabular}[c]{@{}c@{}}\textbf{Field Size} \\ \textbf{(bits)}\end{tabular} &
  \begin{tabular}[c]{@{}c@{}}\textbf{Security} \\ \textbf{(bits)}\end{tabular} &
  \begin{tabular}[c]{@{}c@{}}\textbf{GE}\end{tabular} &
  \begin{tabular}[c]{@{}c@{}}\textbf{Cycles} \end{tabular} &
  \begin{tabular}[c]{@{}c@{}}\textbf{Frequency}\\ \textbf{(MHz)}\end{tabular} &
  \begin{tabular}[c]{@{}c@{}}\textbf{CMOS} \\ \textbf{Tech}\\ \textbf{($\mu$m)}\end{tabular} \\ \midrule

\multicolumn{11}{c}{\textbf{PRIME FIELD}}\\ \hline
\multirow{5}{*}{\rotatebox{90}{\makecell{Weierstrass \\ Curve}}} & P1~\cite{wenger2010low}  &  & 2010 & secp192r1  & 192 & 96 & 11686 & 1377000 &  & 0.18 \\ \cline{2-11}

& P3~\cite{wenger2013lightweight} &  & 2013 & secp160r1 & 160 & 80 & 22537 & 1298000 & 13.56 &0.13 \\  
&  &  &  & secp192r1 & 192   & 96 & 23835 & 1813000 &  & \\
&  &  &  & secp224r1 & 224   & 112 & 25113 & 2469000 &  & \\
&  &  &  & secp224r1 & 224   & 128 & 27244 & 3367000 &  & \\ \cline{2-11}

& P4~\cite{yalccin2016compact} &  & 2016 & secp256r1 & 256 & 128 & 11727 & 6180856 & 16 &0.13 \\

\hline
\multirow{2}{*}{\rotatebox{90}{\makecell{Mont. \\ Curve}}} & P2~\cite{wenger20128}  &  & 2012 &  & 160 & 80 & 20980 & 1300000 &  & 0.13 \\ \cline{2-11}

& P6~\cite{yalccin2016compact} &  & 2016 &  & 160 & 80 & 11366 & 490000 & 10 &0.18 \\  

\\ \hline
\multicolumn{11}{c}{\textbf{BINARY FIELD}}\\ \hline
\multirow{13}{*}{\rotatebox{90}{\makecell{Weierstrass \\ Curve}}} & 
B1~\cite{kumar2006standards}  &  & 2006 & sect113r2 & 113 & 56 & 10112 & 195159 & 13.56 & 0.35 \\ 
&  &  &  & sect131r2 & 131   & 64 & 11969 & 244192 &  & \\
&  &  &  & sect163r2 & 163   & 80 & 15094 & 430654 &  & \\
&  &  &  & sect193r2 & 193   & 96 & 17723 & 564919 &  & \\ \cline{2-11}

& B10~\cite{wenger2013hardware} &  & 2013 & sect163r2 & 163 & 80 & 15282 & 467370 & 1 &0.13 \\ 
&  &  &  &  &    &  & 16121 & 303202 &  & \\
&  &  &  &  &    &  & 16738 & 224222 &  & \\
&  &  &  &  &    &  & 17986  & 182130 &  & \\ \cline{2-11}

& B3~\cite{holler2014hardware} &  & 2013 & sect163r2 & 163 & 80 & 7510 & 12100000 & 13.56 & 0.22 \\ 
&  &  &  &  &    &  & 8920 & 9400000 &  & \\
&  &  &  &  &    &  & 10550 & 7000000 &  & \\
&  &  &  &  &    &  & 10290 & 5100000 &  & \\ 
&  &  &  &  &    &  & 6180 & 2800000 &  & \\ 
\hline

\multirow{6}{*}{\rotatebox{90}{\makecell{Koblitz \\ Curve}}} & 
B2~\cite{azarderakhsh2014efficient}  &  & 2014 & sect163k1 & 163 & 80 & 11571 & 106700 & 1 & 0.065 \\ \cline{2-11}

& B4~\cite{sinha2015lightweight} &  & 2015 & sect283k1 & 283 & 128 & 10204 & 1566000 & 16 & 0.13 \\ \cline{2-11}

& B6~\cite{rovzic20175} &  & 2016 & sect163k1 & 163 & 80 & 10,106 &  & 1.13 & 0.13 \\ 
&  &  &  &  &    &  & 11,383 &  & 0.59 & \\
&  &  &  &  &    &  & 12,236 &  & 0.41 & \\
&  &  &  &  &    &  & 12,863 &  & 0.32 & \\ 
\hline

\multirow{6}{*}{\rotatebox{90}{\makecell{Edwards \\ Curve}}} & 
B5~\cite{koziel2015low}  &  & 2015 &  & 163 & 80 & 11219 & 177707 & & 0.065 \\ 
&  &  &  &  & 233   & 112 & 15177 & 351856 &  & \\
&  &  &  &  & 283   & 128 & 19332 & 512555 &  & \\ \cline{2-11}

& B7~\cite{lara2020lightweight} &  & 2019 & & 163 & 80 & 8,083 & 351591 & 104 & XC6SLX \\ 
&  &  &  &  & 233 & 112 & 12149 & 718805 &  & \\
&  &  &  &  & 251 & 126 & 12240 & 824284 &  & \\ 
\bottomrule
\end{tabular*}
\label{tab:HW_imp_LWECC}
\end{table}
    
    \begin{figure} [H]
        \includegraphics[width=11.5cm, height=6 cm, keepaspectratio]{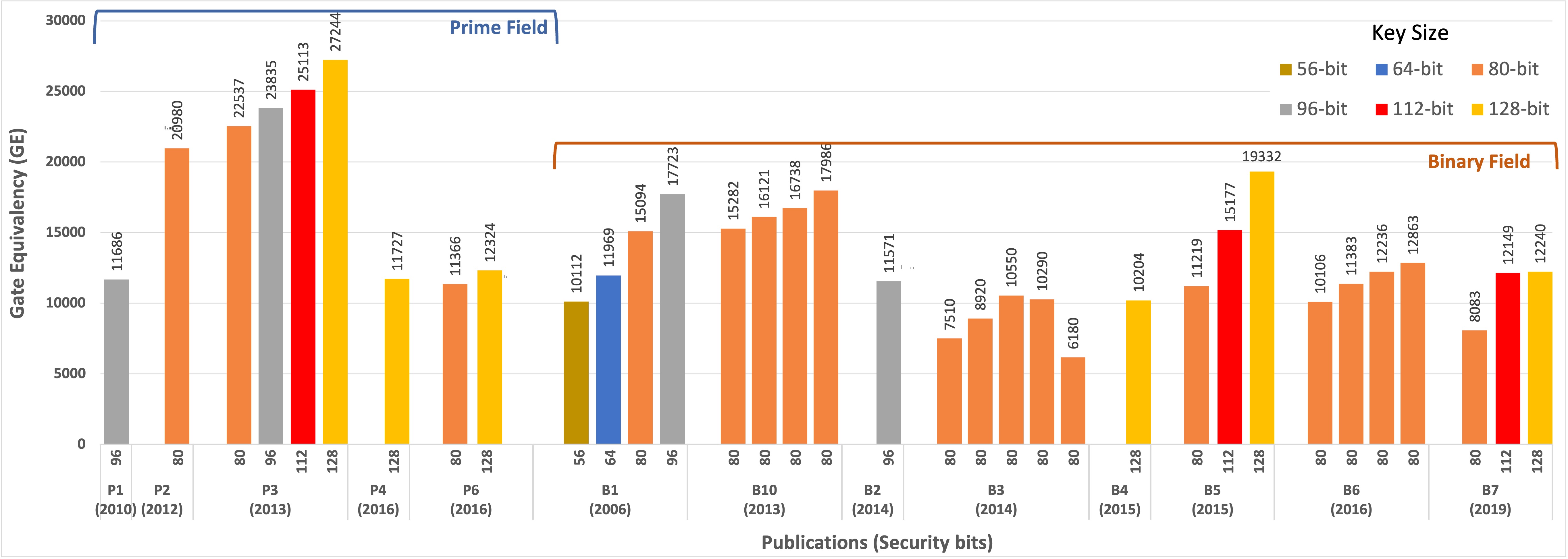}
       ~\caption{Gate Equivalency of LWECC implementations}
        \label{fig:GE_LWECC}
        \Description{The area consumption of LWECC}
    \end{figure}

    \begin{table} [H]
\scriptsize
\caption{Software Implementation of LWECC Implementations.}
\begin{tabular*}{\textwidth}{@{\extracolsep{\fill}}l l c c c c c c c c c c}
\toprule
   & \multicolumn{2}{l}{\textbf{Publication}}  &
  \textbf{Year} &
  \begin{tabular}[c]{@{}c@{}}\textbf{Curve Type}\end{tabular} &
  \begin{tabular}[c]{@{}c@{}}\textbf{Field Size} \\ \textbf{(bits)}\end{tabular} &
  \begin{tabular}[c]{@{}c@{}}\textbf{Security} \\ \textbf{(bits)}\end{tabular} &
  \begin{tabular}[c]{@{}c@{}}\textbf{ROM}\\\textbf{(bytes)}\end{tabular} &
  \begin{tabular}[c]{@{}c@{}}\textbf{RAM}\\\textbf{(bytes)}\end{tabular} &
  \begin{tabular}[c]{@{}c@{}}\textbf{Cycles} \end{tabular} &
  \begin{tabular}[c]{@{}c@{}}\textbf{Frequency}\\ \textbf{(MHz)}\end{tabular} &
  \begin{tabular}[c]{@{}c@{}}\textbf{Hardware} \\ \textbf{Config}\end{tabular} \\ \midrule

\multicolumn{11}{c}{\textbf{PRIME FIELD}}\\ \hline
\multirow{5}{*}{\rotatebox{90}{\makecell{Montgomery \\ Curve}}} & P2~\cite{wenger20128}  &  & 2012 &  & 160 & 80 & 6824 &  & 5545078 & 8 & \multirow{9}{*}{\rotatebox{90}{\makecell{Atmel AVR \\ ATMega-128}}} \\ \cline{2-11}

& P5~\cite{liu2016efficient} &  & 2016 &  & 160  & 80 &  &  & 5527000 & 8 & \\  
&   &  &  &  & 192  & 96 &  &  & 8837000 & 8 & \\ 
&   &  &  &  & 224  & 112 &  &  & 13,255,000 & 8 & \\ 
&   &  &  &  & 256  & 128 &  &  & 18,893,000 & 8 & \\ 

\cline{1-11}
\multirow{4}{*}{\rotatebox{90}{\makecell{MoTE \\ Curve}}} & P7~\cite{liu2016emerging}  &  & 2016 & MoTE P159 & 159 & 80 & 12100 & 446 & 5468000 & 8 &\\ 
&   &  &  & MoTE P191 & 191 & 96 & 12100 & 446 & 8659000 & 8 & \\
&   &  &  & MoTE P223 & 223 & 112 & 12100 & 446 & 12879000 & 8 & \\
&   &  &  & MoTE P255 & 255 & 128 & 12100 & 446 & 18270000 & 8 & \\
\hline

\multirow{4}{*}{\rotatebox{90}{\makecell{Weierstrass \\ Curve}}} & P14~\cite{wenger2013hardware}  &  & 2013 & secp160r1 & 160 & 80 & 4230 & 282 & 5721420 & 1 & MSP430\\ 
&   &  &  & secp192r1 & 192 & 96 & 4846 & 322 & 9100128 & 1 & \\
\cline{2-12}

& P8~\cite{mathur2017secure} &  & 2017 & secp256r1 & 256  &  &  &  & 10912000 & 16 & OpenMote\\ \cline{2-12} 

& P10~\cite{cabana2021low} &  & 2021 & secp192r1 & 192  & 96 &  &  & 2189752 & 6 & igloo AGLE300\\ \hline

Koblitz & P11~\cite{thaparimplementation} &  & 2022 & secp160k1 & 160  & 80 &  &  & 10810680 & 16 & MSP430F5438A\\ \hline

\multirow{4}{*}{\rotatebox{90}{\makecell{Custom \\ (ShortECC)}}} & P9~\cite{sojka2017shortening} &  & 2017 &  & 32  & 16 &  &  & 409120 & 25 & \multirow{3}{*}{\rotatebox{90}{\makecell{MSP430}}}\\ 
 & &  &  &  & 128  & 64 &  &  & 651662120 & 25 & \\ 
 & &  &  &  & 192  & 96 &  &  & 2078311103 & 25 & \\

\\ \hline
\multicolumn{11}{c}{\textbf{BINARY FIELD}}\\ \hline
\multirow{5}{*}{\rotatebox{90}{\makecell{Weierstrass \\ Curve}}} & 
B8~\cite{szczechowiak2008nanoecc}   &  & 2008 & sect163k1 & 163  & 80 & 32100 & 2800 & 8519000 & 8 & \multirow{5}{*}{\rotatebox{90}{\makecell{MSP430 }}}\\ \cline{2-11}

& B9~\cite{gouvea2012efficient}   &  & 2012 & sect163k1 & 163  & 80 & 27800 & 3600 & 2032000 & 8 & \\ \cline{2-11}

& B10~\cite{wenger2013hardware}   &  & 2013 & sect163r2 & 163  & 80 & 4126 & 294 & 7446677 & 1 & \\
&  &  &  & c2tnb191v1 & 191  &  & 3994 & 310 & 8610906 & 96 & \\ \cline{2-12}

& B3~\cite{holler2014hardware}   &  & 2014 & sect163r2 & 163  & 80 & 3050 & 423 & 9700000 &  & 8-bit  Harvard architecture\\

\bottomrule
\end{tabular*}
\label{tab:SW_imp_LWECC}
\end{table}

\end{document}


\appendix

\section{Lightweight Block Ciphers}
The results of the software and hardware implementation obtained from the respective research publications are tabulated in the appendix, accompanied by corresponding bar graphs. 

    \subsection{Substitution Permutation Network (SPN)}

        \input{Tables/02. Hardware Implementation Tables/01. HW_Metrics_SPN_Ciphers}
        \input{Tables/03. Software Implementation Tables/01. SW_Metrics_SPN_Ciphers}

    \begin{figure}
        \includegraphics[width=\textwidth, height=10 cm, keepaspectratio]{Figures/01. Gate Equivalency/01. GE_SPN.jpg}
       ~\caption{Gate Equivalency (GE) of SPN Block Ciphers.}
        \label{fig:GE_SPN}
        \Description{The area consumption of SPN block ciphers}
    \end{figure}

    \begin{figure}
        \includegraphics[width=\textwidth, height=10 cm, keepaspectratio]{Figures/02. Throughput/01. TP_SPN.jpg}
       ~\caption{Throughput of SPN block ciphers}
        \label{fig:Throughput_SPN}
        \Description{Throughput of SPN block ciphers}
    \end{figure}

    \subsection{Feistel Network (FN)}

    \input{Tables/02. Hardware Implementation Tables/02. HW_Metrics_FN_Ciphers}

    In the software implementation performed on Atmel AVR ATtiny45, the code size is specified, but not the RAM. The authors of these works claim that all internal variables were saved in CPU registers, and SRAM was only used to hold plaintext, ciphertext, and the master key. Thus, we omit RAM usage when calculating RANK on this platform.

    \input{Tables/03. Software Implementation Tables/02. SW_Metrics_FN_Ciphers}

    \begin{figure} [H]
        \includegraphics[width=\textwidth, height=11 cm, keepaspectratio]{Figures/01. Gate Equivalency/02. GE_FN.jpg}
       ~\caption{Gate Equivalency (GE) of FN Block Ciphers.}
        \label{fig:GE_FN}
        \Description{The area consumption of FN block ciphers}
    \end{figure}
    
    \begin{figure} [H]
        \includegraphics[width=\textwidth, height=11 cm, keepaspectratio]{Figures/02. Throughput/02. TP_FN.jpg}
       ~\caption{Throughput of FN Block Ciphers.}
        \label{fig:Throughput_FN}
        \Description{Throughput of FN block ciphers}
    \end{figure}
    
\newpage
    \subsection{Generalized Feistel Network (GFN)}

    \input{Tables/02. Hardware Implementation Tables/03. HW_Metrics_GFN_LMD_Ciphers}
    \input{Tables/03. Software Implementation Tables/03. SW_Metrics_GFN_LMD_Ciphers}

    \begin{figure} [H]
        \includegraphics[width=\textwidth, height=12 cm, keepaspectratio]{Figures/01. Gate Equivalency/03. GE_GFN_LMD.jpg}
       ~\caption{Gate Equivalency of GFN and LMD Block Ciphers.}
        \label{fig:GE_GFN_LMD}
        \Description{The area consumption of GFN and LMD block ciphers}
    \end{figure}

    \begin{figure} [H]
        \includegraphics[width=\textwidth, height=11 cm, keepaspectratio]{Figures/02. Throughput/03. TP_GFN_LMD.jpg}
       ~\caption{Throughput of GFN and LMD Block Ciphers}
        \label{fig:Throughput_GFN_LMD}
        \Description{Throughput of GFN and LMD block ciphers}
    \end{figure}
    
\newpage
    \subsection{Addition Rotation eXclusive-or- (ARX)}

    \input{Tables/02. Hardware Implementation Tables/04. HW_Metrics_ARX_Ciphers}
    \input{Tables/03. Software Implementation Tables/04. SW_Metrics_ARX_Ciphers}

    \begin{figure} [H]
        \includegraphics[width=14.3 cm, height=8 cm, keepaspectratio]{Figures/01. Gate Equivalency/04. GE_ARX.jpg}
       ~\caption{Gate Equivalency (GE) of ARX Block Ciphers.}
        \label{fig:GE_ARX}
        \Description{The area consumption of ARX block ciphers}
    \end{figure}

    \begin{figure} [H]
        \includegraphics[width=\textwidth, height=10 cm, keepaspectratio]{Figures/02. Throughput/04. TP_ARX.jpg}
       ~\caption{Throughput of ARX Block Ciphers.}
        \label{fig:Throughput_ARX}
        \Description{Throughput of GFN block ciphers}
    \end{figure}
    
\section{Lightweight Stream Ciphers}

    \subsection{eSTREAM ciphers}

    \input{Tables/02. Hardware Implementation Tables/05. HW_Metrics_eSTREAM_Ciphers}
    \input{Tables/03. Software Implementation Tables/05. SW_Metrics_eSTREAM_Ciphers}

    \subsection{LWAE ciphers}

    \input{Tables/02. Hardware Implementation Tables/06. HW_Metrics_LWAE_Ciphers}

    \begin{figure} [H]
        \includegraphics[width=\textwidth, height=12 cm, keepaspectratio]{Figures/01. Gate Equivalency/05. GE_eSTREAM_LWAE.jpg}
       ~\caption{Gate Equivalency (GE) of eSTREAM and LWAE Stream Ciphers.}
        \label{fig:GE_eSTREAM_LWAE}
        \Description{The area consumption of eSTREAM and LWAE block ciphers}
    \end{figure}

    \begin{figure} [H]
        \includegraphics[width=\textwidth, height=6 cm, keepaspectratio]{Figures/02. Throughput/05. TP_eSTREAM.jpg}
       ~\caption{Throughput of eSTREAM Ciphers.}
        \label{fig:Throughput_eSTREAM}
        \Description{The throughput of eSTREAM ciphers}
    \end{figure}
    
    \subsection{The remaining lightweight stream ciphers}
    \input{Tables/02. Hardware Implementation Tables/07. HW_Metrics_OLWSC_Ciphers}
    \input{Tables/03. Software Implementation Tables/06. SW_Metrics_OLWSC_Ciphers}
    \begin{figure} [H]
        \includegraphics[width=14cm, height=6 cm, keepaspectratio]{Figures/01. Gate Equivalency/06. GE_OLWSC.jpg}
       ~\caption{Gate Equivalency (GE) of Other Lightweight Stream Ciphers.}
        \label{fig:GE_OLWSC}
        \Description{The area consumption of other lightweight stream ciphers}
    \end{figure}
    \begin{figure} [H]
        \includegraphics[width=\textwidth, height=6 cm, keepaspectratio]{Figures/02. Throughput/06. TP_OLWSC.jpg}
       ~\caption{Throughput of Other Lightweight Stream Ciphers.}
        \label{fig:Throughput_OLWSC}
        \Description{The throughput of other lightweight stream ciphers}
    \end{figure}
\section{Lightweight Hybrid Ciphers}
    
    \input{Tables/03. Software Implementation Tables/07. SW_Metrics_LHC}


\section{PRIME and BINARY Field Implementations}
    \input{Tables/02. Hardware Implementation Tables/08. HW_metrics_ECC}
    
    \begin{figure} [H]
        \includegraphics[width=11.5cm, height=6 cm, keepaspectratio]{Figures/01. Gate Equivalency/07. GE_LWECC.jpg}
       ~\caption{Gate Equivalency of LWECC implementations}
        \label{fig:GE_LWECC}
        \Description{The area consumption of LWECC}
    \end{figure}

    \input{Tables/03. Software Implementation Tables/08. SW_Metrics_ECC}